\documentclass[a4paper,11pt]{article}
\usepackage{jheppub}
\makeatletter\gdef\@fpheader{}\makeatother

\usepackage{booktabs}
\usepackage{array}
\usepackage{xspace}
\usepackage{placeins}
\usepackage{tikz}
\usetikzlibrary{arrows.meta}

\newcommand{\pT}{p_{\mathrm{T}}}
\newcommand{\cth}{\cos\theta^{*}}
\newcommand{\acth}{|\cos\theta^{*}|}
\newcommand{\GeV}{\,\text{GeV}}
\newcommand{\TeV}{\,\text{TeV}}
\newcommand{\pb}{\,\text{pb}}
\newcommand{\fb}{\,\text{fb}}
\newcommand{\nb}{\,\text{nb}}
\newcommand{\fL}{f_{\mathrm{L}}}
\newcommand{\epsL}{\varepsilon_{\mathrm{L}}}
\newcommand{\epsT}{\varepsilon_{\mathrm{T}}}
\newcommand{\WL}{W_{\mathrm{L}}}
\newcommand{\WT}{W_{\mathrm{T}}}
\newcommand{\cmax}{c_{\max}}
\newcommand{\NL}{N_{\mathrm{L}}}
\newcommand{\NT}{N_{\mathrm{T}}}
\newcommand{\MGaMC}{\textsc{MadGraph5\_\allowbreak aMC@NLO}\xspace}
\newcommand{\Pythia}{\textsc{Pythia}\,8\xspace}
\newcommand{\Delphes}{\textsc{Delphes}\xspace}
\newcommand{\FastJet}{\textsc{FastJet}\xspace}
\newcommand{\OL}{\textsc{OmniLearned}\xspace}
\newcommand{\rung}[1]{\texttt{#1}}

\title{Pretraining on boosted boson decays for polarization tagging of hadronic $W$ bosons}

\author{Saptaparna Bhattacharya}
\affiliation{Department of Physics and Astronomy, Southern Methodist University, Dallas, Texas, USA}
\emailAdd{saptaparnab@smu.edu}

\abstract{%
Longitudinally polarized weak bosons probe the unitarization of vector-boson scattering. At large boosts, longitudinal and transverse $W$ bosons produce jets with the same mass and two-prong structure, distinguished only by the decay angle. Longitudinal bosons are rare in $W$+jet events at these large boosts. We ask where the polarization information of a boosted $W$ boson lies and how much a tagger can recover. Requiring both decay quarks within the jet truncates the decay-angle distribution at an edge set by $m/(\pT R)$, which, as we show, bounds any hadronic polarization tagger. Within this acceptance, the decay angle carries nearly all the polarization information of a jet in $W$+jet events. Whether pretraining retains the decay angle is not obvious, so we pretrain a point-edge transformer on $10^7$ unpolarized boosted-boson, top-quark and QCD jets and fine-tune it with $10^3$ to $5\times10^4$ labeled jets per class. With the fewest labels, a boosted decision tree on substructure observables performs best; with the most labels, the strongest networks come within 0.003 in AUC of the ideal decay-angle tagger. Pretraining saves a factor of about two to three in labels, and our boson pretraining improves on a generic one by a few thousandths in AUC, more with frozen transformer blocks. Student networks with a few hundred times fewer parameters, distilled with unlabeled jets, match their teacher. More information lies outside the jet. The production angle, reconstructed with the recoiling jet, improves every tagger. In diboson production and vector-boson scattering at matrix-element level, the production kinematics carry more information than the decay angles of both bosons. Trading the generality of a jet tagger for an event-level tagger trained on the process thus brings large gains, which measurements of longitudinal vector-boson scattering could exploit.}


\begin{document}
\maketitle
\flushbottom
\addtocontents{toc}{\protect\enlargethispage{2\baselineskip}}

\section{Introduction}
\label{sec:intro}

At energies well above the electroweak scale, the longitudinal polarization states of the $W$ and $Z$ bosons are equivalent to the Goldstone bosons of electroweak symmetry breaking~\cite{Cornwall:1974km,Lee:1977eg,Chanowitz:1985hj}. In the Standard Model (SM), the Higgs boson cancels the energy-dependent growth of longitudinal vector-boson scattering amplitudes, and any deviation in its couplings to gauge bosons spoils the cancellation in exactly these amplitudes. The polarization of weak bosons in diboson production and vector-boson scattering (VBS) is therefore a direct probe of how these amplitudes are unitarized. The polarization of $W$ bosons at large transverse momentum was first measured in $W$+jets events~\cite{CMS:2011kaj,ATLAS:2012au}. Since then, joint polarization states have been observed in $W^\pm Z$ production~\cite{ATLAS:2022oge}, and evidence has been reported for pairs of longitudinally polarized bosons in $ZZ\to4\ell$ events~\cite{ATLAS:2023zrv}. Polarized cross sections have also been measured in same-sign $W^\pm W^\pm jj$ production~\cite{CMS:2020etf}, where ATLAS found evidence for the production of at least one longitudinally polarized $W$ boson with an observed significance of 3.3 standard deviations~\cite{ATLAS:2025ssWWpol}. All of these measurements use leptonic decays.

Hadronic decays carry about two thirds of the $W$ and $Z$ branching fractions, but at large boost, the two decay quarks merge into a single large-radius jet. The polarization then enters only through the decay angle $\theta^*$, the polar angle of a decay quark in the boson rest frame, measured from the direction of flight of the boson. For a boosted boson, this angle controls how the energy is shared between the two prongs~\cite{De:2020iwq}. Otherwise, a longitudinal and a transverse $W$ boson produce jets with the same mass, the same two-prong structure and very similar constituent multiplicities. Classical substructure observables~\cite{Larkoski:2017jix,Thaler:2010tr,Larkoski:2014gra,Larkoski:2014wba} capture part of this information, while networks acting on the full set of jet constituents, such as the Particle Transformer~\cite{Qu:2022mxj}, can in principle use all of it. A polarization-aware tagger based on Particle Transformers has recently been presented by ATLAS~\cite{Giannini:2026cometa}, and decorrelation techniques have been proposed to prevent a $W/Z$ selection from sculpting the angle used in the measurement~\cite{BOOST2026:decorrelation}.

Such taggers are trained on labeled simulations, but longitudinal bosons are rare in nature: in $W^+$+jet production with a jet $\pT$ above 450\GeV, polarized matrix elements~\cite{BuarqueFranzosi:2019boy} assign 0.0045\pb{} of the leading-order cross section to longitudinal $W$ bosons, compared to 1.62\pb{} for transverse ones, with the helicity defined in the partonic center-of-mass frame. In the laboratory frame, the leading-order fraction is about 0.9\% at 450--550\GeV{} (section~\ref{sec:frame}), and next-to-next-to-leading-order predictions are two to three times larger than leading-order ones~\cite{Gao:2026polEEC}, so that longitudinal bosons make up at most a few percent of these events. With polarized matrix elements, balanced training samples can be generated in simulation. However, a tagger trained on one production process must transfer to others, such as VBS, where data available for calibration are scarce. This makes polarization tagging a natural test for jet foundation models, which are pretrained on large, diverse jet samples and then adapted to downstream tasks with few labels~\cite{Mikuni:2024qsr,Bhimji:2025omnilearned,Birk:2024knn,Golling:2024abg,Vigl:2024ftm,Elsharkawy:2026pretraining}. Whether a pretrained representation retains the decay angle of a boosted boson is not obvious: a classification objective rewards the identification of jet types, for which the decay angle is a nuisance variable, whereas a generative objective has to reproduce how the energy is shared within the jet. Fine-tuning on few labels tests this directly: the fewer the labels, the more the tagger has to rely on what the pretraining has retained.

A foundation model is also expensive to run. Knowledge distillation~\cite{Hinton:2015dark} trains a small student network to reproduce the class probabilities of a large teacher rather than the labels alone; in jet tagging at the LHC, it has been used to bring the accuracy of large networks to models fast enough for real-time selection~\cite{Liu:2023kd}. Distillation offers two further benefits here. First, a teacher fine-tuned on few labels can transfer its knowledge using unlabeled jets, similar to self-training~\cite{Xie:2019noisy}; this is advantageous because polarization labels exist only in simulation, whereas unlabeled jets are plentiful. Second, distilling the teacher into a student restricted to a few jet observables shows how much of the teacher's learned representation those observables capture.

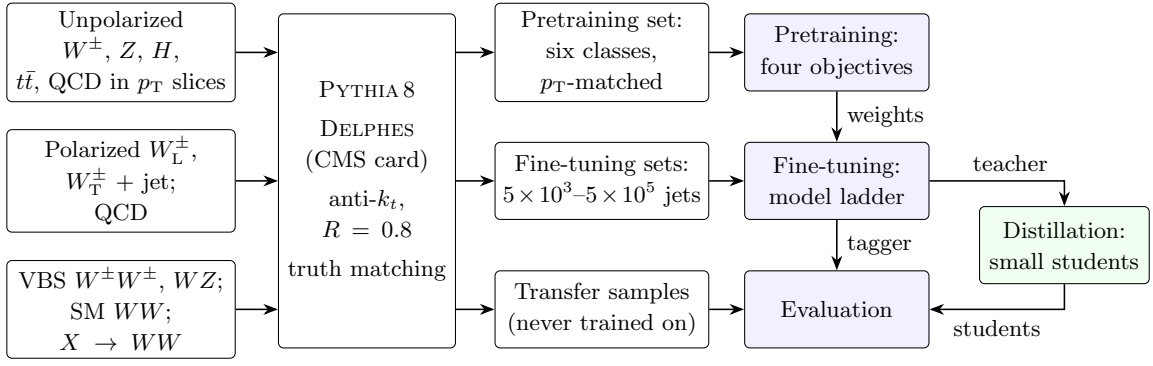
\begin{figure}[t]
\centering
\resizebox{\textwidth}{!}{%
\begin{tikzpicture}[
  box/.style={draw, rounded corners=2pt, align=center, font=\footnotesize, inner sep=3pt, minimum height=1.05cm},
  arr/.style={-{Stealth[length=2mm]}, semithick}]
\node[box, text width=2.9cm] (pt)  at (0, 1.75) {Unpolarized $W^\pm$, $Z$, $H$,\\ $t\bar t$, QCD in $\pT$ slices};
\node[box, text width=2.9cm] (pol) at (0, 0) {Polarized $W^\pm_{\mathrm{L}}$, $W^\pm_{\mathrm{T}}$ + jet;\\ QCD};
\node[box, text width=2.9cm] (tr)  at (0, -1.75) {VBS $W^\pm W^\pm$, $WZ$;\\ SM $WW$; $X\to WW$};
\node[box, text width=2.2cm, minimum height=4.55cm] (sim) at (3.35, 0) {\textsc{Pythia}\,8\\[3pt] \textsc{Delphes}\\ (CMS card)\\[3pt] anti-$k_t$, $R=0.8$\\[3pt] truth matching};
\node[box, text width=2.7cm] (ptset) at (6.55, 1.75) {Pretraining set:\\ six classes, $\pT$-matched};
\node[box, text width=2.7cm] (ftset) at (6.55, 0) {Fine-tuning sets:\\ $10^3$--$5\times10^4$ per class};
\node[box, text width=2.7cm] (trset) at (6.55, -1.75) {Transfer samples\\ (never trained on)};
\node[box, text width=2.3cm, fill=blue!6] (ptrain) at (9.75, 1.75) {Pretraining:\\ four objectives};
\node[box, text width=2.3cm, fill=blue!6] (ftrain) at (9.75, 0) {Fine-tuning:\\ model ladder};
\node[box, text width=2.3cm, fill=blue!6] (eval) at (9.75, -1.75) {Evaluation};
\node[box, text width=2.1cm, fill=green!6] (kd) at (12.85, -0.875) {Distillation:\\ small students};
\draw[arr] (pt) -- (sim.west |- pt);
\draw[arr] (pol) -- (sim.west |- pol);
\draw[arr] (tr) -- (sim.west |- tr);
\draw[arr] (sim.east |- ptset) -- (ptset);
\draw[arr] (sim.east |- ftset) -- (ftset);
\draw[arr] (sim.east |- trset) -- (trset);
\draw[arr] (ptset) -- (ptrain);
\draw[arr] (ftset) -- (ftrain);
\draw[arr] (trset) -- (eval);
\draw[arr] (ptrain) -- node[right, font=\footnotesize]{weights} (ftrain);
\draw[arr] (ftrain) -- node[right, font=\footnotesize]{tagger} (eval);
\draw[arr] (ftrain.east) -| node[pos=0.3, above, font=\footnotesize]{teacher} (kd.north);
\draw[arr] (kd.south) |- node[pos=0.75, below, font=\footnotesize]{students} (eval.east);
\end{tikzpicture}}
\caption{Workflow of the study. Pretraining, fine-tuning and transfer samples pass through the same simulation and jet selection. Polarization labels enter only in fine-tuning; the distilled students learn from the predictions of the fine-tuned teacher, also on unlabeled jets, and the transfer samples are used for evaluation only.}
\label{fig:pipeline}
\end{figure}

In this paper, we ask whether pretraining on boson decays helps. We construct a pretraining dataset of unpolarized boosted $W^+$, $W^-$, $Z$, Higgs-boson and top-quark jets alongside QCD jets. The samples are matched in transverse momentum so that the classes cannot be distinguished by kinematics alone. On this dataset, we pretrain the small configuration of a published point-edge transformer~\cite{Bhimji:2025omnilearned}; of the four pretraining objectives we define, three have been run for this paper. The network never sees a polarization label during pretraining. We then fine-tune the model to separate longitudinal from transverse $W$ bosons. We compare it with the same architecture trained from scratch and with the same network initialized from a published generic pretraining that covers many jet types, including boosted boson decays. The architecture is the same for every network, so any difference comes from the pretraining. Finally, we distill the fine-tuned foundation model into small student networks using both labeled and unlabeled jets. Table~\ref{tab:claims} lists the questions the study addresses and the approach to each. Figure~\ref{fig:pipeline} summarizes the workflow. This paper describes the setup, the physical constraints that limit any hadronic polarization tagger and the polarization information available beyond the decay angle, including what the rest of the event adds in diboson production and vector-boson scattering. We compare the initializations with $10^3$ to $5\times10^4$ labeled jets per class, with three trainings per network on each training set and a common test sample (section~\ref{sec:res-scan}).

\begin{table}[t]
\centering
\small
\begin{tabular}{@{}>{\raggedright}p{0.36\textwidth}>{\raggedright\arraybackslash}p{0.58\textwidth}@{}}
\toprule
Question & Approach \\
\midrule
Does pretraining on boson decays improve on generic pretraining? & fine-tuning our network and the same network from the published generic pretraining on the same labeled jets \\ \addlinespace[4pt]
Is the gain largest with few labels? & the same with $10^3$ to $5\times10^4$ labeled jets per class \\ \addlinespace[4pt]
Does the pretraining objective matter? & fine-tuning from four pretraining objectives, with the transformer blocks fine-tuned or frozen \\ \addlinespace[4pt]
Does the tagger transfer to other processes? & applying the $W$+jet tagger unchanged to VBS $W^\pm W^\pm$ (LL vs TT), VBS $WZ$, SM $WW$ and $X\to WW$; a template fit of $\fL$ in unpolarized $W$+jet events \\ \addlinespace[4pt]
Does the jet selection preserve the decay angle? & checking the truth $\acth$ distribution after a $W/Z$-versus-QCD selection, decorrelated from mass and angle, against eq.~\eqref{eq:envelope} \\ \addlinespace[4pt]
Do small taggers gain from distillation of the foundation model? & distilling the fine-tuned network into small students on labeled and unlabeled jets, and training the same students on the labels alone \\
\bottomrule
\end{tabular}
\caption{Questions the study is designed to answer and the approach to each. This paper answers the first three and the last (sections~\ref{sec:res-scan} and~\ref{sec:res-kd}); section~\ref{sec:res-outlook} discusses how the other two can be addressed with existing samples. The metrics are the area under the ROC curve (AUC) and the rejection of transverse jets, $1/\epsT$, at a longitudinal efficiency $\epsL=0.5$, averaged over several trainings.}
\label{tab:claims}
\end{table}

Section~\ref{sec:physics} reviews the decay-angle distributions, derives the acceptance imposed by jet containment, shows that, in $W$+jet events, the decay angle is nearly all the polarization information a jet carries, bounds the separation that any polarization tagger can reach, and estimates what the rest of the event adds in diboson production and vector-boson scattering. Section~\ref{sec:samples} describes the simulated samples, section~\ref{sec:dataset} the pretraining dataset, and section~\ref{sec:network} the network, the pretraining objectives, the fine-tuning protocol and the distillation. Section~\ref{sec:results} presents the results and section~\ref{sec:conclusions} concludes. Appendix~\ref{app:analytic} outlines the derivations of the analytic results, appendix~\ref{app:sim} documents details of the simulation, appendix~\ref{app:computing} the data pipeline and the computing cost, and appendix~\ref{app:opendata} a check of tagger inputs in ATLAS Open Data.

\section{Polarization in hadronic boson decays}
\label{sec:physics}

\subsection{Decay-angle distributions}

Consider a $W$ boson with helicity $\lambda\in\{-1,0,+1\}$, defined along its direction of flight in a frame specified below, decaying to a pair of massless fermions. Let the decay angle $\theta^*$ be the polar angle of one decay product in the boson rest frame, measured from this direction of flight. The production angle $\Theta$ (section~\ref{sec:production}) is instead the polar angle of the direction of flight itself, measured from the beam axis in the partonic center-of-mass frame. The normalized distribution of the decay angle is
\begin{equation}
\frac{1}{\sigma}\frac{\mathrm{d}\sigma}{\mathrm{d}\cth}
 = \frac{3}{8}\,f_{+}\,(1\pm\cth)^2 + \frac{3}{8}\,f_{-}\,(1\mp\cth)^2 + \frac{3}{4}\,\fL\,(1-\cos^2\theta^*)\,,
\label{eq:helicity}
\end{equation}
with $f_++f_-+\fL=1$; the upper signs apply when $\theta^*$ is measured from the antifermion and the lower signs when it is measured from the fermion, for either charge of the $W$ boson. In a hadronic decay, the quark and the antiquark are not distinguished jet by jet (jet charge and flavor tagging separate them only statistically), so only $\acth$ is accessible. Folding eq.~\eqref{eq:helicity} gives
\begin{equation}
P(c) \equiv \frac{1}{\sigma}\frac{\mathrm{d}\sigma}{\mathrm{d}c}
 = (1-\fL)\,\frac{3}{4}\,(1+c^2) + \fL\,\frac{3}{2}\,(1-c^2)\,, \qquad c=\acth\in[0,1]\,,
\label{eq:folded}
\end{equation}
and the two transverse helicities become indistinguishable. The transverse and longitudinal shapes have $\langle\cos^2\theta^*\rangle=2/5$ and $1/5$, so that for full acceptance
\begin{equation}
\fL = 2 - 5\,\langle\cos^2\theta^*\rangle\,.
\label{eq:fL}
\end{equation}
Whatever the value of $\fL$, the distribution normalized at $c=0$ obeys
\begin{equation}
1-c^2 \;\le\; \frac{P(c)}{P(0)} \;\le\; 1+c^2\,,
\label{eq:envelope}
\end{equation}
because $P(c)/P(0)$ is a convex combination of the two pure shapes, each normalized at $c=0$. A distribution that leaves this band cannot be described by any mixture of polarization states: a selection or a reconstruction has distorted the angle. We refer to eq.~\eqref{eq:envelope} as the mixture envelope and use it as a reference-free test of selections (section~\ref{sec:res-outlook}). The test is one-sided: it can show that a distribution is not a mixture, but a distortion that turns one mixture into another passes unnoticed. When applied bin by bin, this test requires a multiple-comparisons correction, such as setting a significance threshold of $\Phi^{-1}(1-0.05/n_{\mathrm{bins}})$ standard deviations per bin. The tested bins must also be restricted to the range of $c$ that all events can populate, because the containment edge discussed in section~\ref{sec:containment} varies event by event.

\subsection{Reference frame and generator-level validation}
\label{sec:frame}

Polarization fractions depend on the frame in which the helicity axis is defined. We follow the convention of the polarized matrix elements from ref.~\cite{BuarqueFranzosi:2019boy} in \MGaMC~\cite{Alwall:2014hca}, where helicity is defined along the boson direction in the partonic center-of-mass frame. We compute $\cth$ using the generator-level decay quarks in this same frame. For every jet, we also store the angle defined with the boson direction in the laboratory frame; at the transverse momenta considered here, the two definitions differ by 0.05--0.06 in $\cth$ (root mean square).

\begin{figure}[t]
\centering
\includegraphics[width=0.495\textwidth]{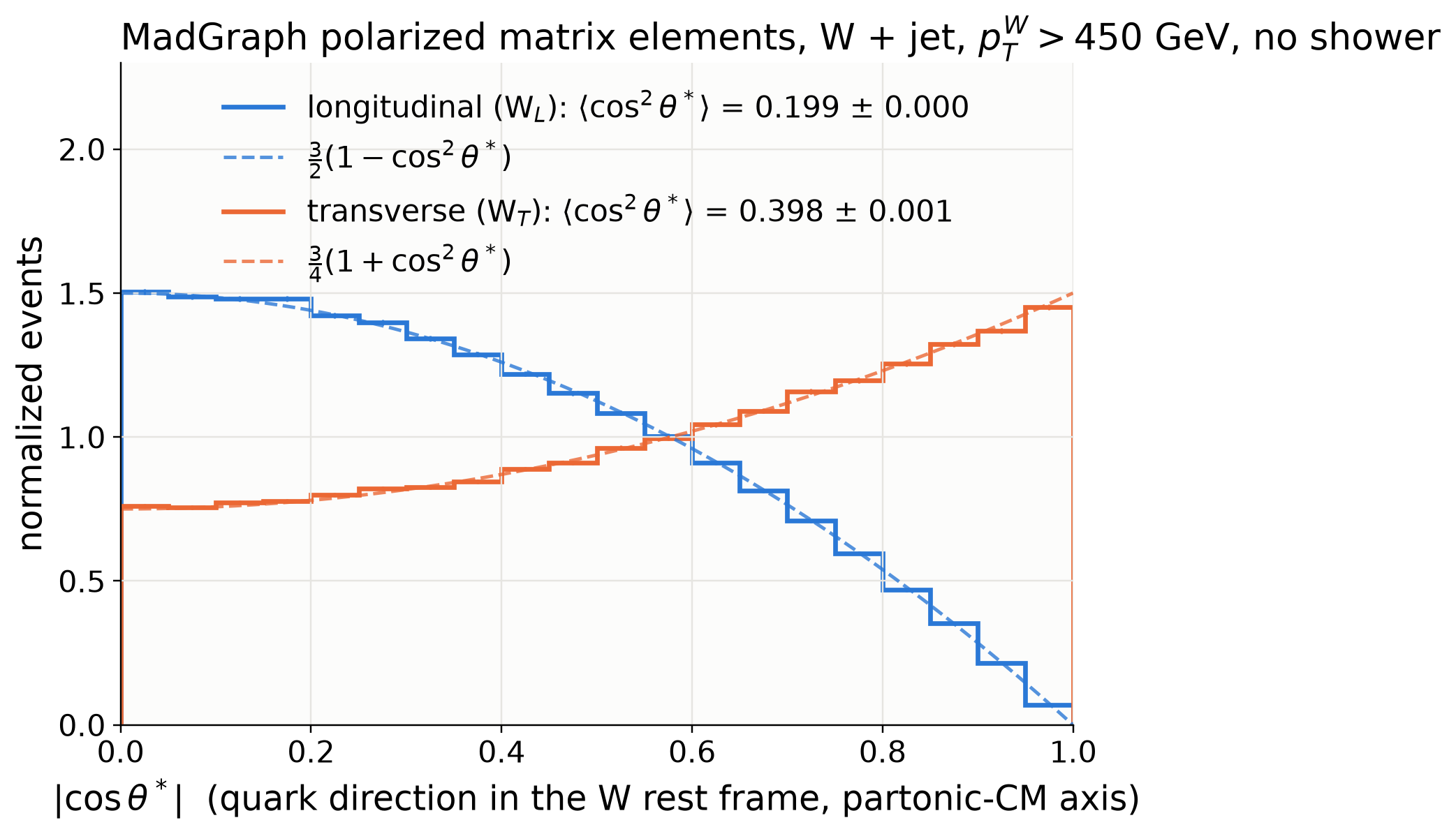}\hfill
\includegraphics[width=0.495\textwidth]{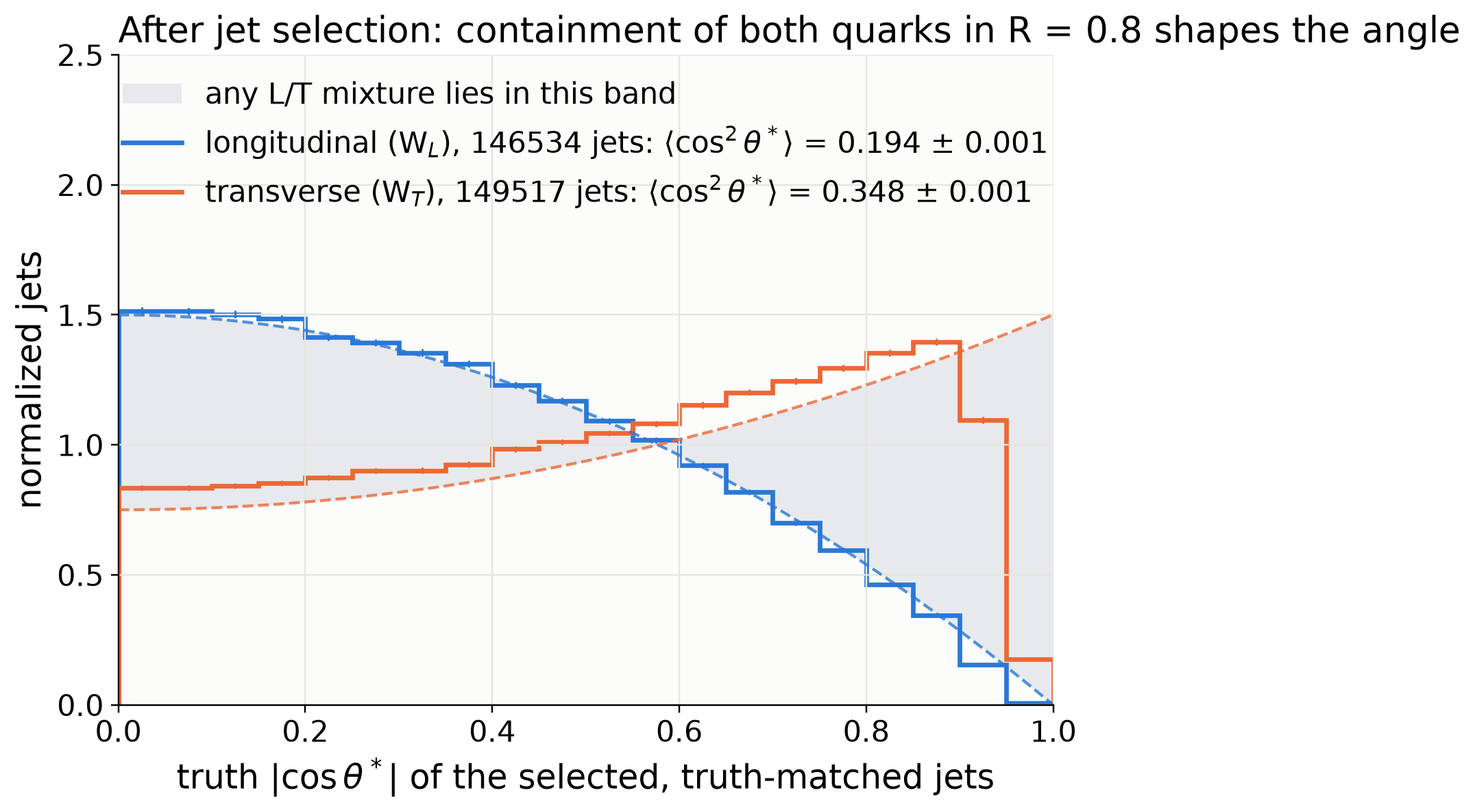}\\[4pt]
\includegraphics[width=0.495\textwidth]{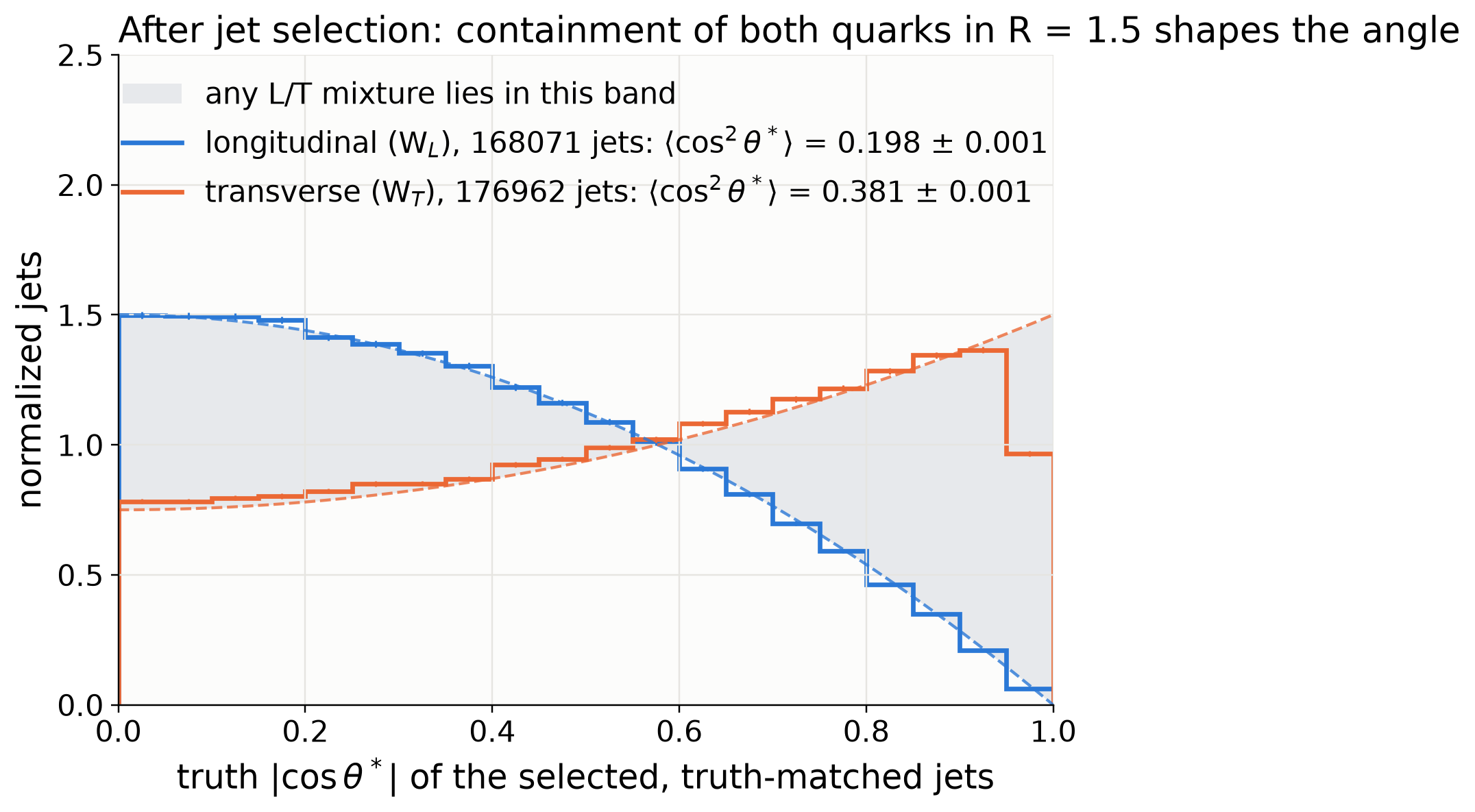}
\caption{Decay angle of helicity-selected $W^+$+jet events. Top left: matrix-element level, $\pT^W>450\GeV$, $3\times10^5$ events per sample, with the shapes of eq.~\eqref{eq:folded} for $\fL=1$ and $\fL=0$ (dashed). Top right and bottom: truth $\acth$ of jets after showering, detector simulation and selection (500--1000\GeV, both decay quarks within $R$ of the jet axis), with $R=0.8$ (top right) and $R=1.5$ (bottom), from the same generated events, before $\pT$ matching; the shaded band is the region accessible to any mixture of the untruncated shapes.}
\label{fig:validation}
\end{figure}

\begin{figure}[tbp]
\centering
\includegraphics[width=\textwidth]{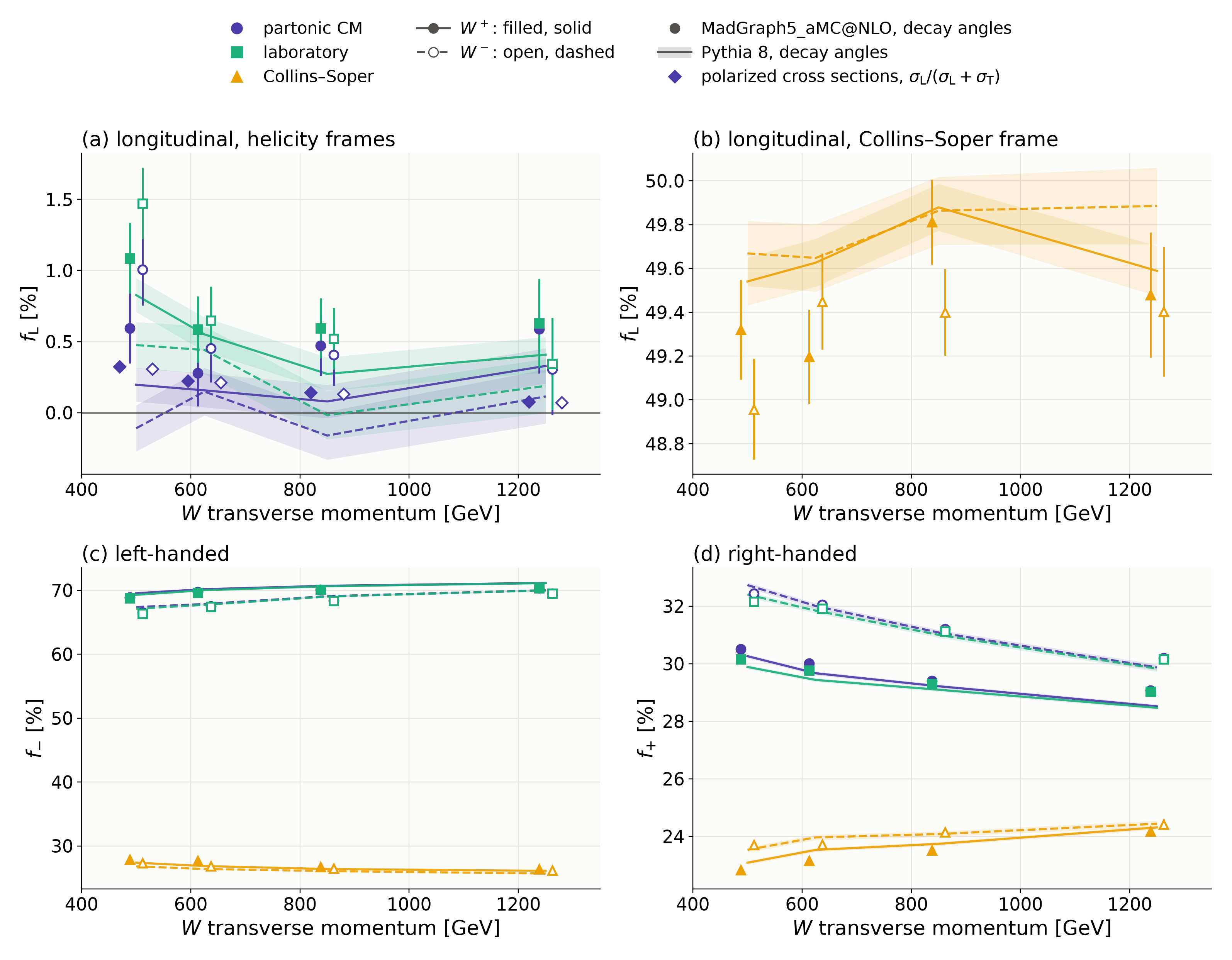}
\caption{Helicity fractions of $W^+$ bosons (filled markers, solid lines) and $W^-$ bosons (open markers, dashed lines) in $W$+jet events at leading order, against the $W$ transverse momentum, in the partonic center-of-mass (purple) and laboratory (green) helicity frames and in the Collins--Soper frame (amber, axis oriented along the longitudinal momentum of the boson): (a) longitudinal fraction in the two helicity frames, (b) longitudinal fraction in the Collins--Soper frame, (c) left-handed and (d) right-handed fractions, with $\theta^*$ measured from the antifermion. Points: decay angles of the unpolarized \MGaMC samples, with statistical uncertainties; diamonds in (a): longitudinal share of the polarized \MGaMC cross sections; lines and bands: decay angles of the \Pythia events and their statistical uncertainty. All are evaluated in the bins 450--550, 550--700, 700--1000 and 1000--1500\GeV. The lines join the bin centers; the markers of each bin are shifted sideways from its center so that they do not overlap, those for $W^+$ to the left and those for $W^-$ to the right. The decay angles of the \MGaMC samples overestimate $\fL$ by about 0.3 percentage points (text). The Collins--Soper axis is nearly perpendicular to the direction of flight (figure~\ref{fig:axes}), so the same, almost purely transverse bosons appear about half longitudinal in (b).}
\label{fig:frames}
\end{figure}

\begin{figure}[tbp]
\centering
\includegraphics[width=\textwidth]{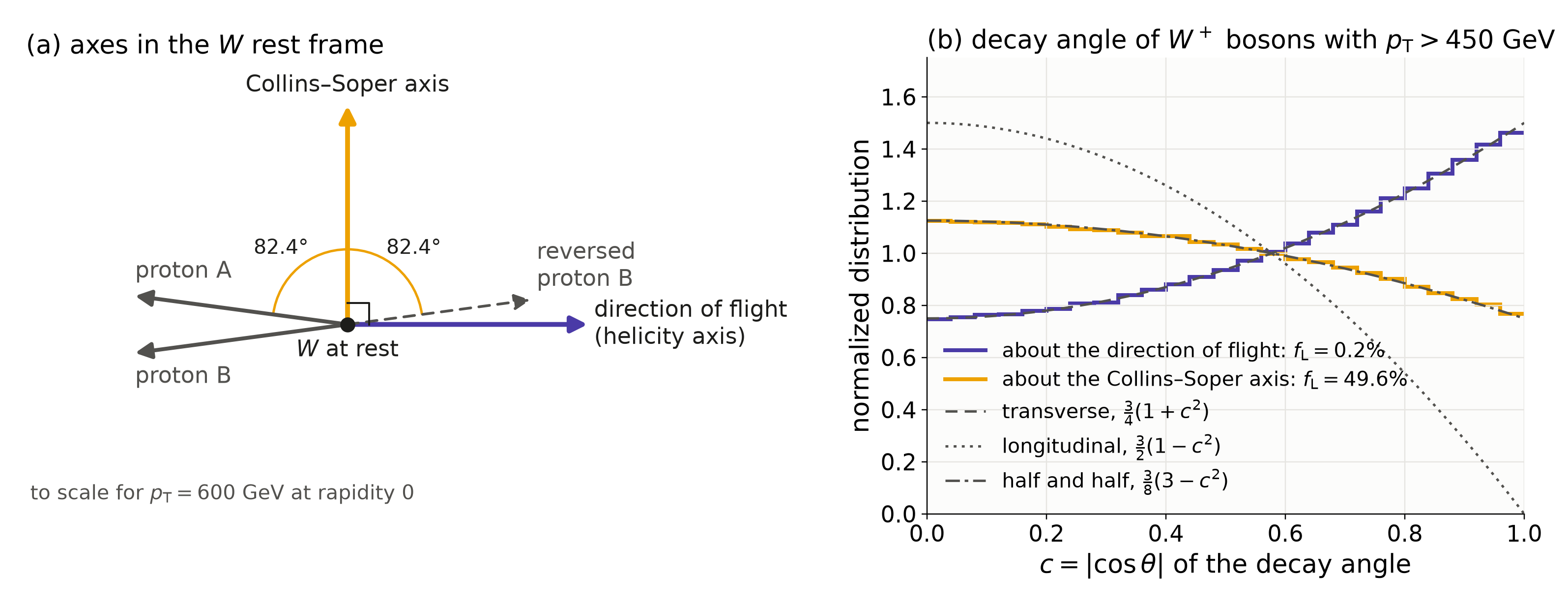}
\caption{Why the bosons of figure~\ref{fig:frames} are almost purely transverse along their direction of flight but about half longitudinal along the Collins--Soper axis. (a) The two axes in the rest frame of a $W$ boson with $\pT=600\GeV$ at zero rapidity, drawn to scale: both protons move within $7.6^\circ$ of the direction opposite to the direction of flight, and the Collins--Soper axis, which bisects the angle between the momentum of proton A and the reversed momentum of proton B, is perpendicular to the direction of flight. (b) Distribution of $c=|\cos\theta|$ for the decay angle measured from the direction of flight (partonic center-of-mass helicity frame) and from the Collins--Soper axis, for the \Pythia $W^+$ bosons with $\pT>450\GeV$ of figure~\ref{fig:frames}, together with the transverse and longitudinal shapes of eq.~\eqref{eq:folded} and their equal mixture; $\fL$ from eq.~\eqref{eq:fL}.}
\label{fig:axes}
\end{figure}

Figure~\ref{fig:validation}~(top left) shows the decay-angle distributions of helicity-selected $W^+$+jet events at matrix-element level for a $W$ transverse momentum above 450\GeV. The means, $\langle\cos^2\theta^*\rangle=0.199$ (longitudinal) and 0.398 (transverse), each with a statistical uncertainty below 0.001, lie slightly below the expected $1/5$ and $2/5$; through eq.~\eqref{eq:fL}, this corresponds to an excess of 0.5 to 1 percentage point in $\fL$, the generator effect discussed below. Figure~\ref{fig:frames} shows how the helicity fractions of $W$ bosons in $W$+jet events depend on the frame at leading order. For each charge, the \MGaMC samples comprise $4.5\times10^6$ unpolarized events and $8\times10^5$ events per helicity, generated with the settings of section~\ref{sec:generation} in slices of the recoil $\pT$, and each slice enters only the bins that lie at least 100\GeV{} above its lower edge. The \Pythia~\cite{Bierlich:2022pfr} events, $1.2\times10^7$ in total with the $W$-boson mass within 15 widths of its pole, as in the \MGaMC samples, include the exact decay correlations at this order. In the partonic frame, the polarized cross sections give $\fL=0.32\%$ at 450--550\GeV, falling to 0.08\% at 1--1.5\TeV{} (0.31\% and 0.07\% for $W^-$), and the decay angles of the \Pythia events agree within their statistical uncertainty of 0.1--0.2 percentage points per bin. With the helicity axis along the boson direction in the laboratory, $\fL$ is larger by 0.5--0.6 points at 450--550\GeV{} and by less than 0.1 above 1\TeV, where the two helicity frames converge. In both, the transverse bosons are about 70\% left-handed ($f_-=69$--70\% for $W^+$ and 67--70\% for $W^-$). In the Collins--Soper frame~\cite{Collins:1977iv}, in contrast, the same bosons are about half longitudinal. Its axis is set by the beams: in the boson rest frame, it bisects the angle between the momentum of one proton and the reversed momentum of the other, and we orient it along the longitudinal momentum of the boson. Seen from a boson with a large transverse momentum, both protons move almost opposite to its direction of flight, so the axis is nearly perpendicular to that direction: on average it lies 84--89$^\circ$ from it in the bins and helicity frames of figure~\ref{fig:frames} (figure~\ref{fig:axes}a; see also section~\ref{sec:production}). A state of helicity $\lambda=\pm1$ along the direction of flight has, along an axis at an angle $\beta$ from it, the components $m=\lambda$, 0 and $-\lambda$ with probabilities $\frac14(1+\cos\beta)^2$, $\frac12\sin^2\beta$ and $\frac14(1-\cos\beta)^2$. For $\beta$ near $90^\circ$, bosons that are transverse along their direction of flight are therefore about half longitudinal along the Collins--Soper axis, and the rest divides almost evenly between $f_-$ and $f_+$. Their decay angles show this directly (figure~\ref{fig:axes}b).

The decay angles of the \MGaMC samples themselves give values of $\fL$ that lie $0.32\pm0.09$ points above the polarized cross sections on average (figure~\ref{fig:frames}a), and the transverse samples give $0.6\pm0.2\%$ instead of zero. Equation~\eqref{eq:fL} magnifies small deviations of a generated angular distribution from the spin-1 shape: removing 0.3\% of the events at $\acth\approx0.85$, for example, raises $\fL$ by half a point. The excess grows by another $0.5\pm0.2$ points just above a generator cut, so that samples generated with the cut on the recoil at the analysis threshold, such as our pretraining sample, overestimate $\fL$ by 1--2 points at 450--550\GeV. The polarized cross sections add up to 3--4\% more than the unpolarized cross section in every bin; their ratio, however, agrees with the \Pythia events. Either way, at these transverse momenta, the $W$ bosons are almost purely transverse.

\subsection{Boosted kinematics and jet containment}
\label{sec:containment}

A boson of mass $m$ and transverse momentum $\pT\gg m$ that decays to two massless partons carrying momentum fractions $z$ and $1-z$ produces an opening angle
\begin{equation}
\Delta R_{q\bar q} \simeq \frac{m}{\pT\sqrt{z(1-z)}}
\label{eq:dR}
\end{equation}
in the rapidity--azimuth plane. In the boosted limit $z=(1+\cth)/2$, so that $z(1-z)=\sin^2\theta^*/4$ and
\begin{equation}
\Delta R_{q\bar q} \simeq \frac{2m}{\pT\,\sin\theta^*}\,.
\label{eq:dRtheta}
\end{equation}
Two consequences follow. First, the energy sharing between the two prongs measures the decay angle directly,
\begin{equation}
p_\theta \equiv \frac{|E_1-E_2|}{|\vec p_V|} = \acth\,,
\label{eq:ptheta}
\end{equation}
which is the basis of the observable of ref.~\cite{De:2020iwq}. For massless partons, this is an identity when $\theta^*$ is measured from the boson direction in the laboratory, since $E_{1,2}=\frac12(E_V\pm|\vec p_V|\cos\theta^*)$; normalizing by $E_1+E_2=E_V$ instead gives $\beta\acth$. In a jet, the two prongs are subjets, and the jet momentum stands in for $\vec p_V$. Second, the axis of a jet containing both partons closely follows the boson direction (their momentum-weighted centroid). As a result, the softer parton lies at a distance $z\,\Delta R_{q\bar q}$ from this axis, where $z\ge1/2$ is the momentum fraction of the harder parton. Requiring both partons to fall within $R$ of the jet axis, $z\,\Delta R_{q\bar q}<R$, and using eq.~\eqref{eq:dR} with $z=(1+\acth)/2$ gives
\begin{equation}
\frac{m}{\pT}\sqrt{\frac{1+\acth}{1-\acth}} < R
\quad\Longleftrightarrow\quad
\acth < \cmax(m,\pT) = \frac{(\pT R)^2-m^2}{(\pT R)^2+m^2}\,.
\label{eq:cmax}
\end{equation}
The stricter condition that the two partons lie within $R$ of each other, $\Delta R_{q\bar q}<R$, would instead yield $[1-(2m/(\pT R))^2]^{1/2}$. The two expressions agree to leading order, $1-2m^2/(\pT R)^2$, and differ appreciably only for heavy bosons at low $\pT$, yielding 0.74 versus 0.62 for the Higgs boson at 400\GeV. Our truth matching (section~\ref{sec:jets}) imposes the first condition, and the matched jets follow it. Among Higgs-boson jets with a boson $\pT$ of 350--450\GeV, 9.7\% lie more than 0.05 above the stricter edge, whereas only 1.3\% lie more than 0.05 above $\cmax$. This edge applies to $\acth$ defined relative to the boson direction in the laboratory frame; in the partonic center-of-mass frame (section~\ref{sec:frame}), it is smeared by the difference between the two definitions. The most asymmetric decays, where the transverse distribution peaks, leave their softer parton outside the jet. Table~\ref{tab:cmax} lists $\cmax$ for $R=0.8$ and $R=1.5$. The edge rises with $\pT$ and falls with the boson mass, decreasing from 0.92 for the $W$ boson to 0.82 for the Higgs boson at 500\GeV{} with $R=0.8$. A larger radius pushes this edge outward to 0.98 and 0.95 for $R=1.5$, which is why we also study jets of this radius (section~\ref{sec:jets}). A hadronic polarization tagger can never use the region beyond $\cmax$. Equation~\eqref{eq:fL} also no longer holds for the truncated distribution, and templates for measuring $\fL$ must be built using the same selection.

\begin{table}[t]
\centering
\begin{tabular}{@{}llcccc@{}}
\toprule
 & & \multicolumn{4}{c}{$\cmax$ at boson $\pT$ of} \\
\cmidrule(l){3-6}
$R$ & Boson & 400\GeV & 500\GeV & 800\GeV & 1000\GeV \\
\midrule
0.8 & $W$ ($m=80.4\GeV$) & 0.88 & 0.92 & 0.97 & 0.98 \\
    & $Z$ ($m=91.2\GeV$) & 0.85 & 0.90 & 0.96 & 0.97 \\
    & $H$ ($m=125\GeV$)  & 0.74 & 0.82 & 0.93 & 0.95 \\
\midrule
1.5 & $W$ & 0.96 & 0.98 & 0.99 & 0.99 \\
    & $Z$ & 0.95 & 0.97 & 0.99 & 0.99 \\
    & $H$ & 0.92 & 0.95 & 0.98 & 0.99 \\
\bottomrule
\end{tabular}
\caption{Containment edge of eq.~\eqref{eq:cmax} for $R=0.8$ and $R=1.5$: the largest $\acth$ for which both decay partons lie within $R$ of the jet axis.}
\label{tab:cmax}
\end{table}

\begin{figure}[t]
\centering
\includegraphics[width=\textwidth]{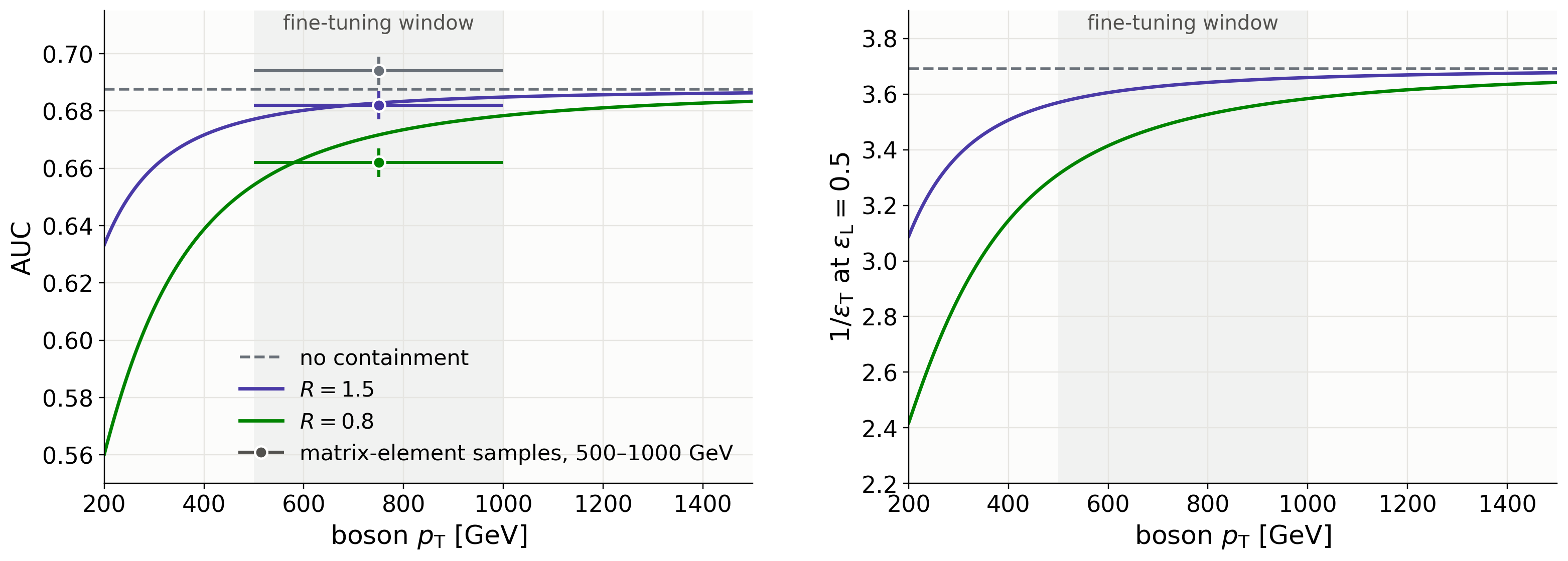}
\caption{Ideal decay-angle tagger for $W$ bosons against the boson $\pT$: AUC (left) and rejection of transverse jets at $\epsL=0.5$ (right), for the shapes of eq.~\eqref{eq:folded} truncated at the containment edge of eq.~\eqref{eq:cmax} with $R=0.8$ and $R=1.5$, and without containment (dashed). The points are the matrix-element values of table~\ref{tab:production}, with horizontal bars over their $\pT$ range of 500--1000\GeV{}; the shaded band is the fine-tuning window.}
\label{fig:ceiling}
\end{figure}

Figure~\ref{fig:ceiling} shows what this edge costs the ideal decay-angle tagger (section~\ref{sec:metrics}) for $W$ bosons; the curves are obtained by truncating the shapes of eq.~\eqref{eq:folded}. The AUC and working points of this tagger have closed-form expressions (appendix~\ref{app:analytic}). With $R=0.8$, the AUC drops from 0.6875 without containment to 0.654 at 500\GeV{} and 0.611 at 300\GeV{}, while the rejection of transverse jets at $\epsL=0.5$ decreases from 3.7 to 3.3 and 2.9, respectively. With $R=1.5$, the AUC remains at 0.66 or above down to 300\GeV{}. The matrix-element values from section~\ref{sec:production}, integrated over 500--1000\GeV{}, agree with these curves.

Figure~\ref{fig:validation}~(top right) shows this effect in the helicity-selected $W^+$+jet samples after showering, detector simulation and jet selection (section~\ref{sec:samples}). The longitudinal jets keep $\langle\cos^2\theta^*\rangle=0.194$, while the transverse ones fall from 0.40 to 0.348, or 0.345 once the two classes are matched in $\pT$ as for the comparisons of section~\ref{sec:res-scan}; with $1.3\times10^5$ matched jets per class, the statistical uncertainty of these means is below 0.001. A single hard cut at $\acth<0.936$, applied to the shapes in eq.~\eqref{eq:folded}, reproduces the transverse value of the matched jets, gives 0.196 for the longitudinal ones, against the measured 0.194, and removes 9\% of the transverse and 0.6\% of the longitudinal jets (appendix~\ref{app:analytic}). Below the edge, the shapes follow eq.~\eqref{eq:folded} but are renormalized; this explains why the transverse distribution lies above the band at $0.5<\acth<0.9$. With $R=1.5$ (figure~\ref{fig:validation}, bottom), the transverse distribution continues almost to $\acth=1$ and more closely follows the untruncated shape. The two classes yield $\langle\cos^2\theta^*\rangle=0.198$ and 0.381 (0.380 after the matching). A hard cut at $\acth<0.977$, which is close to the edges in table~\ref{tab:cmax}, reproduces the transverse value, gives 0.199 for the longitudinal jets, against the measured 0.198, and removes 3\% of the transverse and 0.1\% of the longitudinal jets. The jet mass and the number of constituents for the two classes are nearly indistinguishable in these samples (figure~\ref{fig:jets}); therefore, nearly all of the separation between them must arise from how the energy is shared within the jet.

\subsection{Information beyond the polar decay angle}
\label{sec:production}

Equation~\eqref{eq:folded} integrates over two angles that could also carry polarization information: the production angle $\Theta$ between the boson and the beam axis in the partonic center-of-mass frame (the rest frame of the two incoming partons), and the azimuth $\phi^*$ of the decay quark around the boson direction, measured relative to the plane containing the beam axis and the boson. In proton--proton collisions, the distribution of $\cos\Theta$ is symmetric about zero, so we use $|\cos\Theta|$. The distribution of $\Theta$ depends on the production process. Because $\Theta$ can only be reconstructed together with the recoiling system, a jet tagger observes it at most through the jet pseudorapidity. The azimuth, in contrast, is visible within the jet as the orientation of the two prongs in the rapidity--azimuth plane relative to the beam direction. However, it carries polarization information only through interference between helicity states. For a transverse boson, interference between the $+1$ and $-1$ helicity states modulates the azimuth,
\begin{equation}
\frac{1}{\sigma}\frac{\mathrm{d}\sigma}{\mathrm{d}c\,\mathrm{d}\phi^*} = \frac{3}{8\pi}\left[1+c^2+2|\rho_{+-}|\,(1-c^2)\cos(2\phi^*-\delta)\right], \qquad c=\acth\,,
\label{eq:azimuth}
\end{equation}
where $\rho$ is the spin density matrix of the boson, normalized to $\rho_{++}+\rho_{--}=1$, and $\delta$ is a phase. This modulation survives the exchange of the quark and antiquark ($\theta^*\to\pi-\theta^*$, $\phi^*\to\phi^*+\pi$), and the positivity of $\rho$ limits it to $|\rho_{+-}|\le1/2$, meaning $|\langle\cos2\phi^*\rangle|\le1/4$ (appendix~\ref{app:analytic}). Interference between longitudinal and transverse states, which produces terms proportional to $\sin2\theta^*\cos\phi^*$, is absent from helicity-selected samples by construction.

\begin{figure}[t]
\centering
\includegraphics[width=\textwidth]{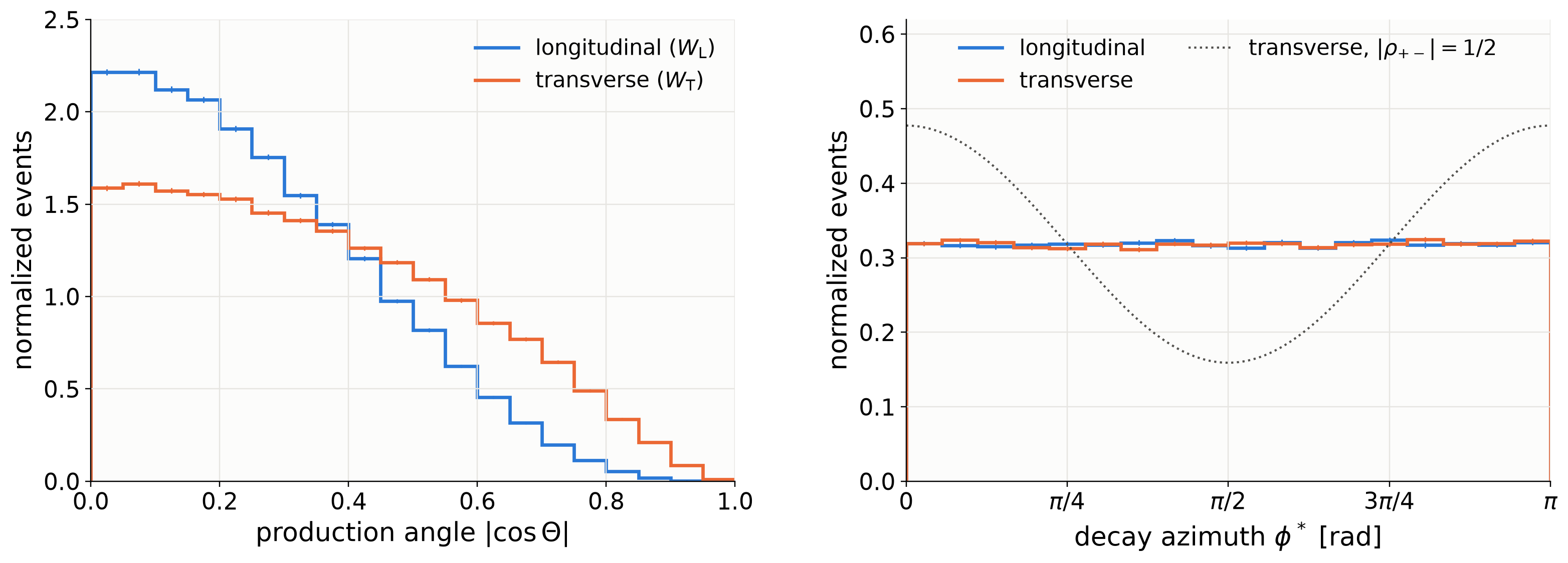}
\caption{Production angle $|\cos\Theta|$ (left) and decay azimuth $\phi^*$ (right) of longitudinal and transverse $W^+$ bosons in $W^+$+jet events at matrix-element level, for a boson $\pT$ of 500--1000\GeV{}, in the polarized samples of $3\times10^5$ events per helicity; for the production angle, the two helicities are matched in $\pT$. The azimuth is folded to $[0,\pi)$. The dotted curve is the largest modulation that eq.~\eqref{eq:azimuth} allows, for $|\rho_{+-}|=1/2$ and $\delta=0$.}
\label{fig:production}
\end{figure}

Table~\ref{tab:production} compares the two polar angles at matrix-element level in the polarized samples of section~\ref{sec:generation}, with $3\times10^5$ events per helicity. Longitudinal bosons are produced more centrally (figure~\ref{fig:production}, left), and the production angle alone separates the helicities with an AUC of 0.60. This information is nearly independent of the decay angle. When combined in a gradient-boosted classifier, the two angles achieve an AUC of 0.722, which is 0.034 higher than the decay angle alone. The gain remains about the same when the containment edge for $R=0.8$ or $R=1.5$ is applied (0.038 and 0.035). In $W$+jet events, the production angle thus provides a larger gain than increasing the jet radius from $R=0.8$ to $R=1.5$, which recovers only 0.018 of the 0.026 in AUC lost to containment. As shown below, the production angle survives detector simulation if the recoiling jet is included.

\begin{table}[t]
\centering
\begin{tabular}{@{}lcccc@{}}
\toprule
Acceptance & Events per class & Decay angle & Production angle & Both \\
\midrule
\multicolumn{5}{@{}l}{\emph{Matrix element}} \\
none & 147,471 & 0.688 & 0.603 & 0.722 \\
$\acth<\cmax$, $R=0.8$ & 140,067 & 0.662 & 0.602 & 0.700 \\
$\acth<\cmax$, $R=1.5$ & 145,292 & 0.680 & 0.603 & 0.715 \\
\midrule
\multicolumn{5}{@{}l}{\emph{Selected jets; $\Theta$ rebuilt from the jet and the leading other jet}} \\
$R=0.8$ & 132,130 & 0.660 & 0.594 & 0.693 \\
$R=1.5$ & 155,519 & 0.677 & 0.591 & 0.707 \\
\bottomrule
\end{tabular}
\caption{Separation of longitudinal from transverse $W^+$ bosons in $W^+$+jet events: AUC of $\acth$ and of $|\cos\Theta|$, both in the partonic center-of-mass frame, and of the two combined with gradient-boosted trees (five-fold cross-validation). Top: matrix-element level, for the polarized samples ($3\times10^5$ events per helicity) restricted to a boson $\pT$ of 500--1000\GeV{} and matched in 50\GeV{} bins of it; the second and third rows keep events whose decay angle in the laboratory frame lies below the containment edge of eq.~\eqref{eq:cmax}. Bottom: the selected jets of the same events after showering and detector simulation (section~\ref{sec:samples}), matched in 50\GeV{} bins of jet $\pT$; the decay angle is the true one, $\Theta$ is rebuilt as described in the text, and the combined AUC is averaged over three assignments of the five folds; against the cross-validated AUC of the true decay angle alone, 0.659 and 0.677, the rebuilt production angle adds 0.034 and 0.030. The statistical uncertainty is about 0.001.}
\label{tab:production}
\end{table}

The azimuth carries no information in these samples (figure~\ref{fig:production}, right). With $1.5\times10^5$ events per helicity in this $\pT$ range, $\langle\cos2\phi^*\rangle=0.003\pm0.002$ for transverse and $0.000\pm0.002$ for longitudinal bosons, and an unpolarized sample of $2\times10^4$ events gives $-0.010\pm0.007$. Adding $\cos2\phi^*$ and $\sin2\phi^*$ to the decay angle changes the cross-validated AUC by 0.001, from 0.687 to 0.688, within its statistical uncertainty. This behavior follows from the kinematics of $W$+jet production at large $\pT$. In the Collins--Soper frame~\cite{Collins:1977iv}, leading-order production yields angular coefficients $A_0=A_2$ (the Lam--Tung relation~\cite{Lam:1980uc}) that approach 1 at large $\pT$, and our unpolarized sample has $A_0=1.03\pm0.03$ and $A_2=1.05\pm0.05$. With $A_0=A_2=1$, the decay distribution becomes $1+(\hat n\cdot\hat x)^2$, where $\hat n$ is the decay direction and $\hat x$ is the $x$-axis of the Collins--Soper frame. At these transverse momenta, $\hat x$ points almost exactly along the boson direction ($|\hat x\cdot\hat z_{\mathrm{hel}}|=0.996$ on average). The distribution is therefore $1+\cos^2\theta^*$ about the helicity axis. This indicates that the bosons are transverse without linear polarization, and their longitudinal fraction is small, as found in section~\ref{sec:frame}.

After showering, detector simulation and selection, little production information reaches the jet. In the selected jets of these samples, matched in $\pT$, the jet $|\eta|$ alone separates longitudinal from transverse jets with an AUC of 0.519 (0.518 for $R=1.5$). Added to the true decay angle in a cross-validated gradient-boosted classifier, it raises the AUC by 0.002 for both radii (from 0.659 and 0.677), close to the 0.001 expected in the binormal approximation (appendix~\ref{app:analytic}) for an independent variable of this strength. In $W$+jet events, the polar decay angle therefore carries all the polarization information of a jet to within 0.005 in AUC; on the test jets of section~\ref{sec:res-scan}, the jet mass, the number of constituents and the jet $\pT$ together add 0.003 to the cross-validated AUC of the true decay angle for $R=0.8$ and less than 0.001 for $R=1.5$. The ideal decay-angle tagger (section~\ref{sec:metrics}) is therefore the ceiling for any jet tagger on these samples that cannot distinguish the quark prong from the antiquark prong (section~\ref{sec:limits}). The true angle reaches an AUC of 0.660 with $R=0.8$ and 0.677 with $R=1.5$, for all selected jets as for the test jets (table~\ref{tab:scan}). A tagger trained on $W$+jet events also has no incentive to learn the azimuth, which can carry linear polarization in other processes. This lack of azimuthal bias makes $W$+jet a suitable training process for a tagger intended to transfer across production modes.

\begin{figure}[tbp]
\centering
\includegraphics[width=\textwidth]{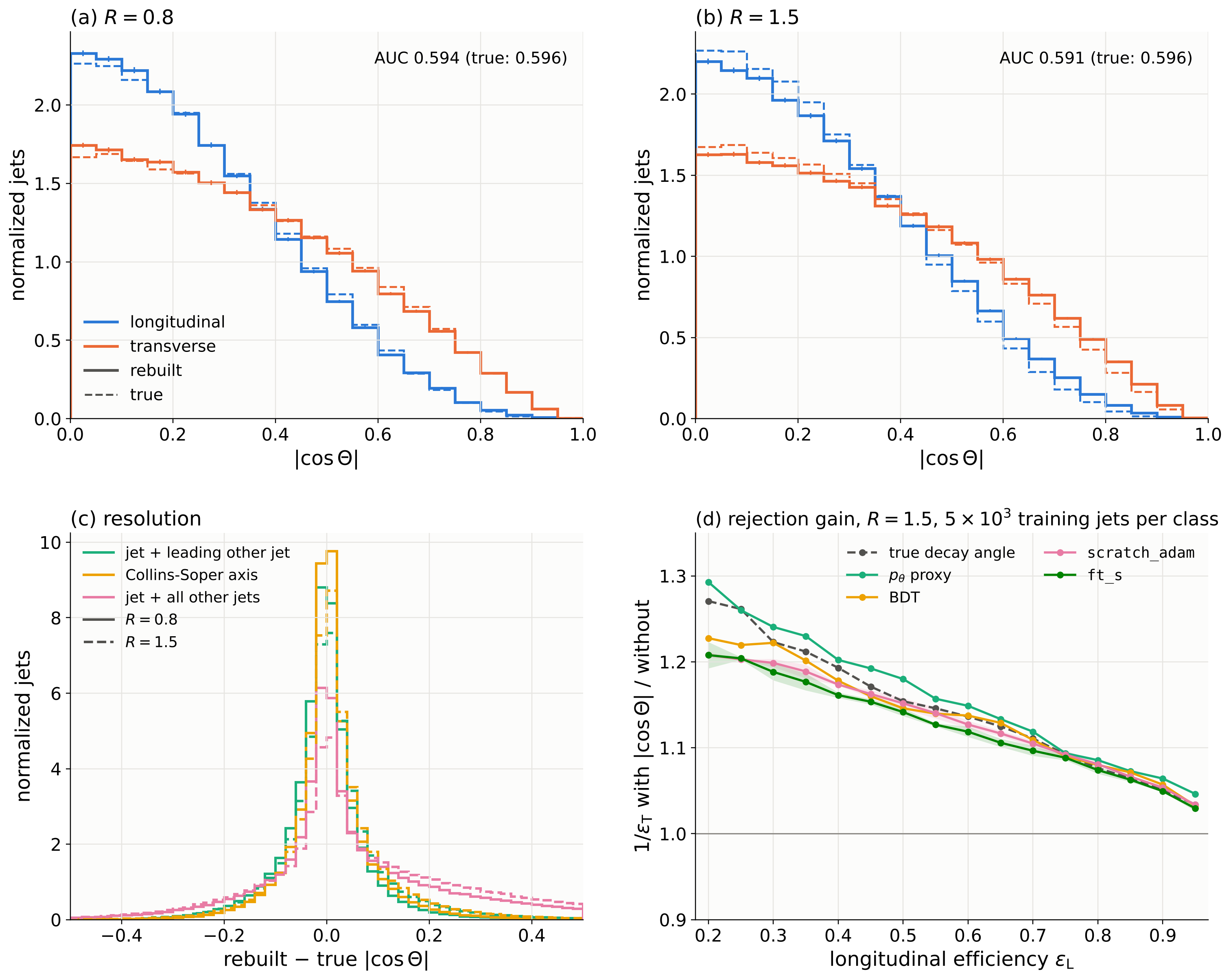}
\caption{Production angle after showering, detector simulation and jet selection, (a--c) for all selected jets and (d) on the test jets of section~\ref{sec:res-scan}. (a, b) $|\cos\Theta|$ of longitudinal and transverse jets, rebuilt from the selected jet and the leading other jet (solid, with statistical uncertainties) and true (dashed), for $R=0.8$ and $R=1.5$, with the two classes matched in jet $\pT$; the AUC of the rebuilt and of the true angle is given. (c) Difference between the rebuilt and the true $|\cos\Theta|$ for the three ways of rebuilding it, for $R=0.8$ (solid) and $R=1.5$ (dashed). (d) Ratio of the transverse rejection with and without the rebuilt $|\cos\Theta|$ for the proxy and for the BDT and two networks trained on $5\times10^3$ jets per class with $R=1.5$, and for the true decay angle, at fixed longitudinal efficiency; with $|\cos\Theta|$, the tagger output and the angle are combined in a cross-validated classifier on the test jets, with one assignment of the five folds. For the networks, the lines show the mean of the three trainings and the bands their standard deviation.}
\label{fig:prodangle-detector}
\end{figure}

The production angle itself survives detector simulation, provided the tagger sees the recoil. We rebuild it from the selected jet and the leading other jet in the event, which is clustered with the same radius and required to have $\pT>30\GeV$ and $|\eta|<5$. Both jets are then boosted along the beam axis to the frame in which their sum has no longitudinal momentum, and $\Theta$ is the polar angle of the jet in that frame. A leading other jet exists for every selected jet. The rebuilt $|\cos\Theta|$ tracks the true value with a root-mean-square difference of 0.092 for $R=0.8$ and 0.109 for $R=1.5$ (figure~\ref{fig:prodangle-detector}). It carries almost all the information of the true angle (table~\ref{tab:production}). When added to the true decay angle in a cross-validated classifier, it raises the AUC by 0.034 for $R=0.8$ and by 0.030 for $R=1.5$, relative to the classifier using the decay angle alone (0.659 and 0.677). By comparison, the true production angle raises the AUC by 0.036 and 0.033 on the same jets. Summing all other jets instead of taking only the leading one incorporates initial-state radiation, which is not part of the $2\to2$ system and shifts the rebuilt angle to larger values (figure~\ref{fig:prodangle-detector}c); this approach recovers only 0.026 and 0.021 in AUC. The Collins--Soper axis~\cite{Collins:1977iv} yields nearly the same result as the longitudinal boost (0.033 and 0.029). A tagger that sees the whole event could therefore exceed the jet-level ceiling by about 0.03, which is larger than the gain from increasing the radius from $R=0.8$ to $R=1.5$. However, such a tagger would also learn the relation between the production angle and the helicity of the training process. Because this relation differs in other processes, such as VBS, an event-level tagger must be trained on polarized samples of the specific process to which it will be applied (section~\ref{sec:diboson}).

When added in the same way to the output of the taggers of section~\ref{sec:res-scan}, which see only the jet, the rebuilt production angle improves each of them by about as much as it improves the true decay angle. On the test jets, it raises the cross-validated AUC of the BDT and the networks trained on $5\times10^3$ labeled jets per class by 0.032--0.034 for $R=0.8$ and by 0.028--0.032 for $R=1.5$, to 0.670--0.683 and 0.687--0.691. With $10^3$ labeled jets per class, the gains are 0.032--0.049, and with $5\times10^4$, 0.027--0.033; the weaker taggers gain more. These gains are measured relative to the AUC of the tagger alone. They are conservative because the combination is trained with five-fold cross-validation solely on the test jets; with this many jets, however, the same classifier trained on the tagger output alone loses only about 0.001 in AUC. Measured in the same way, the true decay angle gains 0.034 for $R=0.8$ and 0.030 for $R=1.5$. For a variable independent of the tagger output, the binormal approximation (appendix~\ref{app:analytic}) predicts gains of 0.024--0.044 and 0.020--0.028, between 0.004 and 0.011 less than measured, for the true decay angle as for the trained taggers. The taggers therefore extract no information about the production angle, and their order changes little. With the production angle included, the best taggers exceed the true decay angle alone on the same jets: with $5\times10^3$ labeled jets per class, the BDT and the network fine-tuned from \rung{pt\_cls} reach 0.682 and 0.683 for $R=0.8$, against 0.660, and the generically pretrained network 0.691 for $R=1.5$, against 0.677. In terms of transverse rejection, the production angle adds 7--29\% for every tagger at longitudinal efficiencies between 0.2 and 0.8, and 14--19\% at $\epsL=0.5$ (figure~\ref{fig:prodangle-detector}d). At $\epsL=0.5$, the rejection of the BDT trained on $5\times10^3$ jets per class with $R=1.5$ rises from 3.33 to 3.82, and that of the true decay angle from 3.55 to 4.10.

\subsection{How well polarization can be tagged}
\label{sec:limits}

Even a perfect measurement of the decay angle separates longitudinal from transverse bosons only weakly. The likelihood ratio of the folded shapes in eq.~\eqref{eq:folded}, $2(1-c^2)/(1+c^2)$, reaches its maximum of 2 at $c=0$. Therefore, no selection on $\acth$ can more than double the longitudinal fraction of a sample. A perfect measurement of $\acth$ yields an AUC of $11/16=0.6875$. This drops to 0.660 and 0.677 with the containment edges found for the selected jets with $R=0.8$ and $R=1.5$, respectively (section~\ref{sec:containment}). Table~\ref{tab:limits} shows how much additional information could improve performance, assuming perfectly measured angles and no detector effects. Appendix~\ref{app:analytic} outlines how these limits are derived.

The sign of $\cth$, measured relative to the quark, separates the two transverse helicities. Their distributions, $(1\pm\cth)^2$ in eq.~\eqref{eq:helicity}, fold into the same shape, but the sign adds information whenever the transverse bosons are not an equal mixture of the two helicity states. In our samples, they are not. Measured relative to the up-type quark, the transverse test jets of section~\ref{sec:res-scan} have $\langle\cth\rangle=0.164\pm0.002$ for $R=0.8$ and $0.175\pm0.002$ for $R=1.5$, and the quark lies in the forward hemisphere in 63\% of the jets for both radii. In contrast, the longitudinal jets are symmetric, with $\langle\cth\rangle=0.000\pm0.002$ for both radii. Accounting for the containment edges, this corresponds to transverse bosons that are 68\% left-handed (appendix~\ref{app:analytic}). This agrees with predictions and measurements for $W$ bosons at large transverse momentum~\cite{Bern:2011ie,CMS:2011kaj,ATLAS:2012au}. For a 68:32 mixture, perfectly identifying the quark prong would raise the AUC of the ideal tagger by 0.012 without containment and by 0.013--0.014 with it, while the likelihood ratio would remain below 2.2. The energy-sharing proxy cannot use this information, and the BDT sees only the overall jet charge, which reflects the charge of the boson. The networks see the charges and impact parameters of the constituents. They could, in principle, learn to identify the quark prong either from the charm quark present in about half of the hadronic decays, or from the charges of the individual prongs ($+2/3$ for the up-type quark and $+1/3$ for the down-type antiquark). However, these charge assignments separate the prongs only statistically. The ideal decay-angle tagger is therefore the ceiling for jet taggers that cannot distinguish the quark prong from the antiquark prong. The networks of section~\ref{sec:res-scan} identify the quark prong in part, but none of them exceeds the AUC of this tagger.

The production angle adds approximately 0.03 to the AUC (section~\ref{sec:production}), but only for a tagger that sees the recoil. Beyond that, the separation grows substantially only if all transverse bosons share a single helicity (yielding an AUC of $4/5$ for the signed angle) or if there is a second boson in the event. For two perfectly measured bosons, the separation of a pair of longitudinal bosons from a pair of transverse bosons reaches an AUC of 0.77 without the sign and 0.784 with it, assuming our specific mixture of helicities. The decay of a single boson thus never yields an AUC near 0.9. The decays of two bosons can reach this level only if every quark is perfectly identified and all transverse bosons share a single helicity ($\frac32-\frac{24\pi^2}{385}=0.885$ without containment). Achieving larger values for a single boson requires information from outside its decay. This could come from the production process (section~\ref{sec:diboson}), or from differences between the two training samples (such as their transverse-momentum spectra) that a tagger might learn instead of the polarization itself. This is why we evaluate all taggers on samples matched in transverse momentum and compare them against the ideal tagger on the same jets.

\begin{table}[t]
\centering
\begin{tabular}{@{}lccc@{}}
\toprule
 & \multicolumn{3}{c}{Containment} \\
\cmidrule(l){2-4}
Information & none & $R=0.8$ & $R=1.5$ \\
\midrule
\multicolumn{4}{@{}l}{\emph{One boson}} \\
$\acth$ & 0.688 & 0.660 & 0.677 \\
$\cth$, quark identified & 0.700 & 0.673 & 0.690 \\
$\acth$ and $|\cos\Theta|$ & 0.722 & 0.700 & 0.715 \\
$\cth$, quark identified, one transverse helicity & 0.800 & 0.781 & 0.793 \\
\midrule
\multicolumn{4}{@{}l}{\emph{Two bosons, both longitudinal against both transverse}} \\
$\acth$ & 0.769 & 0.729 & 0.754 \\
$\cth$, quarks identified & 0.784 & 0.746 & 0.769 \\
$\cth$, quarks identified, one transverse helicity & 0.885 & 0.864 & 0.877 \\
\bottomrule
\end{tabular}
\caption{Largest AUC for separating longitudinal from transverse $W$ bosons with perfectly measured angles and without detector effects. The decay-angle rows use the shapes of eq.~\eqref{eq:helicity}, $\frac34(1-\cos^2\theta^*)$ for longitudinal bosons and $\frac38(1\pm\cth)^2$ for the two transverse helicities, mixed 68:32 as in the transverse test jets or of one helicity only; with $\acth$, the quark and the antiquark are not told apart. Containment truncates $\acth$ at the edges found for the selected jets, 0.936 with $R=0.8$ and 0.977 with $R=1.5$ (section~\ref{sec:containment}). The row with the production angle is the matrix-element result of table~\ref{tab:production}, with the edge of eq.~\eqref{eq:cmax} applied event by event and a statistical uncertainty of about 0.001. Two bosons decay independently, with the same mixture of helicities, and their likelihood ratios multiply. Without containment, the first row and the two rows for one transverse helicity have the closed forms $11/16$, $4/5$ and $\frac32-\frac{24\pi^2}{385}$; the other decay-angle entries are integrated numerically (appendix~\ref{app:analytic}).}
\label{tab:limits}
\end{table}

\subsection{Event-level information in diboson production and vector-boson scattering}
\label{sec:diboson}

In $W$+jet events, the production angle adds about 0.03 to the AUC (section~\ref{sec:production}). To estimate how much a tagger that sees the whole event can gain when the event contains two bosons, we generate semileptonic diboson events at leading order with the polarized matrix elements of ref.~\cite{BuarqueFranzosi:2019boy}: $W^\pm Z$ and $W^+W^-$ production in which one $W$ boson decays to an electron or a muon and a neutrino and the other boson decays to quarks, with both bosons longitudinal (LL) or both transverse (TT), and $W^\pm Z$ production in which only the $Z$ boson is polarized ($Z_{\mathrm{L}}$ against $Z_{\mathrm{T}}$, with the $W$ boson unpolarized). We also generate the same final states in electroweak vector-boson scattering, $W^\pm W^\pm jj$ (LL and TT) and $W^\pm Z jj$ (LL and TT, and $Z_{\mathrm{L}}$ against $Z_{\mathrm{T}}$), with two tagging jets of $\pT>30\GeV$ and $|\eta|<5$, a dijet mass above 500\GeV{} and a rapidity separation above 2.5 between the two hardest tagging jets, which must lie in opposite hemispheres.\footnote{The built-in rapidity-gap requirement of \MGaMC{} acts on the two hardest jets of the event, including decay products that are exempt from the other jet requirements; for a boosted boson, these are often its two decay quarks. It then rejects symmetric decays and distorts the decay angle: in our first $W^\pm W^\pm jj$ and $W^\pm Z jj$ samples with two longitudinal bosons, it raised $\langle\cos^2\theta^*\rangle$ of the hadronic boson from 0.20 to 0.26. We therefore apply the requirement to the tagging jets only, with a user-defined cut, which restores 0.20.} The quark pair must have $\pT>300\GeV$; with this requirement, LL pairs make up 27\% of the LL and TT cross section in $W^\pm Z$ production and 6\% in $W^+W^-$ production, against 4\% and 3\% in $W^\pm Z jj$ and $W^\pm W^\pm jj$ production. We keep events with a charged lepton of $\pT>25\GeV$ and $|\eta|<2.5$, a neutrino of $\pT>25\GeV$, and a hadronic boson with $|\eta|<2$ whose decay is contained in a jet of radius $R=0.8$ (eq.~\eqref{eq:cmax}). The decay angles are defined as in section~\ref{sec:frame}. A gradient-boosted classifier, trained with five-fold cross-validation on about $5\times10^4$ events per class, then measures how well the classes separate with nested sets of information (table~\ref{tab:diboson}): the folded decay angle of the hadronic boson, which is what a jet tagger can use; both decay angles, where the charge of the lepton signs the leptonic one; the production kinematics alone, from the momenta of the two bosons and, in vector-boson scattering, of the tagging jets; and all of them together, with the true neutrino or with its longitudinal momentum reconstructed from the $W$-boson mass.

\begin{table}[t]
\centering
\footnotesize
\setlength{\tabcolsep}{4pt}
\begin{tabular}{@{}lcccccc@{}}
\toprule
 & Hadronic & Both decay & Production & Decay angles & Reconstructed & Quark \\
Process, classes & decay angle & angles & only & and production & neutrino & identified \\
\midrule
$W^\pm Z$, LL vs TT & 0.624 & 0.696 & 0.900 & 0.916 & 0.894 & 0.920 \\
$W^\pm Z$, $Z_{\mathrm{L}}$ vs $Z_{\mathrm{T}}$ & 0.627 & 0.686 & 0.879 & 0.894 & 0.873 & 0.899 \\
$W^+W^-$, LL vs TT & 0.635 & 0.695 & 0.726 & 0.780 & 0.770 & 0.806 \\
\midrule
$W^\pm W^\pm jj$, LL vs TT & 0.639 & 0.691 & 0.906 & 0.921 & 0.915 & 0.925 \\
$W^\pm Z jj$, LL vs TT & 0.626 & 0.679 & 0.882 & 0.901 & 0.896 & 0.904 \\
$W^\pm Z jj$, $Z_{\mathrm{L}}$ vs $Z_{\mathrm{T}}$ & 0.631 & 0.633 & 0.814 & 0.832 & 0.829 & 0.837 \\
\bottomrule
\end{tabular}
\caption{Separation of longitudinal from transverse bosons in semileptonic diboson production (upper rows) and vector-boson scattering (lower rows) at matrix-element level, for a hadronic boson with $\pT>300\GeV$ whose decay is contained in a jet of radius $R=0.8$. The first column is the AUC of the folded decay angle of the hadronic boson itself. The others are cross-validated AUCs of a gradient-boosted classifier using the two decay angles (the leptonic one signed by the charge of the lepton), the momenta of the two bosons (and of the tagging jets in vector-boson scattering), both, both with the longitudinal momentum of the neutrino reconstructed from the $W$-boson mass, and both with the quark prong of the hadronic decay identified. LL: both bosons longitudinal; TT: both transverse; for $Z_{\mathrm{L}}$ vs $Z_{\mathrm{T}}$, only the $Z$ boson is polarized. With $R=1.5$, the values are higher by 0.02--0.04 in the first two columns and by up to 0.02 in the others; without the containment requirement, the hadronic decay angle reaches 0.684--0.687. Each entry uses about $5\times10^4$ events per class, and the statistical uncertainty is about 0.002.}
\label{tab:diboson}
\end{table}

In diboson production, the two decay angles together reach an AUC of about 0.70 in both processes, more than the 0.62--0.64 of the hadronic decay angle alone. Without the lepton and containment requirements, they reach 0.77, as table~\ref{tab:limits} predicts for two bosons (0.769); the lepton requirements remove leptons emitted backward in the $W$ rest frame, where the transverse distribution peaks. The production kinematics add far more in $W^\pm Z$ production. On their own, they separate LL from TT pairs with an AUC of 0.90, and the polar angle of the $Z$ boson in the diboson rest frame alone reaches 0.90: longitudinal pairs are produced at large angles to the beam, whereas transverse pairs are produced mostly at small angles. With the decay angles, the AUC rises to 0.92, or to 0.89 with the reconstructed neutrino. Even with the $W$ boson unpolarized, the polarization of the $Z$ boson is tagged with an AUC of 0.89 (0.87 with the reconstructed neutrino). In $W^+W^-$ production, the production kinematics carry less (0.73), the event-level AUC is 0.78 (0.77 with the reconstructed neutrino), and identifying the quark prong adds another 0.03, against 0.004 in $W^\pm Z$ production. Adding the momenta of all final-state particles leaves these values unchanged, so the variables capture the information in the events. These are upper limits: showering, detector effects and the reconstruction of the hadronic boson as a jet lower them, and LL against TT is the most favorable comparison, since mixed states lie in between.

In vector-boson scattering, the decay angles carry about as much information as in diboson production: 0.63--0.64 for the hadronic decay angle and 0.68--0.69 for both decay angles of LL and TT pairs. The production kinematics again carry far more, 0.91 in $W^\pm W^\pm jj$ and 0.88 in $W^\pm Z jj$ production, but here the information lies mostly in the tagging jets. Their rapidity separation alone reaches 0.78 and 0.80, whereas the polar angle of the hadronic boson in the diboson rest frame, which carries most of the information in $W^\pm Z$ production, reaches only 0.56--0.60. With the decay angles, the AUC rises to 0.92 and 0.90, and to 0.83 for the $Z$ boson alone with the $W$ boson unpolarized; the reconstructed neutrino costs less than 0.01, and identifying the quark prong adds at most 0.005. In each process the gain is large, but it comes from different variables.

The gain from the production kinematics is large but specific to the process. The decay-angle distributions of eq.~\eqref{eq:helicity} hold for any process, so a jet tagger trained in one process measures the same quantity in another. The production kinematics, in contrast, depend on the process and on the couplings that produce the bosons, which are the couplings a polarization measurement in vector-boson scattering aims to test. An event-level tagger must therefore be trained and calibrated on polarized samples of the process to which it is applied, and a change in the production dynamics changes its response as well as the polarization fractions. Going from a generic jet tagger to a process-specific event-level tagger thus brings large gains, at the risk of losing generality.

\section{Simulated samples}
\label{sec:samples}

\subsection{Event generation}
\label{sec:generation}

Events are generated at leading order in QCD for proton--proton collisions at $\sqrt{s}=13\TeV$ using \MGaMC version 3.6.4~\cite{Alwall:2014hca}. The $W$, $Z$ and Higgs bosons are decayed to quark pairs within \MGaMC, which retains the full spin correlations.

\paragraph{Pretraining samples.} Table~\ref{tab:ptsamples} lists the pretraining samples. Each sample is generated in slices of the transverse momentum of the recoiling parton (or of the top quark for $t\bar t$), with equal numbers of events per slice, so that the selected jets populate the whole range of 300--1500\GeV. Because \MGaMC applies no transverse-momentum requirements to a resonance inside a decay chain, the top quarks are generated stable and then decayed by \Pythia, with the intermediate $W$ boson forced to decay hadronically. The $H\to b\bar b$ sample uses the HEFT model because the five-flavor model with massless $b$ quarks sets the $b$-quark Yukawa coupling to zero. The production comprises $4\times10^{6}$ events for each of the $W^+$, $W^-$, $Z$ and $H$ samples (four slices of ten runs of $10^{5}$ events), $3\times10^{6}$ events for $t\bar t$ and $2\times10^{6}$ for QCD, for a total of about $2.1\times10^{7}$ events. Two top-up samples are generated for the eight-class variant described in section~\ref{sec:construction}: $Z\to b\bar b$+jet events (because $b\bar b$ pairs make up only about a fifth of hadronic $Z$ decays) and additional QCD dijets (because this class is split in two). Each top-up sample contains $4\times10^{6}$ events in the same slices. Both are showered and passed through the detector simulation like the other samples. No polarized, VBS or diboson samples are used in pretraining.

\begin{table}[t]
\centering
\small
\setlength{\tabcolsep}{3pt}
\begin{tabular}{@{}llllr@{}}
\toprule
Class & Process & Truth match & Slices [GeV] & $\sigma$ (one slice) \\
\midrule
$W^+$ & $pp\to W^+j$, $W^+\to q\bar q'$ & $W$, two quarks in jet & recoil 250, 450, 700, 1000 & 1.40\pb \\
$W^-$ & $pp\to W^-j$, $W^-\to q\bar q'$ & $W$, two quarks in jet & as $W^+$ & 0.74\pb \\
$Z$ & $pp\to Zj$, $Z\to q\bar q$ (incl.\ $b\bar b$) & $Z$, two quarks in jet & as $W^+$ & 0.93\pb \\
$H$ & $pp\to Hj$, $H\to b\bar b$ (HEFT) & $H$, two quarks in jet & as $W^+$ & 14.8\fb \\
top & $pp\to t\bar t$, $t\to bq\bar q'$ & $t$, three quarks in jet & top $\pT$ 300, 500, 800 & 1.38\pb \\
QCD & $pp\to jj$ & none; two leading jets & as $W^+$ & 0.96\nb \\
\midrule
$Z$ & $pp\to Zj$, $Z\to b\bar b$ (top-up) & $Z$, two quarks in jet & as $W^+$ & 0.205\pb \\
QCD & $pp\to jj$ (top-up) & none; two leading jets & as $W^+$ & as QCD \\
\bottomrule
\end{tabular}
\caption{Pretraining samples. The generation slices are given by their lower edges; the last one is open-ended. The cross sections are leading-order values for the 450--700\GeV{} recoil slice (for $t\bar t$, a top quark above 500\GeV), determined from a first production of $10^4$ events per sample; for the $Z\to b\bar b$ top-up sample, the cross section is taken from its ten production runs in that slice. The two top-up samples are used only in the eight-class variant (section~\ref{sec:construction}).}
\label{tab:ptsamples}
\end{table}

\paragraph{Polarized fine-tuning samples.} The fine-tuning samples consist of $W^\pm$+jet events with the $W$ boson in a definite helicity state, generated using the polarized matrix elements of ref.~\cite{BuarqueFranzosi:2019boy} (\texttt{w+\char`\{0\char`\}} and \texttt{w+\char`\{T\char`\}}, along with their $W^-$ analogs), as well as QCD dijet events. The recoiling parton is required to have $\pT>450\GeV$ and $|\eta|<5$, while generator-level cuts on the decay quarks are disabled to keep the decay distributions unbiased. Any mixture of $W^+$ and $W^-$ bosons must be the same in both classes: because the jets carry the boson charge (section~\ref{sec:properties}), an unequal mixture would let the charge stand in for the polarization. The polarized jets in this paper are $W^+$ jets from samples of $3\times10^5$ events per helicity, in 15 files of $2\times10^4$ events each (section~\ref{sec:res-scan}).

\paragraph{Transfer and closure samples.} The transfer samples, which are never used in training, comprise electroweak $W^\pm W^\pm jj$ production generated as $\WL\WL$ and $\WT\WT$ (since \MGaMC does not combine polarized and unpolarized identical bosons in a single process); $W^\pm Zjj$ production with either a longitudinal or a transverse $Z$ boson; SM $W^+W^-$ production; and a heavy scalar $X$ with $m_X=1\TeV$ decaying to $W^+W^-$, whose $W$ bosons are longitudinal. A generator-level selection retains events in which the quark pair whose invariant mass is closest to $m_W$ has $\pT>300\GeV$. Unpolarized $W$+jet events are used for a closure test (section~\ref{sec:res-outlook}).

\subsection{Showering and detector simulation}

Parton showering, hadronization and multiple parton interactions are simulated with the Monash tune~\cite{Skands:2014pea} of \Pythia~\cite{Bierlich:2022pfr}, and the detector response is modeled with \Delphes~3.5.1~\cite{deFavereau:2013fsa} and its CMS card. Both steps run in a single pass over chunks of $2\times10^{4}$ events. Because the default CMS card sets the impact-parameter uncertainties to zero, we extend it with a track-smearing module that applies CMS-like resolutions to the transverse and longitudinal impact parameters (appendix~\ref{app:sim}). The jet constituents are the energy-flow candidates (tracks, photons and neutral hadrons). No additional pile-up interactions are simulated.

\subsection{Jets and truth matching}
\label{sec:jets}

Jets are clustered from energy-flow candidates with $\pT>0.5\GeV$ using the anti-$k_t$ algorithm~\cite{Cacciari:2008gp} with a radius parameter of $R=0.8$, as implemented in \FastJet~\cite{Cacciari:2011ma}, and are required to satisfy $|\eta|<2$. In samples containing a resonance, the hard-process $W$, $Z$, Higgs boson or top quark, along with its decay quarks, is identified in the generator record. The identification follows the chain of copies created by the parton shower and, for top quarks, the decay through the intermediate $W$ boson. We take the highest-$\pT$ jet within $\Delta R<0.8$ of the resonance and keep it if its axis lies within $\Delta R<0.8$ of every decay quark. Up to two resonances are matched per event, which matters only for the $t\bar t$ sample. In the QCD samples, the two leading jets are kept. For the eight-class variant (section~\ref{sec:construction}), each QCD jet is assigned the flavor of the hard-process parton (quark or gluon) closest to its axis within $\Delta R<0.8$; jets without such a parton are discarded. The pretraining selection requires $300<\pT<1500\GeV$, whereas the fine-tuning selection requires $500<\pT<1000\GeV$.

The fraction of generated events that yields a selected jet rises over the first three slices and then levels off. For $W^+$ events, it is 0.31, 0.83, 0.94 and 0.92 in the four recoil slices, and for the Higgs boson 0.27, 0.80, 0.93 and 0.91. These values reflect the jet $\pT$ threshold and the containment requirement of eq.~\eqref{eq:cmax}. The production yields $1.82\times10^{7}$ selected jets with $R=0.8$.

\begin{figure}[t]
\centering
\includegraphics[width=\textwidth]{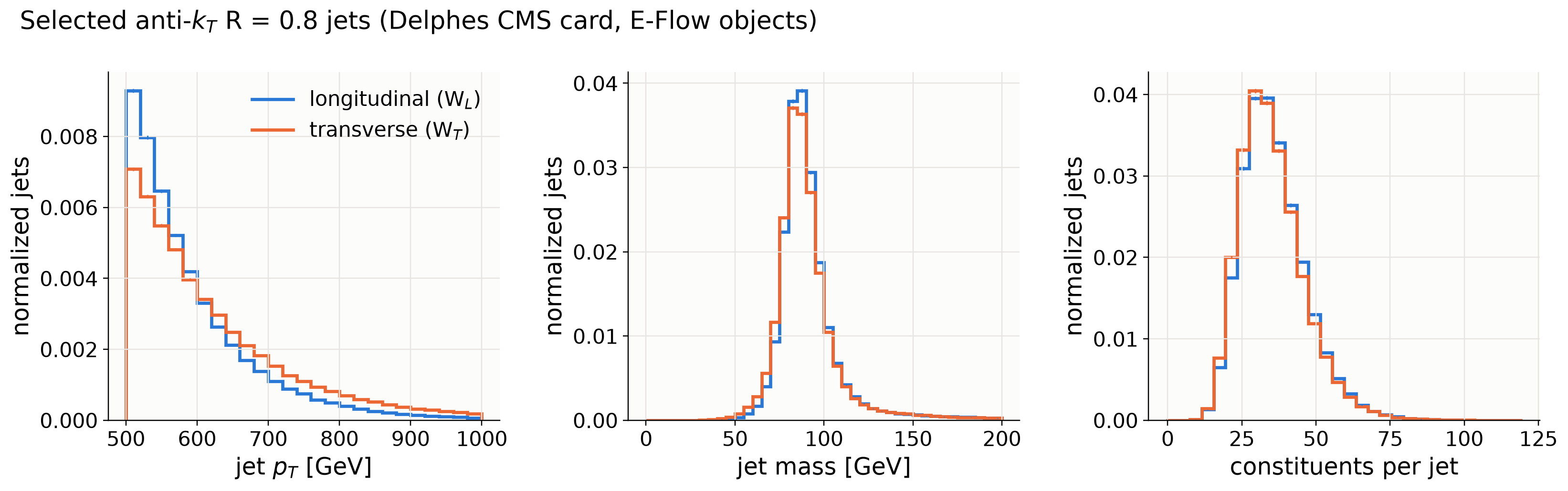}\\[6pt]
\includegraphics[width=\textwidth]{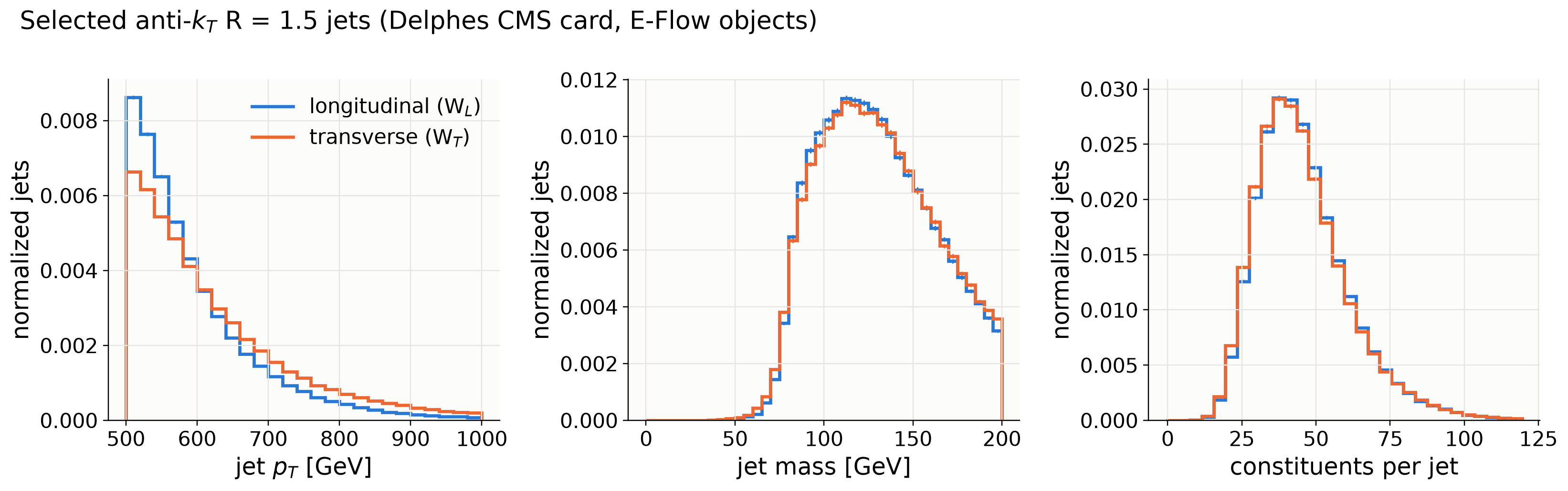}
\caption{Transverse momentum (left), mass (middle) and number of constituents (right) of the selected, truth-matched jets in the samples of helicity-selected $W^+$+jet events, with $R=0.8$ (top) and $R=1.5$ (bottom), before $\pT$ matching. Each distribution is normalized to unit area in the range shown; for $R=1.5$, the mass distribution extends beyond 200\GeV. The longitudinal jets have a softer $\pT$ spectrum because the longitudinal cross section falls more steeply with $\pT$. The comparisons in section~\ref{sec:res-scan} use $\pT$-matched jets.}
\label{fig:jets}
\end{figure}

A second jet definition uses $R=1.5$ in the clustering, in the resonance and parton matching and in the containment requirement. These large-radius jets are not groomed, because grooming removes soft prongs at wide angles and with them the asymmetric decays that carry the polarization. The price is a larger contribution from the underlying event. The containment edge moves outward (table~\ref{tab:cmax}), and three-prong top-quark jets are contained from about $2m_t/R\approx230\GeV$, instead of $430\GeV$ for $R=0.8$. Figure~\ref{fig:jets} compares the two radii using the polarized $W^+$+jet samples. With $R=1.5$, the mass of the ungroomed jets, which peaks near 85\GeV{} for $R=0.8$, spreads from 80\GeV{} to beyond 200\GeV, and the jets contain more constituents. Grooming restores this peak: the soft-drop mass peaks at the same value for both radii (figure~\ref{fig:substructure}).

\subsection{Jet representation}

Each jet is represented by its 128 highest-$\pT$ constituents (zero-padded if necessary), with the nine features listed in table~\ref{tab:features}. The particle-identification index distinguishes charged hadrons, electrons and muons of either charge, photons and neutral hadrons, as in the published pretrained models~\cite{Bhimji:2025omnilearned}. The impact parameters are those of the charged tracks and are set to zero for neutral candidates. For each jet, we also store its kinematics and constituent multiplicity, the generator-level boson $\pT$, the value of $\cth$ in both reference frames of section~\ref{sec:frame}, the boson identity, the helicity label and the generator weight. For the regression objective of section~\ref{sec:objectives}, the directions and energy shares of the decay quarks are stored as well.

\begin{table}[t]
\centering
\small
\begin{tabular}{@{}ll@{}}
\toprule
Feature & Definition \\
\midrule
$\Delta\eta$, $\Delta\phi$ & pseudorapidity and azimuth relative to the jet axis \\
$\log\pT$, $\log E$ & transverse momentum and energy in GeV; $\pT=0$ marks padding \\
PID index & position in $(-211,-13,-11,0,11,13,22,130,211)$ \\
$\tanh d_0$, $\sigma(d_0)$ & transverse impact parameter and its uncertainty (tracks) \\
$\tanh d_z$, $\sigma(d_z)$ & longitudinal impact parameter and its uncertainty (tracks) \\
\bottomrule
\end{tabular}
\caption{Constituent features, following the input format of the published pretrained models of ref.~\cite{Bhimji:2025omnilearned}.}
\label{tab:features}
\end{table}

\section{The pretraining dataset}
\label{sec:dataset}

\subsection{Construction}
\label{sec:construction}

The pretraining dataset has six classes: $W^+$, $W^-$, $Z$, $H$, top and QCD. We keep $W^+$ and $W^-$ as separate classes because the jets carry the boson charge through the charges of their constituents; telling the two apart is a task in its own right.

An eight-class variant splits two of these classes: $Z$ bosons by decay flavor into $Z\to q\bar q$ (with $q=u,d,s,c$) and $Z\to b\bar b$, and QCD jets by the initiating parton into quark and gluon jets (section~\ref{sec:jets}). Both splits require the network to resolve differences within what the six-class objective treats as a single class: the heavy-flavor content that $Z\to b\bar b$ shares with Higgs-boson jets and the different radiation patterns of quarks and gluons. They thus test whether a finer classification keeps more of the variation within a class (section~\ref{sec:res-outlook}). In the full sample of $2.9\times10^6$ matched $Z$-boson jets, $b\bar b$ pairs make up $(21.99\pm0.02)\%$. This matches the 22.0\% expected from the generator-level cross sections of the two $Z$ samples in the 450--700\GeV{} slice (table~\ref{tab:ptsamples}) and is close to the leading-order value of 21.9\%. The higher fraction of 22.8\% seen in the first two converted files was an artifact of their $\pT$ slice. The fraction falls with the jet $\pT$, from 24.0\% at 300--450\GeV{} to 18.8\% at 1000--1500\GeV. This dependence presumably arises because $b$ jets lose part of their transverse momentum to neutrinos from semileptonic decays, which moves them to lower jet $\pT$. Applying the same construction to $R=1.5$ jets yields a third dataset.

Files are assigned to the training, validation and test splits in a 70:10:20 ratio within each sample and generation slice, so that no random sequence of the shower or detector simulation is shared between splits. Because the network's training code (section~\ref{sec:network}) does not accept per-jet weights, the classes are matched in jet $\pT$ by resampling rather than reweighting. With $N_c(b)$ denoting the number of training jets of class $c$ in $\pT$ bin $b$, a jet of class $c$ in bin $b$ is kept with probability
\begin{equation}
a_c(b) = \frac{\min_{c'} N_{c'}(b)}{N_c(b)}\,,
\label{eq:accept}
\end{equation}
so that each class keeps as many jets per bin as the scarcest class has in that bin. Among all choices that give identical spectra for all classes, this one keeps the largest number of jets. We use twelve bins with edges at 300, 350, 400, 450, 500, 600, 700, 800, 900, 1000, 1150, 1300 and 1500\GeV. The same acceptance probabilities are applied to the validation and test splits, which are then reduced to equal class counts by a random draw across all files of each class. Table~\ref{tab:splits} gives the resulting dataset sizes. The final proportions are 72:8:20 rather than 70:10:20, because the QCD sample, generated with half as many runs per slice, divides into whole files less evenly (18:2:5 files per slice instead of 35:5:10), and the validation and test splits of the other classes are reduced to match the QCD counts. Each class keeps between 74\% and 80\% of its selected jets.

\begin{table}[t]
\centering
\begin{tabular}{@{}lrrrr@{}}
\toprule
 & Training & Validation & Test & Total \\
\midrule
Jets per class & 1,679,229 & 186,798 & 466,672 & 2,332,699 \\
All six classes & 10,075,374 & 1,120,788 & 2,800,032 & 13,996,194 \\
\bottomrule
\end{tabular}
\caption{Size of the pretraining dataset after $\pT$ matching and class balancing.}
\label{tab:splits}
\end{table}

A second variant of the dataset adds nine regression targets per jet: the direction relative to the jet axis ($\Delta\eta$, $\Delta\phi$) and the energy share of up to three decay quarks, ordered by energy and set to zero if absent. The datasets are stored in HDF5 files in chunks of eight jets, small enough that reading the data does not limit the training throughput (appendix~\ref{app:computing}).

\subsection{Properties of the pretraining jets}
\label{sec:properties}

\begin{figure}[t]
\centering
\includegraphics[width=\textwidth]{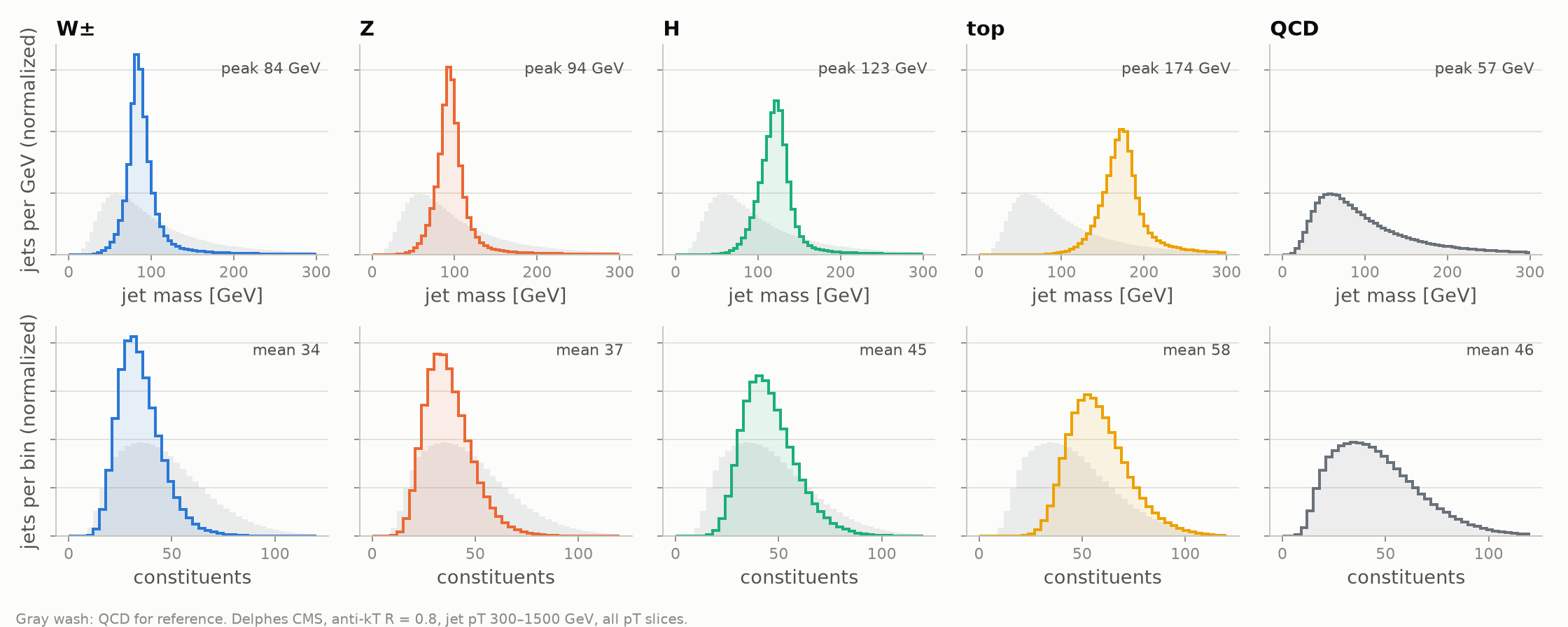}
\caption{Jet mass (top) and number of constituents (bottom) of the pretraining classes, with $W^+$ and $W^-$ combined. Each distribution is normalized to unit area, and the gray band shows QCD jets for reference. The peak positions, estimated to within about 1\GeV, and the mean multiplicities are quoted in each panel.}
\label{fig:mass}
\end{figure}

\begin{figure}[t]
\centering
\includegraphics[width=\textwidth]{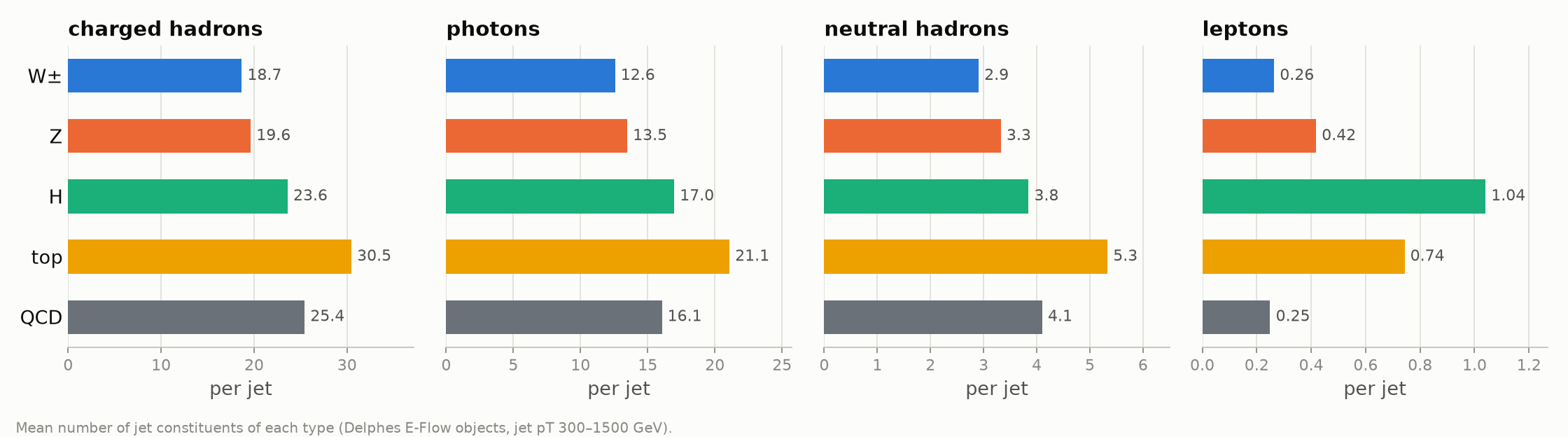}
\caption{Mean number of constituents per jet for each particle type, from the energy-flow candidates of the pretraining classes.}
\label{fig:composition}
\end{figure}

\begin{figure}[t]
\centering
\includegraphics[width=\textwidth]{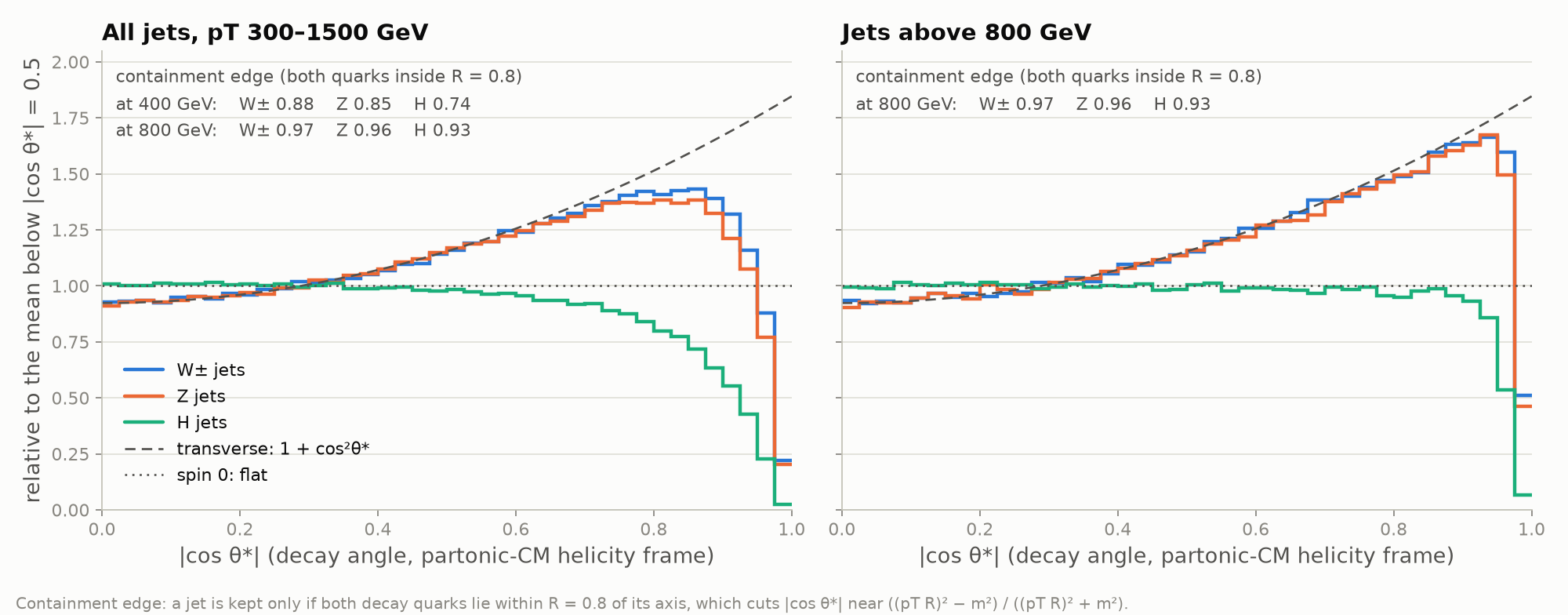}
\caption{Truth $\acth$ in the partonic center-of-mass frame (section~\ref{sec:frame}) of the selected $W^\pm$, $Z$ and Higgs-boson jets, each normalized to its mean below $\acth=0.5$. Left: all jets (300--1500\GeV); right: jets above 800\GeV. The dashed curve shows the transverse shape $1+\cos^2\theta^*$ and the dotted line the flat distribution of a spin-0 decay. Each panel quotes the containment edges of eq.~\eqref{eq:cmax}.}
\label{fig:costheta}
\end{figure}

\begin{figure}[t]
\centering
\includegraphics[width=\textwidth]{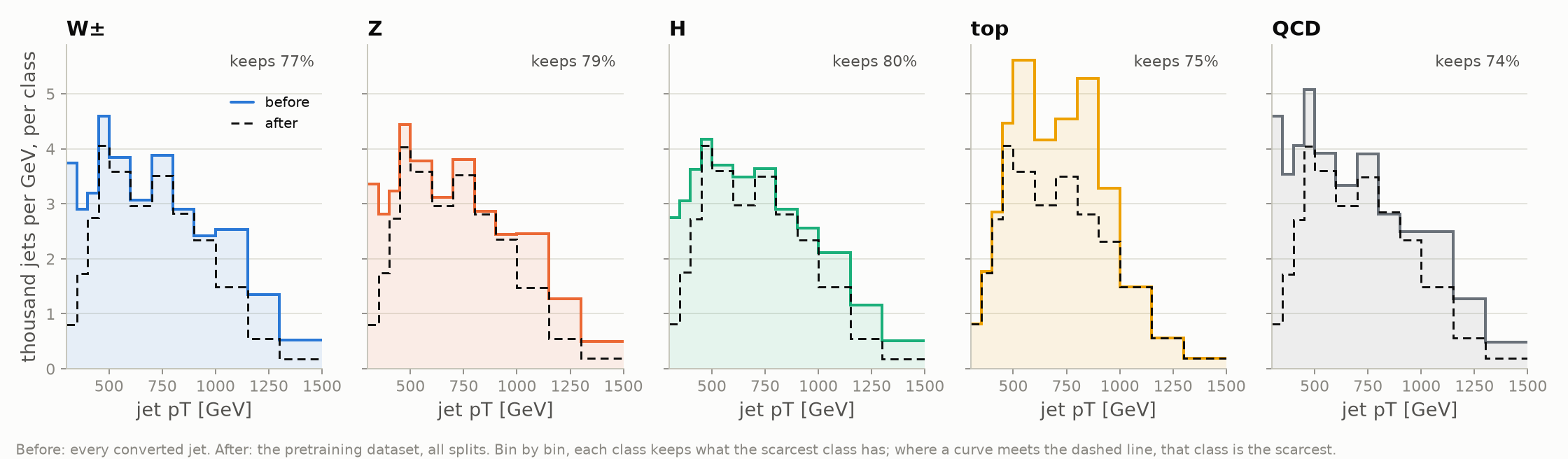}
\caption{Jets per class per GeV before (solid: all selected jets) and after (dashed: pretraining dataset, all splits) $\pT$ matching, with $W^+$ and $W^-$ averaged. Each panel quotes the fraction of jets kept. Where a class curve touches the dashed line, that class is the scarcest in that bin.}
\label{fig:matching}
\end{figure}

Figures~\ref{fig:mass}--\ref{fig:matching} characterize the six classes using a subset of 40 files per class, spread over all generation slices. The jet mass distributions (figure~\ref{fig:mass}) peak at 84 ($W$), 94 ($Z$), 123 ($H$) and 174\GeV{} (top). The $W$ and $Z$ peaks lie about 3\GeV{} above the boson masses, a shift consistent with the contributions of the underlying event and of detector smearing to the mass of ungroomed $R=0.8$ jets. The Higgs-boson peak lies 2\GeV{} below the boson mass because neutrinos from semileptonic $b$- and $c$-hadron decays escape detection. The mean number of constituents is 34 ($W$), 37 ($Z$), 45 ($H$), 46 (QCD) and 58 (top). Higgs-boson jets have about as many constituents as QCD jets, a large fraction of which are gluon jets, and three-prong top-quark jets have the most.

The constituent composition (figure~\ref{fig:composition}) largely tracks the total multiplicity for hadrons and photons, whereas the lepton content traces the presence of heavy flavor. There are 1.04 electrons or muons per jet for $H\to b\bar b$, 0.74 for top quarks and 0.42 for $Z$ bosons, compared to 0.26 for $W$ bosons and 0.25 for QCD jets. The charges of the constituents also carry the boson charge, which is why $W^+$ and $W^-$ are kept in separate classes.

Figure~\ref{fig:costheta} shows the decay-angle distributions of the selected $W$, $Z$ and Higgs-boson jets. The $W$ and $Z$ jets follow the transverse shape $1+\cos^2\theta^*$, as expected from section~\ref{sec:frame}. The Higgs-boson jets follow the flat distribution of a spin-0 decay up to the containment edge. Above 800\GeV, where $\cmax\ge0.93$ for all three bosons, the Higgs-boson distribution is flat up to $\acth\approx0.9$, while the $W$ and $Z$ distributions follow $1+\cos^2\theta^*$ up to about 0.93. In the inclusive sample, Higgs-boson jets at low $\pT$ can only populate small $\acth$ values ($\cmax=0.57$ at 300\GeV{} and 0.74 at 400\GeV). This kinematic restriction depletes the inclusive distribution from $\acth\approx0.5$ onward. The acceptance therefore correlates the class labels of the pretraining set with the decay angle, in a way that depends on the boson mass and on $\pT$.

Figure~\ref{fig:matching} shows the jet $\pT$ spectra before and after matching. Top-quark jets are the scarcest class below 450\GeV, where their three decay quarks rarely fit in one jet (which requires $\pT\gtrsim2m_t/R\approx430\GeV$), and again above 1\TeV, where the highest $t\bar t$ generation slice runs out. In between, Higgs-boson, $W$ and QCD jets take turns as the scarcest class. As a result, 65\% of the matched jets lie in the fine-tuning window of 500--1000\GeV, while about 5\% lie below 400\GeV{} and another 5\% above 1150\GeV. Each of these tails still holds about $9\times10^{4}$ training jets per class. The steps at 450--500\GeV{} and 700--800\GeV{} stem from generating equal numbers of events per slice. Since they are common to all classes, they carry no class information.

\section{Network, pretraining and fine-tuning}
\label{sec:network}

\subsection{Architecture}

All transformer networks in this study use the small configuration of the point-edge transformer PET~v2, as implemented in version 0.2.0 of the \OL package~\cite{Bhimji:2025omnilearned}. It has eight transformer blocks, eight attention heads and a latent dimension of 128, with about $3\times10^{6}$ trainable parameters. The constituent kinematics $(\Delta\eta,\Delta\phi,\log\pT,\log E)$ are embedded together with local features built from the ten nearest neighbors of each constituent in the $(\eta,\phi)$ plane. The pairwise interaction terms enter the global attention as a bias, the particle-identification index is embedded via a lookup table, and the four impact-parameter features are embedded through a small multilayer perceptron. This is the configuration of the published checkpoints, so fine-tuning from their generic pretraining loads every pretrained layer except the output layer and the time embedding used by the generative objective. That generic pretraining used more than $10^9$ jets from seven datasets~\cite{Bhimji:2025omnilearned}. Among them is JetClass~\cite{Qu:2022mxj}, whose classes include $W\to q\bar q'$, $Z\to q\bar q$, Higgs-boson and top-quark jets with transverse momenta of 500--1000\GeV, simulated with \Delphes. The generic baseline has therefore already seen boosted boson decays, mixed into a much broader dataset, so the comparison with it tests a focused boson pretraining against a broad one rather than against no pretraining.

\subsection{Pretraining objectives}
\label{sec:objectives}

We define four pretraining objectives on the dataset described in section~\ref{sec:dataset}:
\begin{itemize}
\item \rung{pt\_cls}: classification of the six process classes;
\item \rung{pt\_gen}: label-free generative training applying the flow-matching objective of ref.~\cite{Bhimji:2025omnilearned} to the constituent kinematics;
\item \rung{pt\_both}: both objectives combined, which is the recipe of the generic pretraining with only the data replaced;
\item \rung{pt\_reg}: regression of the nine decay-quark targets defined in section~\ref{sec:dataset}.
\end{itemize}
The last objective comes close to learning the decay angle itself, so we treat it as a separate variant rather than as evidence for boson pretraining as such. Each run trains for 30 epochs of 2,000 steps with 128 jets per GPU on four NVIDIA A100 GPUs, about $3\times10^{7}$ jets or three passes over the training set. The learning rate of $3\times10^{-4}$ follows a cosine schedule to zero, and the weight decay is 0.1. For \rung{pt\_cls}, an epoch (including 2,000 validation steps) takes 403\,s, and the full run takes 3.4 hours. The training and validation losses converge to 0.498 and 0.499, respectively, compared with $\ln6\approx1.79$ for a uniform guess. The generative run, \rung{pt\_gen}, takes about as long and reaches a flow-matching loss of 1.935 on the validation set. The combined run, \rung{pt\_both}, takes 820\,s per epoch, 6.9 hours in total. The generative part of its loss converges to a similar value (1.939 on the validation set), while the classification part ends at 0.733, well above the 0.499 of \rung{pt\_cls}. The regression pretraining \rung{pt\_reg} and the eight-class variant (section~\ref{sec:construction}), which is to be pretrained with \rung{pt\_cls} at both radii and with \rung{pt\_gen} at $R=1.5$, have not yet been run.

\subsection{Fine-tuning and baselines}
\label{sec:finetuning}

For the polarization task, each network is trained as a two-class classifier to distinguish longitudinal from transverse $W$-boson jets. Including a third class of QCD jets would make the validation loss used for checkpoint selection nearly blind to the polarization (appendix~\ref{app:analytic}). Rejecting QCD jets is instead the task of a separate $W/Z$-versus-QCD selection (section~\ref{sec:res-outlook}). Following the recipe of ref.~\cite{Bhimji:2025omnilearned}, fine-tuning from a pretrained network uses a learning rate of $5\times10^{-6}$ for the transformer blocks and a rate five times larger for the input embeddings, the interaction layers and the output layer. We use the Lion optimizer~\cite{Chen:2023lion} with a weight decay of 0.1, one warm-up epoch and a batch size of 128 jets on a single GPU. Pretrained networks are fine-tuned either fully, with all weights updated as above, or frozen, with the transformer blocks kept at their pretrained weights so that only the input embeddings, the interaction layers and the output layer are trained. Because the package's command-line interface has no option to freeze layers, the frozen variants run the transformer blocks at a negligible learning rate of $10^{-9}$ and the other layers at $5\times10^{-4}$. Networks trained from scratch use the AdamW optimizer~\cite{Loshchilov:2017bfa} with a learning rate of $3\times10^{-4}$, a weight decay of 0.05 and a linear warm-up over about 400 optimizer steps, rounded up to between one and ten whole epochs; with the Lion optimizer at a learning rate of $5\times10^{-4}$ and a weight decay of 0.5, as in the recipe of ref.~\cite{Bhimji:2025omnilearned}, every one of our trainings from scratch diverged within 10 to 40 epochs. The recipe prescribes ten epochs (15 when training from scratch). This suffices for large training sets but not for small ones, because the number of optimizer steps per epoch scales with the number of training jets. Every network is therefore trained for at least $4\times10^{3}$ optimizer steps (100 epochs for a training set of $5\times10^3$ jets), and the checkpoint with the lowest validation loss is kept. Before fine-tuning from our own pretraining, we remove the optimizer state from the checkpoint because the number of classes changes from six (or eight) to two.

Two baselines without neural networks complete the comparison. The first is the energy-sharing proxy $p_\theta$ of eq.~\eqref{eq:ptheta}, which requires no training and is computed from the two exclusive-$k_t$ subjets of the jet. The second is a boosted decision tree (BDT) trained with XGBoost~\cite{Chen:2016btl} (up to 500 trees of depth 4) on 21 substructure observables, including the jet mass, the soft-drop mass~\cite{Larkoski:2014wba}, $N$-subjettiness ratios~\cite{Thaler:2010tr}, $D_2$~\cite{Larkoski:2014gra}, $p_\theta$ and the jet charge. Table~\ref{tab:ladder} lists all models compared in this study. Each neural model is trained three times on each of six nested, class-balanced training sets of $10^3$ to $5\times10^4$ jets per class, which share one test set (section~\ref{sec:res-scan}).

\begin{table}[t]
\centering
\small
\begin{tabular}{@{}lll>{\raggedright\arraybackslash}p{0.25\textwidth}@{}}
\toprule
Model & Pretraining & Polarization training & What the comparison isolates \\
\midrule
$p_\theta$ proxy & none & none & the decay angle alone \\
BDT & none & XGBoost, 21 observables & the classical ceiling \\
\rung{scratch\_adam} & none & full network & the network without pretraining \\
\rung{ft\_s} & generic, published & full; frozen & generic pretraining (baseline) \\
\rung{ft\_ours} $\leftarrow$ \rung{pt\_cls} & ours, six classes & full; frozen & boson identity (main model) \\
\rung{ft\_ours} $\leftarrow$ \rung{pt\_gen} & ours, generative & full; frozen & unlabeled boson jets alone \\
\rung{ft\_ours} $\leftarrow$ \rung{pt\_both} & ours, both & full; frozen & the generic recipe on our data \\
\rung{ft\_ours} $\leftarrow$ \rung{pt\_reg} & ours, decay quarks & full & decay-aware pretraining \\
\midrule
PFN, EFN students & none (teacher) & distilled; labels only & what the teacher passes on \\
observable student & none (teacher) & distilled; labels only & what observables capture \\
\bottomrule
\end{tabular}
\caption{Models compared in this study. Frozen: only the input embeddings, the interaction layers and the output layer are trained, and the transformer blocks keep their pretrained weights. PFN, EFN: particle-flow and energy-flow networks. All fine-tuned and pretrained models share the architecture described in section~\ref{sec:network}; the distilled student networks are described in section~\ref{sec:distillation}. A fine-tuned medium-size network may be added in the future to probe the effect of model size.}
\label{tab:ladder}
\end{table}

\subsection{Knowledge distillation}
\label{sec:distillation}

The teacher is a fine-tuned network (section~\ref{sec:res-kd}). The student sees the teacher only through its class probabilities $p_t$ for each jet, which we obtain with the evaluation step of the \OL package. A student with logits $z$ is trained on the loss
\begin{equation}
\mathcal{L} = \alpha\,T^2\,\mathrm{KL}\!\left(p_t^{(T)} \,\middle\|\, \mathrm{softmax}(z/T)\right) + (1-\alpha)\,\mathrm{CE}(y, z)\,, \qquad p_t^{(T)} \propto p_t^{1/T}\,,
\label{eq:kd}
\end{equation}
where the softened probabilities $p_t^{(T)}$ are obtained by dividing the teacher's logits by the temperature $T$, and the cross entropy with the true label $y$ applies only to labeled jets~\cite{Hinton:2015dark}. We use $T=3$ and $\alpha=0.5$. The transfer set consists of the labeled training jets, either alone or together with a larger pool of jets with their labels removed, which the teacher scores. Because the teacher has seen its own training jets, its predictions on them are sharper than on new jets, which limits what it can pass on through them; the unlabeled jets do not have this limitation.

We distill the teacher into three student networks. First, a particle-flow network (PFN)~\cite{Komiske:2018cqr} applies a multilayer perceptron with layers of 64, 64 and 32 nodes to the features of each constituent, with the particle-identification index one-hot encoded. It then sums the outputs over the constituents and maps the sum through a second perceptron (64, 64) to the two classes, with $1.4\times10^4$ parameters in total. Second, its infrared- and collinear-safe counterpart, an energy-flow network (EFN), processes only the constituent directions weighted by their transverse-momentum fractions ($1.3\times10^4$ parameters). Third, an observable network ($5\times10^3$ parameters) uses eight jet observables computed from the constituents: the logarithm of the jet $\pT$, the jet mass, the constituent multiplicity, the energy asymmetry $p_\theta$ of two prongs found with a $\pT$-weighted two-means split, the separation of the prongs, $\tau_{21}$ computed with the same axes, $p_{\mathrm{T}}^{D}$ and the jet charge with $\kappa=0.5$. A variant adds the jet $|\eta|$ to test for production information, of which section~\ref{sec:production} finds little at truth level. Each student is also trained on the labels alone, as a baseline without a teacher. The students are trained with the Adam optimizer~\cite{Kingma:2014vow} for at least $4\times10^3$ steps, and the state with the lowest cross entropy on the labeled validation jets is kept. For each student, in addition to the metrics of section~\ref{sec:metrics}, we quote the number of parameters and the inference time per jet on one CPU core. An implementation in FPGA firmware with \texttt{hls4ml}~\cite{Duarte:2018ite} is left for future work.

\subsection{Metrics}
\label{sec:metrics}

For every model, the longitudinal score is
\begin{equation}
s = \frac{p_{\mathrm{L}}}{p_{\mathrm{L}}+p_{\mathrm{T}}}\,,
\label{eq:score}
\end{equation}
where $p_{\mathrm{L}}$ and $p_{\mathrm{T}}$ are the predicted probabilities of the two polarization classes. For the two-class networks (section~\ref{sec:finetuning}), $s=p_{\mathrm{L}}$. The BDT and the distilled students are evaluated in the same way; for the proxy, a smaller $p_\theta$ counts as more longitudinal. We quote the AUC and the rejection $1/\epsT$ of transverse jets at longitudinal efficiencies of $\epsL=0.5$ and $0.3$. The quoted uncertainties are the standard deviation over the trainings; for the difference between two taggers, the standard errors of the two means and the statistical uncertainty from bootstrap resampling of the test jets are added in quadrature.

As a reference, we use an ideal tagger that measures the true $\acth$ of each jet and nothing else. Its likelihood ratio, $2(1-c^2)/(1+c^2)$ from eq.~\eqref{eq:folded}, falls monotonically with $c=\acth$, so the ideal tagger simply cuts on $c$. Its AUC and working points have closed forms (appendix~\ref{app:analytic}). For the untruncated shapes, the ideal tagger has an AUC of $11/16=0.6875$ and a rejection $1/\epsT=3.7$ (6.5) at $\epsL=0.5$ (0.3). With the containment edge at $c=0.936$ found for the selected $R=0.8$ jets (section~\ref{sec:containment}), these values drop to an AUC of 0.660 and rejections of 3.4 (5.9); with the edge at $c=0.977$ found with $R=1.5$, they are 0.677 and 3.6 (6.3). A tagger that does better must use information other than $\acth$, such as the identity of the quark prong (section~\ref{sec:limits}) or the production kinematics. In $W$+jet events, little such information reaches the jet (section~\ref{sec:production}): the jet pseudorapidity, through which the production kinematics could enter, adds less than 0.005 to the AUC of the true decay angle, and the decay azimuth adds nothing. On our samples, the ideal tagger is therefore the ceiling for every jet tagger that does not identify the quark prong, including the distilled students. On the test jets, the true decay angle reaches the same AUCs, 0.660 and 0.677.

\section{Results}
\label{sec:results}

This section reports the six-class evaluation of the pretraining (section~\ref{sec:res-pretraining}), a comparison of the initializations and of two baselines without neural networks with $10^3$ to $5\times10^4$ labeled jets per class (section~\ref{sec:res-scan}), and the distillation into small students with and without unlabeled jets (section~\ref{sec:res-kd}). Section~\ref{sec:res-outlook} discusses the remaining questions.

\subsection{Pretraining}
\label{sec:res-pretraining}

\begin{figure}[!htb]
\centering
\includegraphics[width=0.65\textwidth]{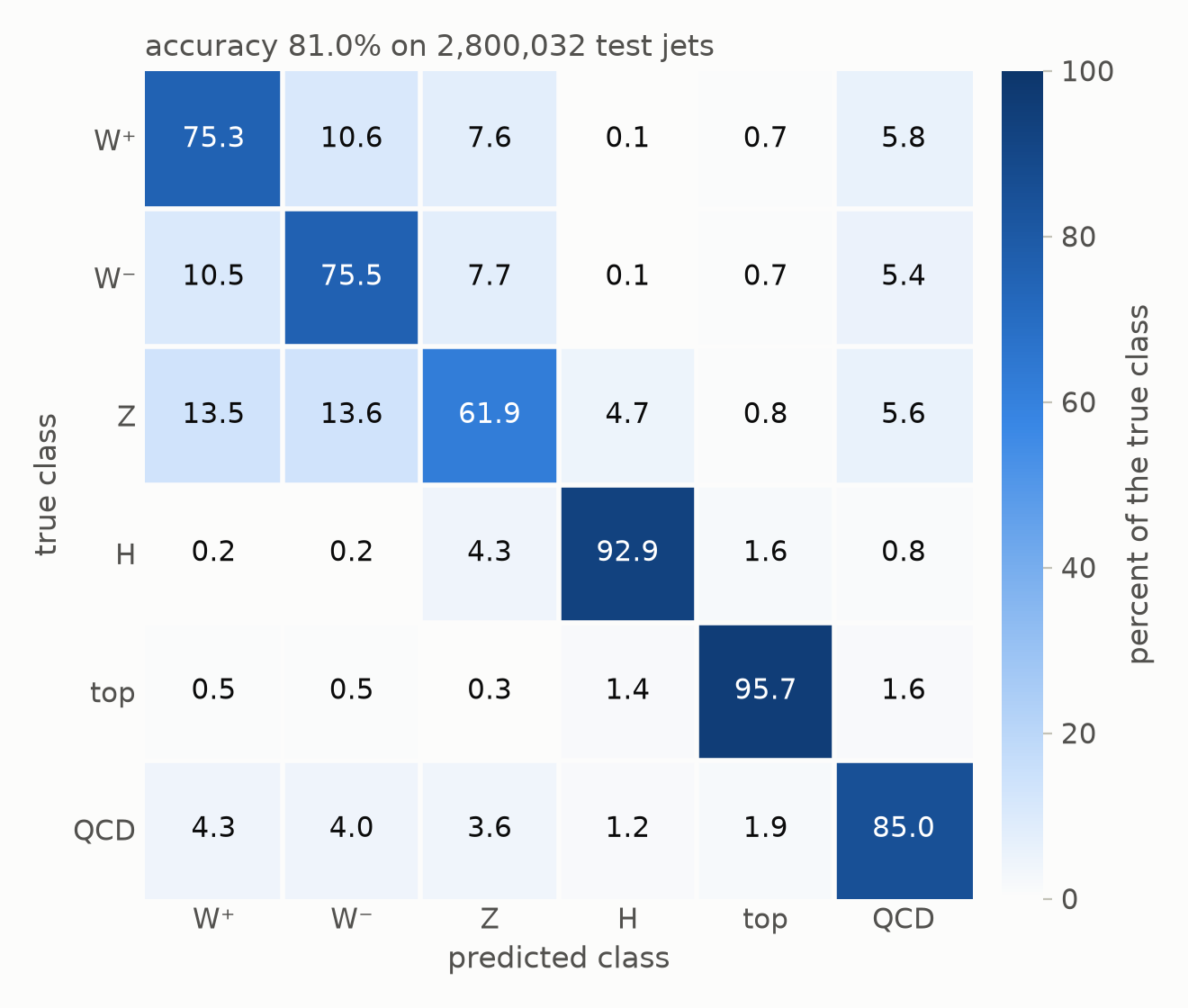}
\caption{Six-class confusion matrix of \rung{pt\_cls} on the test split ($2.8\times10^6$ jets). Each row shows the percentages of the jets of a given true class that are assigned to each class by the largest predicted probability.}
\label{fig:confusion}
\end{figure}

The process-classification pretraining \rung{pt\_cls} assigns 81.0\% of the $2.8\times10^6$ test jets to their true class (figure~\ref{fig:confusion}). Top-quark and Higgs-boson jets are recognized best, at 95.7\% and 92.9\%, followed by QCD jets at 85.0\%; most misclassifications occur among the electroweak bosons. The jet-mass peaks of $W$ and $Z$ bosons lie 10\GeV{} apart and overlap across most of their width (figure~\ref{fig:mass}). Of the $Z$-boson jets, 27\% are misclassified as $W$ bosons (split evenly between the two charges), whereas only 7.6\% of the $W^+$ and 7.7\% of the $W^-$ jets are misclassified as $Z$ bosons. This asymmetry presumably arises because the $Z$ class lies between the two $W$ classes in jet charge, so that a $Z$-boson jet whose charge fluctuates in either direction resembles a $W$ boson. The network also distinguishes $W^+$ from $W^-$ jets, which differ primarily in charge: of the $W$-boson jets assigned to either $W$ class, 88\% are assigned correctly. The mutual misclassification between $Z$ and Higgs-boson jets (4.7\% and 4.3\%) presumably involves $Z\to b\bar b$ decays. These make up about a fifth of hadronic $Z$ decays and share the heavy-flavor signature of $H\to b\bar b$. Between 5.4\% and 5.8\% of the $W$ and $Z$ jets are misidentified as QCD jets, while 12\% of the QCD jets are mistaken for $W$ or $Z$ bosons.

\begin{figure}[!htb]
\centering
\includegraphics[width=\textwidth]{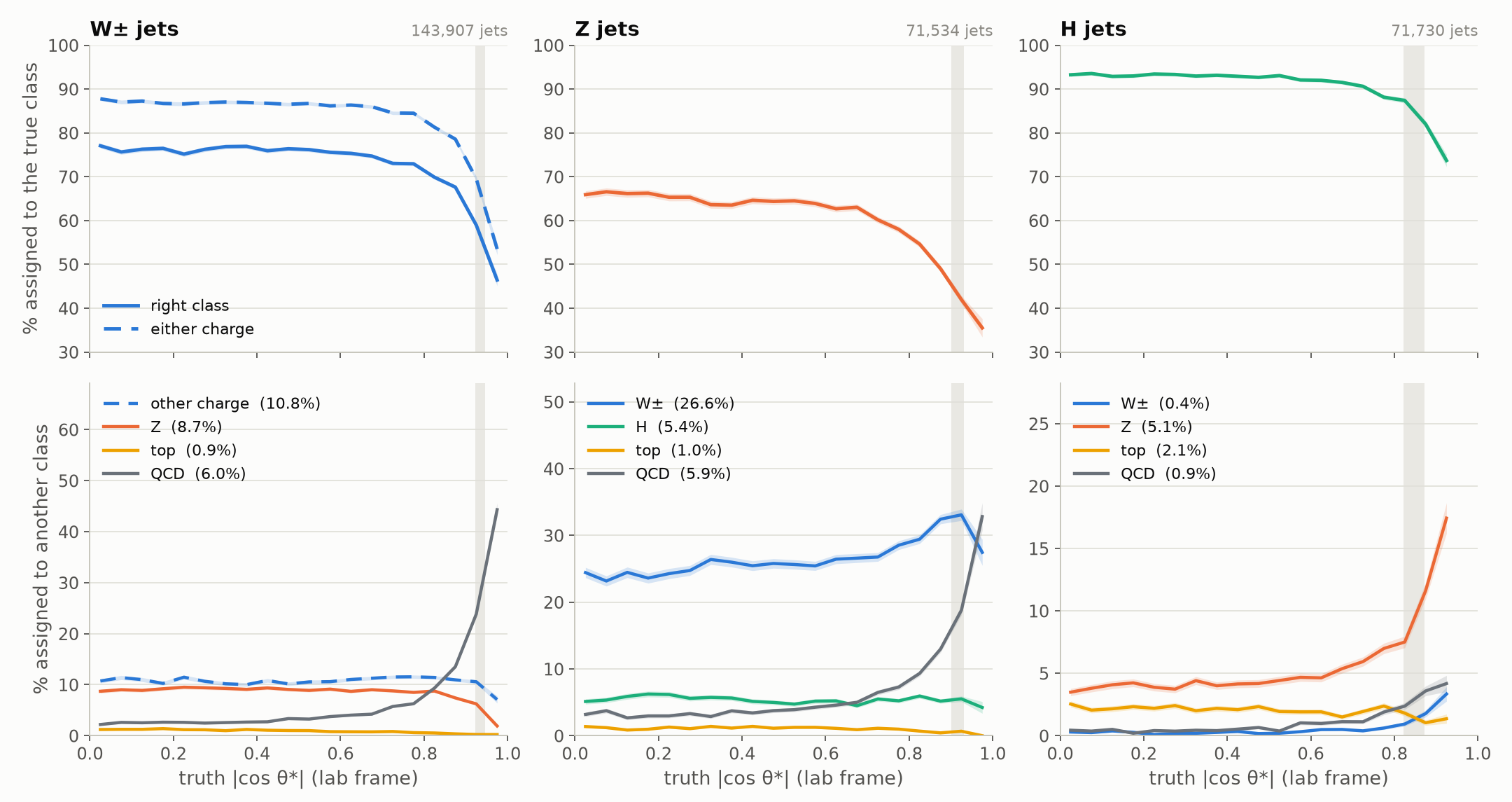}
\caption{Assignments of \rung{pt\_cls} as a function of the true $\acth$ in the laboratory frame, for $W^\pm$ (left), $Z$ (middle) and Higgs-boson jets (right) with $500<\pT<600\GeV$ in the test split. Top: fraction assigned to the true class (for $W^\pm$, solid lines: the correct charge; dashed lines: either charge). Bottom: fractions assigned to the other classes, with the angle-integrated rates in the legend. The shaded bands show the binomial standard errors, and the gray vertical bands mark the containment edge of eq.~\eqref{eq:cmax} between 500 and 600\GeV. Bins with fewer than 200 jets are not shown.}
\label{fig:confusion-angle}
\end{figure}

These misclassifications depend on the decay angle. Figure~\ref{fig:confusion-angle} shows them as a function of the true $\acth$ in the laboratory frame, in which the containment edge applies (section~\ref{sec:containment}), for jets with $500<\pT<600\GeV$. In this range the edge moves little: from 0.92 to 0.95 for the $W$ boson, from 0.90 to 0.93 for the $Z$ boson and from 0.82 to 0.87 for the Higgs boson. The inclusive sample shows the same pattern. Up to $\acth\approx0.6$, the classification rates vary by at most two percentage points. Toward the containment edge, jets migrate to lighter classes. The fraction of $W$-boson jets assigned to either $W$ class falls from 87\% below $\acth=0.3$ to 67\% between 0.9 and 1.0, while the fraction misidentified as QCD jets rises from 2.5\% to 27\%. Correct $Z$-boson assignments fall from 66\% to 41\% (the jets migrate to the $W$ and QCD classes), and correct Higgs-boson assignments fall from 93\% to 72\% (mostly to the $Z$ class). Two effects contribute to this behavior. First, near the edge, the softer quark lies close to the jet boundary, reducing the median jet mass between the first and last bins from 85.5 to 77.3\GeV{} for $W$ bosons and from 121 to 94\GeV{} for Higgs bosons. Among the Higgs-boson jets in the last bin that retain a mass of 105--135\GeV, only 3.3\% are misidentified as $Z$ bosons. Second, a very asymmetric decay resembles the soft wide-angle splitting that typically gives a QCD jet its mass. Even for jet masses between 70 and 95\GeV, the fraction of $W$-boson jets taken for QCD jets rises from 1.1\% to 17\%. The charge assignment, however, does not depend on the decay angle: the two $W$ classes exchange 10--11\% of their jets in every bin. Where the decay angle does not help the classification, the pretraining objective gives the network no reason to retain it. In the polarization task, however, the frozen networks from \rung{pt\_cls} and \rung{pt\_both} lead the frozen generic network by 0.004--0.005 in AUC with $2\times10^4$ and $5\times10^4$ labeled jets per class (section~\ref{sec:res-scan}), so their features retain more of the decay angle than the generic ones.

\subsection{Polarization tagging with $10^3$ to $5\times10^4$ labeled jets per class}
\label{sec:res-scan}

The polarized samples contain $3\times10^5$ $W^+$+jet events per helicity in 15 files per class (section~\ref{sec:generation}). Six files per class provide the training jets, two the validation jets and seven the test jets, so that no random sequence of the generation, shower or detector simulation is shared between the splits. The jets of each radius are matched in $\pT$ by resampling, as in section~\ref{sec:construction}. From the training split, we draw nested, class-balanced training sets of $10^3$, $2\times10^3$, $5\times10^3$, $10^4$, $2\times10^4$ and $5\times10^4$ jets per class, each contained in the larger ones, and from the validation split, in the same way, a validation set as large as each training set, up to the $1.8\times10^4$ ($R=0.8$) and $2.1\times10^4$ ($R=1.5$) jets per class that this split holds. All taggers are evaluated on the same test jets, 61,282 per class with $R=0.8$ and 72,317 with $R=1.5$, on which the statistical uncertainty of an AUC is 0.0015 (appendix~\ref{app:analytic}) and the jet $\pT$ alone separates the two classes with an AUC of 0.496. We train nine networks with $R=0.8$: one from scratch and, from the generic pretraining and from each of our three pretrainings, one fine-tuned with all weights and one with frozen transformer blocks. With $R=1.5$, we train three: one from scratch and two from the generic pretraining, fully fine-tuned and frozen (the boson pretrainings use $R=0.8$ jets). Each network is trained three times on every training set, 216 trainings in all. \OL fixes no random seed, so the trainings differ in their random initialization and in the order of the training jets. One BDT is trained on each training set; the proxy and the true decay angle need no training. Differences between two taggers are quoted with a total uncertainty that adds in quadrature the statistical uncertainty, from bootstrap resampling of the test jets on which both are evaluated, and the standard errors of the two means over the trainings.

\begin{figure}[tbp]
\centering
\includegraphics[width=\textwidth]{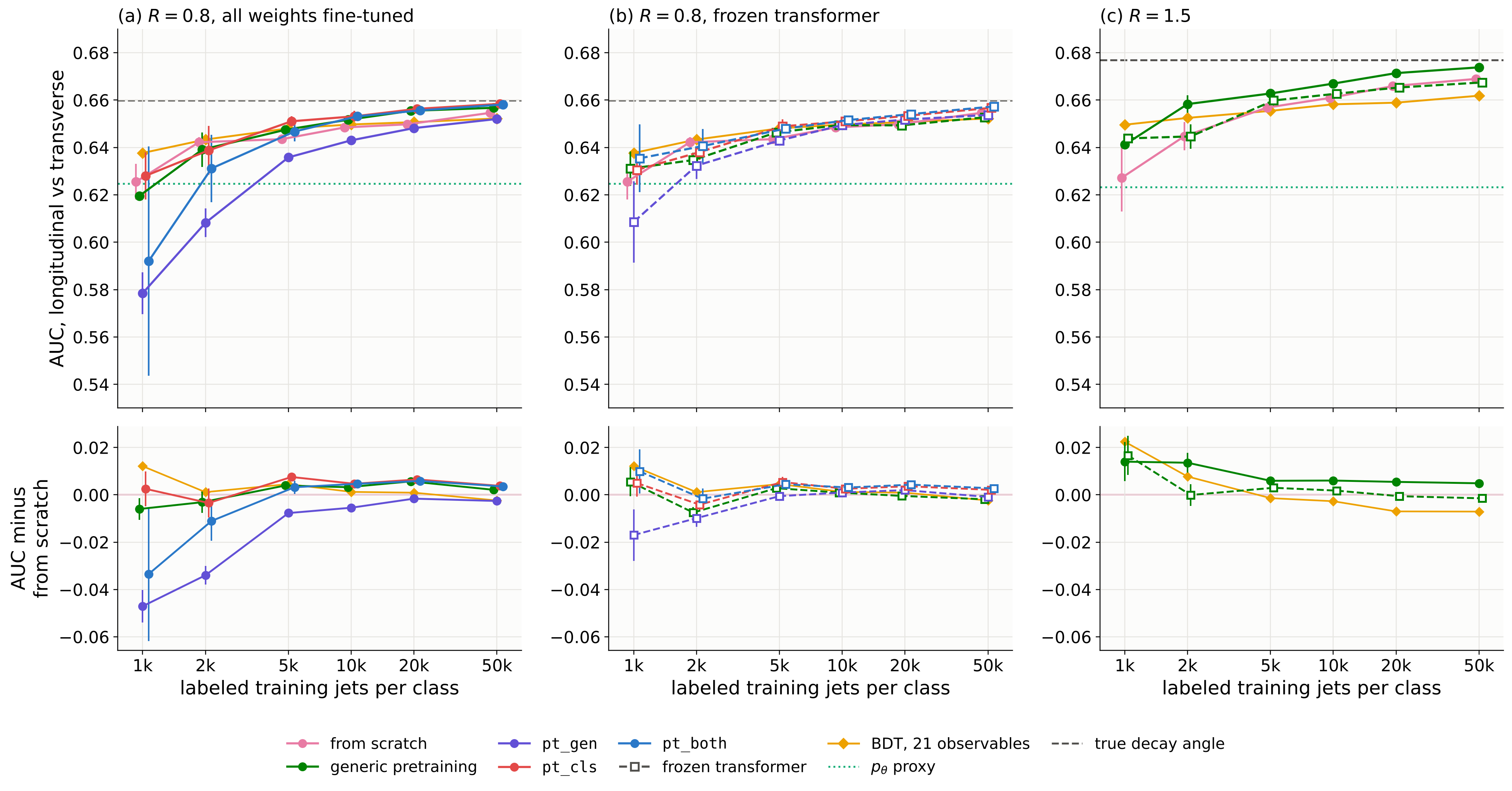}
\caption{AUC against the number of labeled training jets per class on the test jets of the label scan: (a) $R=0.8$, networks with all weights fine-tuned, (b) $R=0.8$, networks with frozen transformer blocks (open squares, dashed lines), with training from scratch and the BDT repeated, and (c) $R=1.5$. The markers show the mean of the three trainings and the error bars their standard deviation; the statistical uncertainty of each AUC, 0.0015, is smaller than most markers. The dashed lines mark the true decay angle and the dotted lines the proxy. The lower panels show the gain over training from scratch with the same labels.}
\label{fig:scan}
\end{figure}

Figure~\ref{fig:scan} and table~\ref{tab:scan} show the AUC of every tagger against the number of labeled jets. With $5\times10^4$ labeled jets per class, every network comes close to the ideal decay-angle tagger. For $R=0.8$, the best one, fine-tuned from \rung{pt\_cls}, reaches 0.658, $0.0014\pm0.0010$ below the true decay angle on the same jets, and for $R=1.5$ the generically pretrained network reaches 0.674, $0.0031\pm0.0010$ below it. Their transverse rejections at $\epsL=0.5$, 3.36 and 3.56, equal those of the true decay angle, 3.34 and 3.55, within their statistical uncertainty of about 1\%. The weakest networks with this many labels stay 0.008 ($R=0.8$) and 0.009 ($R=1.5$) below the true decay angle. With $10^3$ labeled jets per class, in contrast, the BDT is the best tagger for both radii: the best network falls $0.002\pm0.008$ below it with $R=0.8$ and $0.006\pm0.001$ with $R=1.5$. The BDT improves least with more labels, by 0.015 ($R=0.8$) and 0.012 ($R=1.5$) from $10^3$ to $5\times10^4$ jets per class. The networks overtake it with $5\times10^3$ ($R=0.8$) or $2\times10^3$ ($R=1.5$) jets per class, and with $5\times10^4$, the best network leads it by $0.006\pm0.001$ and $0.012\pm0.001$. With few labels, single trainings of the networks scatter. With $10^3$ and $2\times10^3$ jets per class, the run-to-run spread reaches 0.017, and 0.048 for the network fully fine-tuned from \rung{pt\_both}, whose three trainings on $10^3$ jets per class end at 0.537, 0.627 and 0.612; from $10^4$ jets per class on, it is at most 0.002.

\begin{table}[t]
\centering
\footnotesize
\setlength{\tabcolsep}{4pt}
\begin{tabular}{@{}lcccccc@{}}
\toprule
 & \multicolumn{6}{c}{Labeled training jets per class} \\
\cmidrule(l){2-7}
Model & $10^3$ & $2\times10^3$ & $5\times10^3$ & $10^4$ & $2\times10^4$ & $5\times10^4$ \\
\midrule
\multicolumn{7}{@{}l}{$R=0.8$} \\
$p_\theta$ proxy & \multicolumn{6}{c}{0.625} \\
BDT, 21 observables & 0.638 & 0.643 & 0.648 & 0.650 & 0.651 & 0.652 \\
\rung{scratch\_adam} & 0.626(8) & 0.642(2) & 0.644(2) & 0.648(1) & 0.650(1) & 0.655(1) \\
\rung{ft\_s} & 0.620(1) & 0.639(7) & 0.647(1) & 0.652(1) & 0.655(1) & 0.657(1) \\
\rung{ft\_s}, frozen & 0.631(7) & 0.635(3) & 0.646(1) & 0.649(1) & 0.649(2) & 0.653(1) \\
\rung{ft\_ours} $\leftarrow$ \rung{pt\_gen} & 0.578(9) & 0.608(6) & 0.636(1) & 0.643(1) & 0.648(1) & 0.652(1) \\
\rung{ft\_ours} $\leftarrow$ \rung{pt\_cls} & 0.628(10) & 0.639(10) & 0.651(2) & 0.653(2) & 0.656(1) & 0.658(1) \\
\rung{ft\_ours} $\leftarrow$ \rung{pt\_both} & 0.592(48) & 0.631(14) & 0.647(4) & 0.653(1) & 0.656(1) & 0.658(1) \\
\rung{ft\_ours} $\leftarrow$ \rung{pt\_gen}, frozen & 0.609(17) & 0.632(6) & 0.643(1) & 0.649(1) & 0.652(1) & 0.654(1) \\
\rung{ft\_ours} $\leftarrow$ \rung{pt\_cls}, frozen & 0.631(6) & 0.638(3) & 0.649(3) & 0.651(2) & 0.653(1) & 0.657(1) \\
\rung{ft\_ours} $\leftarrow$ \rung{pt\_both}, frozen & 0.635(14) & 0.641(7) & 0.648(1) & 0.652(2) & 0.654(1) & 0.657(1) \\
True decay angle & \multicolumn{6}{c}{0.660} \\
\midrule
\multicolumn{7}{@{}l}{$R=1.5$} \\
$p_\theta$ proxy & \multicolumn{6}{c}{0.623} \\
BDT, 21 observables & 0.650 & 0.652 & 0.655 & 0.658 & 0.659 & 0.662 \\
\rung{scratch\_adam} & 0.627(14) & 0.645(6) & 0.657(2) & 0.661(1) & 0.666(1) & 0.669(1) \\
\rung{ft\_s} & 0.641(1) & 0.658(4) & 0.663(1) & 0.667(1) & 0.671(1) & 0.674(1) \\
\rung{ft\_s}, frozen & 0.644(1) & 0.645(5) & 0.660(2) & 0.663(1) & 0.665(1) & 0.667(1) \\
True decay angle & \multicolumn{6}{c}{0.677} \\
\bottomrule
\end{tabular}
\caption{AUC of longitudinal against transverse $W^+$ jets on the test jets of the label scan (61,282 jets per class with $R=0.8$ and 72,317 with $R=1.5$), against the number of labeled training jets per class. Each network is trained three times; the AUC is the mean, with the standard deviation of the three trainings (the run-to-run spread) in units of the last digit in parentheses. One BDT is trained per training set; the proxy and the true decay angle, the ideal decay-angle tagger, need no training. The statistical uncertainty of an AUC is about 0.0015, and that of a difference between two taggers on the same test jets 0.0003--0.0017. The boson pretrainings use $R=0.8$ jets, so their fine-tunes exist only for $R=0.8$.}
\label{tab:scan}
\end{table}

Pretraining helps modestly, and only once a few thousand labels are available. Fully fine-tuned, the generically pretrained network falls $0.006\pm0.005$ below training from scratch with $10^3$ jets per class for $R=0.8$, but leads it by 0.002--0.006 from $5\times10^3$ jets per class on. For $R=1.5$, it leads by 0.005--0.006 from $5\times10^3$ jets per class on and by 0.014 with $10^3$ and $2\times10^3$, where the trainings from scratch scatter most. Interpolated linearly in $\log N$ between the measured training-set sizes, training from scratch needs 1.7 to 3.2 times as many labels to reach the AUC of the generically pretrained network trained on $10^3$ to $10^4$ jets per class (from $5\times10^3$ on for $R=0.8$). With $5\times10^4$ labels, it does not reach the AUC of the pretrained network trained on $2\times10^4$. Keeping the transformer blocks frozen helps only with the fewest labels. With $10^3$ jets per class, the frozen generic network leads the fully fine-tuned one by $0.012\pm0.004$ ($R=0.8$) and $0.003\pm0.001$ ($R=1.5$), but with $2\times10^4$ and $5\times10^4$ jets per class, full fine-tuning is better by 0.004--0.006, and the frozen network falls 0.0015--0.0019 behind training from scratch.

Pretraining on boson jets changes this picture only slightly when the network is fully fine-tuned. The network from \rung{pt\_cls} is then the best one for $R=0.8$ from $5\times10^3$ labeled jets per class on, ahead of the generically pretrained network by $0.0036\pm0.0015$ with $5\times10^3$ jets per class and by 0.0008--0.0017 (0.9 to 1.9 standard deviations) with more; training from scratch would need $2.5\times10^4$ jets per class, five times as many, to reach its AUC with $5\times10^3$. With frozen transformer blocks, the comparison isolates what the pretrained representation retains, and here the boson pretrainings are clearly ahead: with $2\times10^4$ and $5\times10^4$ labeled jets per class, the frozen networks from \rung{pt\_cls} and \rung{pt\_both} lead the frozen generic network by 0.004--0.005, or 3 to 9 standard deviations. The objective matters. Fully fine-tuned, the generative objective alone gives the weakest network at every training-set size, from $0.041\pm0.005$ below the generic network with $10^3$ labels to $0.005\pm0.001$ with $5\times10^4$; frozen, it matches or slightly exceeds the frozen generic network from $10^4$ labels on. Combined with the classification objective (\rung{pt\_both}, the recipe of the generic pretraining), it behaves like \rung{pt\_cls} from $10^4$ labels on, but its full fine-tunes on fewer labels are the most unstable of all.

\begin{figure}[tbp]
\centering
\begin{minipage}{0.92\textwidth}
\includegraphics[width=\linewidth]{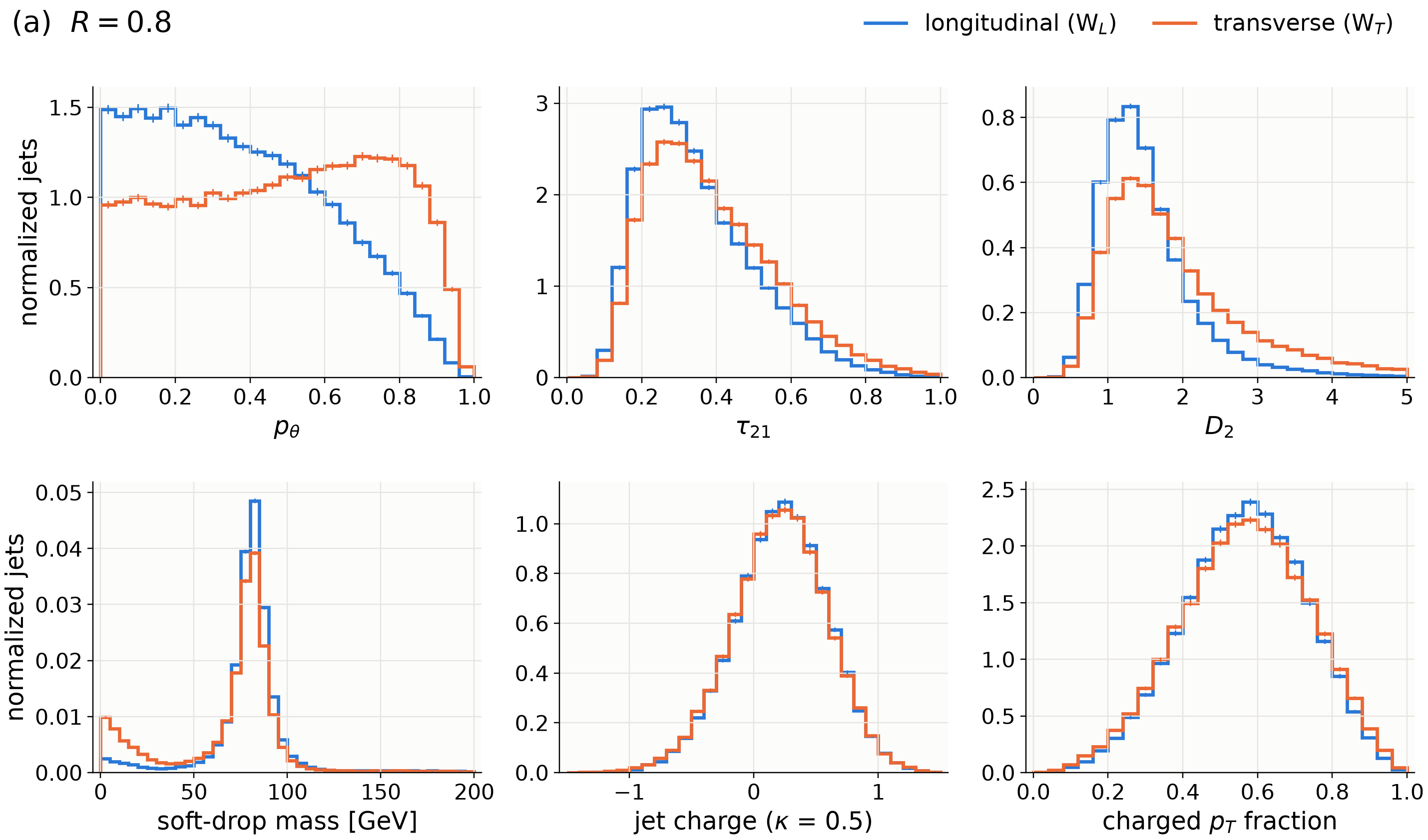}\par\vspace{10pt}
\includegraphics[width=\linewidth]{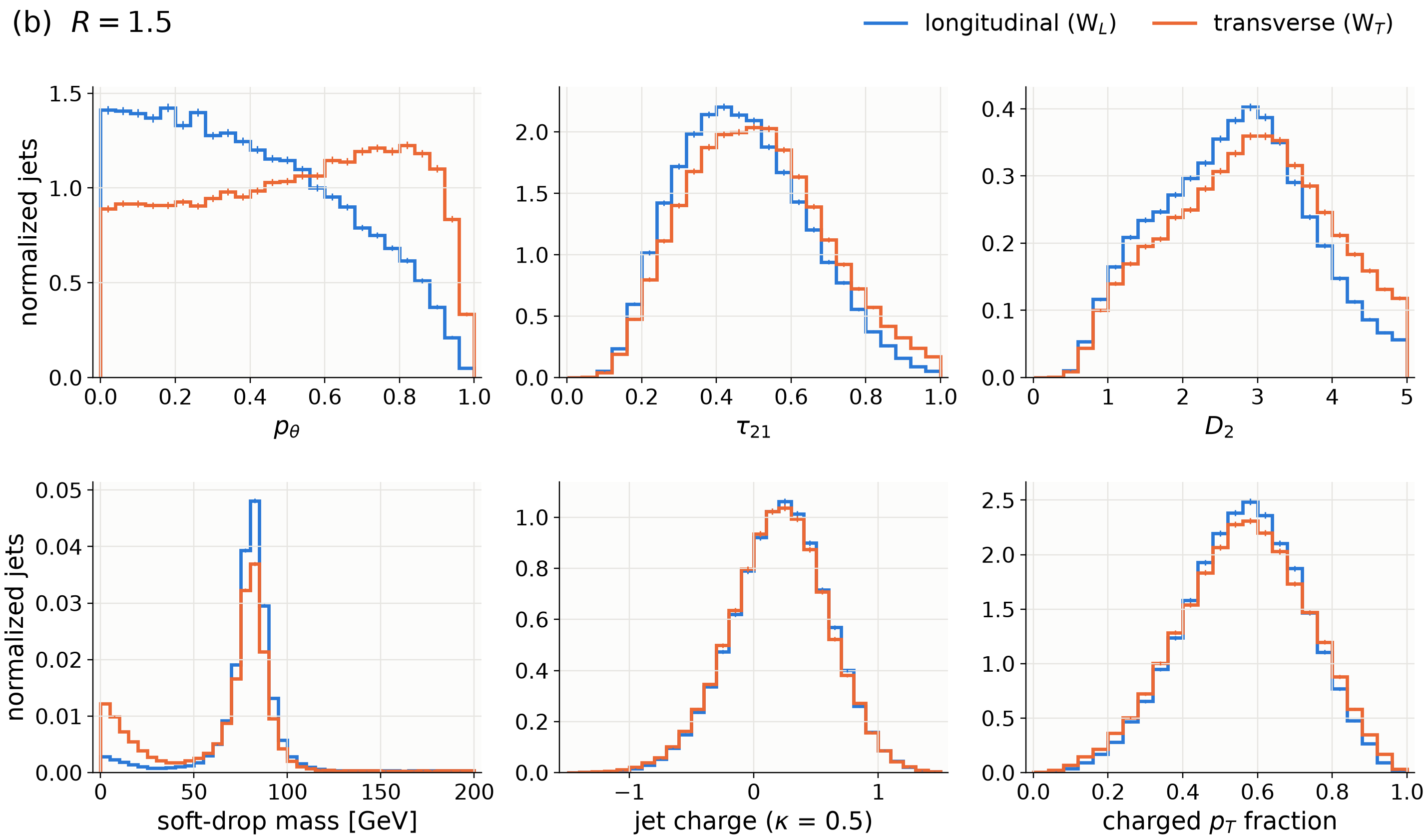}
\end{minipage}
\caption{Six of the BDT inputs for the longitudinal and transverse test jets of the label scan, with $R=0.8$ (a) and $R=1.5$ (b): the energy sharing $p_\theta$ of the two exclusive-$k_t$ subjets (eq.~\eqref{eq:ptheta}), the $N$-subjettiness ratio $\tau_{21}$, $D_2$, the soft-drop mass ($\beta=0$, $z_{\mathrm{cut}}=0.1$), the jet charge ($\kappa=0.5$) and the charged fraction of the jet $\pT$. Each distribution is normalized to unit area in the range shown, and the error bars are statistical; with $R=1.5$, the $D_2$ distribution continues beyond 5.}
\label{fig:substructure}
\end{figure}

The larger radius raises the AUC of the trained taggers as it raises the ceiling. With $5\times10^4$ labeled jets per class, the generically pretrained network gains 0.017 from $R=0.8$ to $R=1.5$, as much as the true decay angle (0.017), training from scratch 0.014 and the BDT 0.010, while the proxy does not improve (0.625 and 0.623), presumably because the wider jet collects more of the underlying event, which smears the energies of its two subjets and also shifts $\tau_{21}$ and $D_2$ to larger values. From $5\times10^3$ labeled jets per class on, the most important input of the BDT is $D_2$ with $R=0.8$ and the soft-drop mass ($\beta=0$, $z_{\mathrm{cut}}=0.1$) with $R=1.5$. The soft-drop mass falls when the softer prong fails the grooming condition, roughly for $\acth>0.8$, a region that the larger radius populates up to its edge. As a result, transverse jets reach low soft-drop masses far more often than longitudinal ones, and more often with $R=1.5$ than with $R=0.8$ (figure~\ref{fig:substructure}).

\begin{figure}[tbp]
\centering
\includegraphics[width=\textwidth]{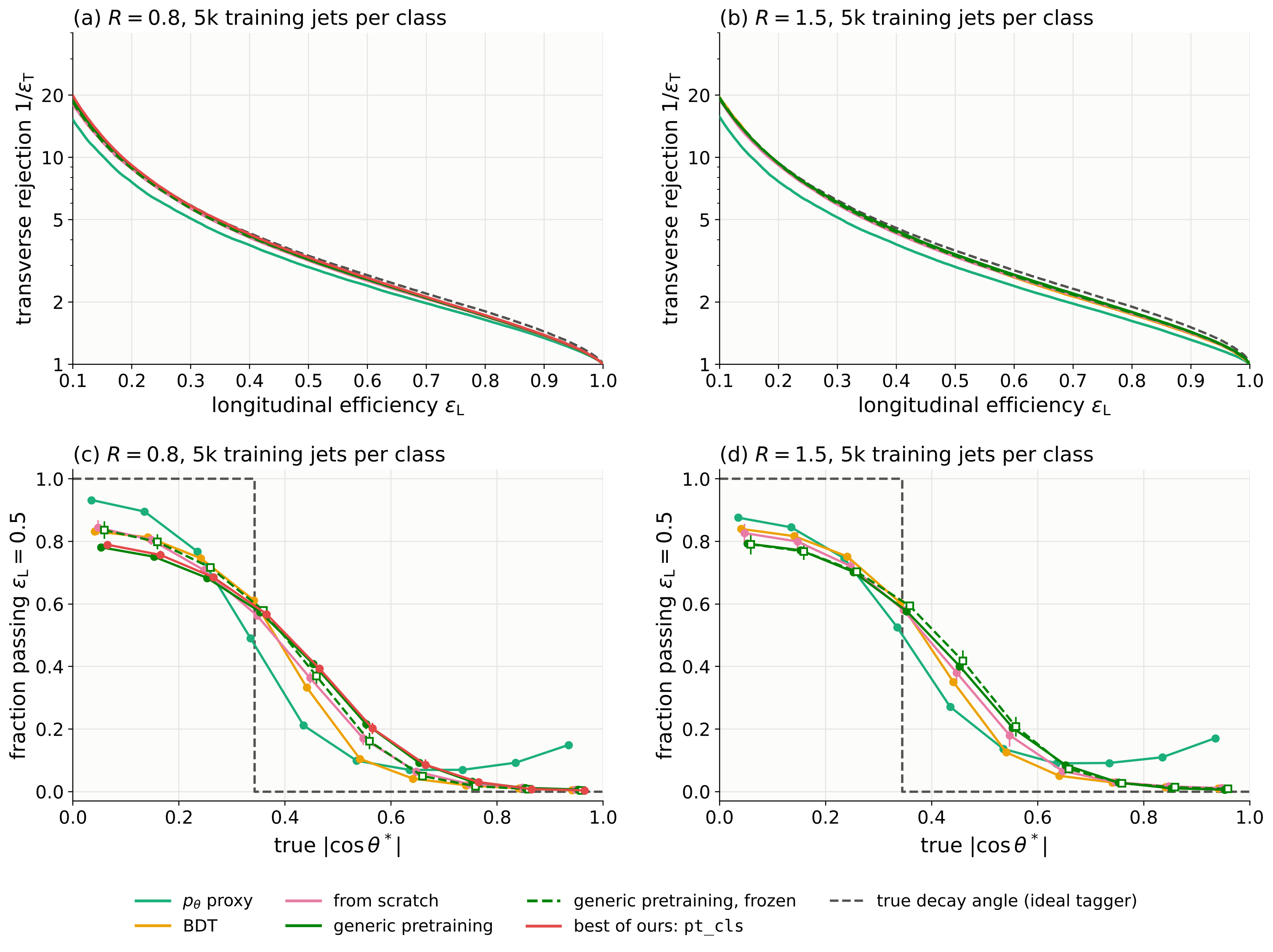}
\caption{Taggers trained on $5\times10^3$ labeled jets per class, on the test jets of the label scan, for $R=0.8$ (left) and $R=1.5$ (right). (a, b) Transverse rejection against longitudinal efficiency; for the networks, the line is the mean of the three trainings and the band spans their range. (c, d) Fraction of the test jets that pass the working point of each tagger at $\epsL=0.5$, in bins of the true $\acth$ in the partonic center-of-mass frame; the error bars are the standard deviation of the three trainings for the networks and binomial for the proxy and the BDT, and the points of different taggers are shifted slightly apart. The dashed lines show the true decay angle, the ideal decay-angle tagger, which in (c, d) passes every jet below the median $\acth$ of the longitudinal jets.}
\label{fig:scan-curves}
\end{figure}

Figure~\ref{fig:scan-curves} shows, for $5\times10^3$ labeled jets per class, the ROC curves of the taggers and, more instructively, how each of them responds to the decay angle: the fraction of test jets that pass its working point at $\epsL=0.5$, as a function of the true $\acth$. For the ideal decay-angle tagger, this curve is a step at the median $\acth$ of the longitudinal jets, 0.34 for both radii. The proxy comes closest to the step in the core, accepting 91\% of the jets with $\acth<0.2$ for $R=0.8$ and 86\% for $R=1.5$. However, it also passes 12\% and 14\% of the most asymmetric decays ($\acth>0.8$), presumably because the softer prong then carries too little energy for the two exclusive-$k_t$ subjets to measure the energy sharing reliably. The BDT rejects these jets, passing 0.7\% and 1.3\% of them, at the cost of a softer turn-on in the core (82\% and 83\% below 0.2). With $10^3$ labeled jets per class, the networks turn on more gradually still, accepting only 62--77\% ($R=0.8$) and 68--71\% ($R=1.5$) of the most symmetric decays ($\acth<0.1$). With more labels, they approach the BDT in the core and reject the asymmetric decays as cleanly: with $5\times10^4$ labeled jets per class, they accept 84--88\% of the jets below 0.1 and pass at most 1\% of those above 0.8.

With $6\times10^4$ and $7\times10^4$ test jets per class, the response curves of the longitudinal and the transverse jets, taken separately, also test whether a tagger responds to the polarization only through the decay angle. For the proxy, they nearly agree ($\chi^2$ of 7 and 17 for 10 bins with $R=0.8$ and $R=1.5$). For the BDT and every network, they do not ($\chi^2$ of 28--292, averaged over the trainings): at the same $\acth$ below 0.6, these taggers pass 1.2--4.9 percentage points more longitudinal than transverse jets on average. Two effects contribute. First, the networks identify the quark prong in part, and more so with more labels. Among the longitudinal jets, whose quark and antiquark are emitted symmetrically, the networks pass fewer jets at the same $\acth$ when the up-type quark is the harder prong, the configuration favored by the mostly left-handed transverse bosons (section~\ref{sec:limits}). With $5\times10^4$ labeled jets per class, the difference is 11--13 percentage points for the networks fully fine-tuned from the generic pretraining, from \rung{pt\_cls} and from \rung{pt\_both}, 5 points for training from scratch, 1.5--7 points for the frozen networks and 1.2 points for the network fully fine-tuned from \rung{pt\_gen}; with $10^3$ labeled jets per class, it is at most 5 points. For the proxy and the BDT, it is at most 1.5 points, with the opposite sign. Second, longitudinal and transverse jets of the same $\pT$ and $\acth$ differ slightly in their kinematics: the longitudinal bosons carry 7--8\GeV{} less transverse momentum, and their jets are 1.0\GeV{} heavier with $R=0.8$ but 1.4\GeV{} lighter with $R=1.5$. Presumably, the steeper spectrum of the longitudinal bosons (figure~\ref{fig:jets}) lets more jets with an upward fluctuation of the jet response migrate into a given bin of jet $\pT$. Neither effect carries much information: in a cross-validated classifier on the test jets, the jet mass, the number of constituents and the jet $\pT$ add 0.003 ($R=0.8$) and less than 0.001 ($R=1.5$) to the AUC of the true $\acth$, and the sign of $\cth$, which a perfect identification of the up-type quark would provide, adds 0.013 for both radii, close to the 0.014 and 0.013 of table~\ref{tab:limits}. Both are properties of the training samples, whose helicity mixture and $\pT$ spectra depend on the production process, so a tagger that uses them may respond differently in other processes; the transfer tests therefore check the response at a fixed decay angle (section~\ref{sec:res-outlook}).

\subsection{Distillation with unlabeled jets}
\label{sec:res-kd}

The teacher is the generically pretrained network fine-tuned on $5\times10^3$ labeled jets per class with $R=0.8$, the one of its three trainings with the lowest validation loss; it was fixed before the scan, as the baseline of table~\ref{tab:ladder}. On the test jets, it reaches an AUC of 0.648, as its three trainings do on average (0.647); the network fine-tuned from \rung{pt\_cls} on the same labels reaches 0.651. The transfer set consists of the same labeled training jets, alone or together with a pool of $2\times10^5$ unpolarized $W^+$ jets with $500\le\pT<1000\GeV$ from the pretraining production. Each student is trained three times in each mode, and table~\ref{tab:kd} lists the results.

\begin{table}[t]
\centering
\footnotesize
\setlength{\tabcolsep}{4pt}
\begin{tabular}{@{}llccccc@{}}
\toprule
 & & & CPU time & & \multicolumn{2}{c}{$1/\epsT$} \\
\cmidrule(l){6-7}
Model & Training & Parameters & per jet & AUC & $\epsL=0.3$ & $\epsL=0.5$ \\
\midrule
Teacher: \rung{ft\_s} & labels & $3\times10^6$ & -- & 0.648 & 5.9 & 3.22 \\
\midrule
PFN & labels only & $1.4\times10^4$ & 55\,$\mu$s & 0.639(1) & 5.6 & 3.13 \\
PFN & distilled, labeled jets & $1.4\times10^4$ & 54\,$\mu$s & 0.644(1) & 5.7 & 3.20 \\
PFN & distilled, with the pool & $1.4\times10^4$ & 55\,$\mu$s & 0.647(1) & 5.8 & 3.24 \\
EFN & labels only & $1.3\times10^4$ & 48\,$\mu$s & 0.644(1) & 5.7 & 3.16 \\
EFN & distilled, labeled jets & $1.3\times10^4$ & 48\,$\mu$s & 0.646(1) & 5.7 & 3.19 \\
EFN & distilled, with the pool & $1.3\times10^4$ & 49\,$\mu$s & 0.646(1) & 5.7 & 3.19 \\
Observables & labels only & $4.9\times10^3$ & 0.2\,$\mu$s & 0.643(1) & 5.7 & 3.20 \\
Observables & distilled, labeled jets & $4.9\times10^3$ & 0.2\,$\mu$s & 0.645(1) & 5.8 & 3.22 \\
Observables & distilled, with the pool & $4.9\times10^3$ & 0.2\,$\mu$s & 0.645(1) & 5.7 & 3.23 \\
Observables and $|\eta|$ & distilled, with the pool & $4.9\times10^3$ & 0.2\,$\mu$s & 0.645(2) & 5.8 & 3.20 \\
\midrule
BDT, 21 observables & labels & -- & -- & 0.648 & 5.8 & 3.25 \\
$p_\theta$ proxy & -- & -- & -- & 0.625 & 5.1 & 2.95 \\
\bottomrule
\end{tabular}
\caption{Distillation of the generically pretrained network fine-tuned on $5\times10^3$ labeled jets per class with $R=0.8$ into small students, each trained three times: on the labels alone, distilled on the labeled training jets alone, or distilled with a pool of $2\times10^5$ unlabeled $W^+$ jets in addition. The AUC is the mean of the three trainings, with their standard deviation in units of the last digit in parentheses, and the rejections are means as well; on the 61,282 test jets per class, the statistical uncertainty of an AUC is 0.0015, and that of the difference between two students 0.0005--0.0012. The CPU time per jet is that of the forward pass on one core of an AMD EPYC 7763 processor, in batches of 1,024 jets; for the observable networks, it excludes the computation of the observables, and the teacher was not timed. The BDT trained on the same labels and the proxy are shown for comparison.}
\label{tab:kd}
\end{table}

Every student gains from the teacher. Distilled on the labeled jets alone, the PFN gains $0.005\pm0.001$ in AUC over the same network trained on the labels alone, and the EFN and the observable network gain $0.002\pm0.001$ each. With the pool, the PFN gains another $0.003\pm0.001$, $0.008\pm0.001$ in total, and reaches 0.647, within 0.001 of its teacher, whereas the EFN and the observable network gain nothing more. The distilled students thus match the BDT trained on the same labels (0.648) to within 0.003 and exceed the transformer trained from scratch on them (0.644), with 200 to 600 times fewer parameters than their teacher. On one CPU core, the PFN and the EFN take 48--55\,$\mu$s per jet and the observable network 0.2\,$\mu$s once its inputs are computed. Adding the jet $|\eta|$ to the observable network changes nothing ($0.0001\pm0.0012$), so the teacher passes on no production information, as section~\ref{sec:production} predicts. The unlabeled jets help the student that sees the most detail of each constituent; for the others, distillation on the labeled jets already transfers what the teacher adds.

\subsection{The remaining questions}
\label{sec:res-outlook}

The label scan answers three of the questions of table~\ref{tab:claims} for the jets of $W^+$+jet events. Pretraining on boson decays improves on generic pretraining, by a few thousandths in AUC when the network is fully fine-tuned and by 0.004--0.005 when its transformer blocks are frozen and $2\times10^4$ or more labeled jets per class are available. The gain is not largest with the fewest labels: with $10^3$ labeled jets per class, the BDT is the best tagger and single trainings of the networks scatter by up to 0.05 in AUC. The objective matters: classification retains the decay angle best and the generative objective least. The distillation answers the last question: a pool of unlabeled jets helps the particle-flow student, and every distilled student matches its teacher. Larger training sets would change little, since the best networks come within 0.003 of the ideal decay-angle tagger with $5\times10^4$ labeled jets per class, and a jet tagger can exceed it only by identifying the quark prong, which is worth at most about 0.014 (section~\ref{sec:limits}). A more stable recipe for full fine-tuning on few labels, for instance a smaller learning rate or a frozen first stage, might let the pretrained networks lead with $10^3$ labeled jets per class as well, and the regression objective \rung{pt\_reg} and the eight-class variant remain to be compared.

The tagger trained on $W$+jet events can then be applied unchanged to samples that never enter the training. The matrix-element samples of section~\ref{sec:diboson}, with longitudinal and transverse bosons in VBS $W^\pm W^\pm jj$ and $W^\pm Zjj$ production and in $W^\pm Z$ and $W^+W^-$ production, need only showering and detector simulation, which takes a few core-hours; the decay of a heavy resonance into $WW$ needs one more sample of similar size. Because the networks respond slightly differently to longitudinal and transverse jets of the same decay angle, through the quark prong and small kinematic differences between the samples (section~\ref{sec:res-scan}), their response at a fixed $\acth$ can be compared between the training and the transfer samples. In addition, a template fit with shapes from the polarized samples can measure the longitudinal fraction in the unpolarized $W^+$+jet events of the pretraining production, to be compared with the generator value. The $W/Z$-versus-QCD classifier, fine-tuned from the same pretrained network on the pretraining jets and decorrelated from both the jet mass and $p_\theta$, can be tested against the mixture envelope of eq.~\eqref{eq:envelope}: after the selection, the true $\acth$ distribution of $W$ jets must remain a mixture of the truncated shapes. New polarized samples are needed only for a fine-tuning window of 300--500\GeV{} with $R=1.5$ (section~\ref{sec:conclusions}).

\section{Conclusions}
\label{sec:conclusions}

We have built a pretraining set of $1.0\times10^{7}$ unpolarized jets from boosted $W^\pm$, $Z$ and Higgs bosons, top quarks and QCD, with identical transverse-momentum spectra over 300--1500\GeV. We pretrained a point-edge transformer on it with three objectives and, after fine-tuning it to separate longitudinal from transverse $W$ bosons, compared it with the same network after generic pretraining and trained from scratch, with $10^3$ to $5\times10^4$ labeled jets per class, three trainings per network and training set, and test samples of $6\times10^4$ ($R=0.8$) and $7\times10^4$ ($R=1.5$) jets per class. Our main conclusions are as follows.

\begin{enumerate}
\item \textbf{Jet containment sets a ceiling for any hadronic polarization tagger.} Requiring both decay quarks inside the jet removes the most asymmetric decays, $\acth>\cmax$ of eq.~\eqref{eq:cmax}, the region in which transverse decays are most distinctive. For $W$ bosons at 500\GeV, $\cmax=0.92$ with $R=0.8$. An ideal tagger with access to the true decay angle therefore reaches an AUC of only about 0.66 with $R=0.8$ and 0.68 with $R=1.5$ for jets between 500 and 1000\GeV, compared to 0.6875 without containment. Even without containment, the separation is weak. The likelihood ratio of $\acth$ never exceeds 2, so no selection can more than double the longitudinal fraction of a sample, and the decay angles of two bosons per event together stay below an AUC of 0.9 (table~\ref{tab:limits}). The same acceptance correlates the process labels of a boson pretraining set with the decay angle.

\item \textbf{Within a jet, the decay angle is all there is.} In $W$+jet events, the jet pseudorapidity adds less than 0.005 to the AUC of the true decay angle, and the decay azimuth adds nothing, because the transverse bosons are not linearly polarized. The ideal decay-angle tagger is therefore the ceiling for jet taggers that cannot tell the quark prong from the antiquark prong. Because the transverse bosons are mostly left-handed, identifying that prong perfectly would raise the AUC by about 0.013.

\item \textbf{The rest of the event carries more information.} The production angle, rebuilt from the jet and the leading other jet after detector simulation, raises the AUC of the ideal decay-angle tagger by 0.034 with $R=0.8$ and by 0.030 with $R=1.5$, to 0.69 and 0.71. This is nearly as much as the true production angle adds (gains measured against a cross-validated classifier on the decay angle alone). Adding the production angle raises the AUC of every trained tagger by 0.03--0.05 and the transverse rejection at $\epsL=0.5$ by 14--19\%, so that the best taggers exceed the ideal decay-angle tagger alone. An event-level tagger can therefore exceed the jet-level ceiling. In diboson events at matrix-element level, the production kinematics raise the AUC for longitudinal against transverse pairs from 0.62--0.64 for the hadronic decay angle to 0.78 in $W^+W^-$ and 0.92 in $W^\pm Z$ production, and in vector-boson scattering to 0.90 in $W^\pm Z jj$ and 0.92 in $W^\pm W^\pm jj$ production, where the information lies mostly in the tagging jets. These gains are specific to the process: such a tagger must be trained on polarized samples of the process to which it will be applied, and it gives up the generality of a tagger that uses only the decay.

\item \textbf{With $5\times10^4$ labeled jets per class, the best networks come within 0.003 of the ideal decay-angle tagger.} They fall 0.001 ($R=0.8$) and 0.003 ($R=1.5$) short of the true decay angle in AUC and equal its transverse rejection at $\epsL=0.5$. With few labels, the BDT on 21 observables is the best tagger: no network is ahead of it with $10^3$ labeled jets per class, and the networks overtake it with $2\times10^3$ to $5\times10^3$. The proxy $p_\theta$, which requires no training, stays 0.035 ($R=0.8$) and 0.054 ($R=1.5$) below the true decay angle; it resolves the angle best for symmetric decays but passes some of the most asymmetric ones. At a fixed decay angle, the networks also respond to the identity of the quark prong, the more so the more labels they see, and to small kinematic differences between the two samples, neither of which is worth more than 0.014 in AUC.

\item \textbf{Pretraining helps modestly, once a few thousand labels are available.} Fully fine-tuned, the generically pretrained network leads training from scratch by 0.002--0.006 in AUC from $5\times10^3$ labeled jets per class on, which training from scratch makes up only with 1.7 to 3.2 times as many labels; with $10^3$ labels, single trainings scatter by more than the gain. Pretraining on boson decays with a classification objective adds a few thousandths more. Fully fine-tuned, the network from \rung{pt\_cls} is the best one for $R=0.8$ from $5\times10^3$ labels on, 0.001--0.004 above the generic network. With frozen transformer blocks, the networks from \rung{pt\_cls} and \rung{pt\_both} lead the frozen generic network by 0.004--0.005 with $2\times10^4$ labels or more, so our pretraining keeps more of the decay angle in its representation. The generative objective alone gives the weakest network, and frozen transformer blocks help only with the fewest labels.

\item \textbf{Small students can stand in for the teacher.} Distilled with a pool of $2\times10^5$ unlabeled jets, a particle-flow network with 200 times fewer parameters than its teacher, trained with $5\times10^3$ labeled jets per class, gains 0.008 in AUC over the same network trained on the labels alone and comes within 0.001 of its teacher. An energy-flow network and a network on eight jet observables gain 0.001--0.003, all of it from distillation on the labeled jets. The students take 48--55\,$\mu$s per jet on one CPU core, or 0.2\,$\mu$s for the network on eight jet observables, and match the BDT trained on the same labels.
\end{enumerate}

Two questions of table~\ref{tab:claims} remain: whether the tagger transfers to vector-boson scattering and other processes, and whether a decorrelated $W/Z$-versus-QCD selection preserves the decay angle. Both can be answered with samples that already exist: the unpolarized jets of the pretraining production and the matrix-element samples of section~\ref{sec:diboson} after detector simulation (section~\ref{sec:res-outlook}). A more stable recipe for full fine-tuning on few labels could also let the pretrained networks lead where the BDT is still the best tagger.

Several extensions are natural. Using $R=1.5$ jets makes a fine-tuning window of 300--500\GeV{} practical, as the containment edge for $W$ bosons remains at 0.94 even down to 300\GeV. This lower range is closer to the kinematics of SM vector-boson scattering. An event-level tagger trained on the target process could add the production kinematics on top of any jet tagger (section~\ref{sec:diboson}). Diboson processes could be added to the pretraining, at the expense of a fully independent transfer test, and the distilled students could run in FPGA firmware for real-time selection~\cite{Duarte:2018ite,Liu:2023kd}. These conclusions should eventually be checked with full detector simulation, pile-up and higher-order event generators. On data, a first look at ATLAS Open Data (appendix~\ref{app:opendata}) finds that the simulation describes most distributions within the statistical uncertainties. It also shows that applying such taggers would need a data format that keeps the jet constituents, together with constraints on the large-radius jet mass scale and resolution from a top control region.

\acknowledgments

We thank the ATLAS Collaboration for the Open Data release used in appendix~\ref{app:opendata}. This research used resources of the National Energy Research Scientific Computing Center, a DOE Office of Science User Facility supported by the Office of Science of the U.S.\ Department of Energy under Contract No.\ \mbox{DE-AC02-05CH11231} using NERSC award \mbox{HEP-ERCAP0037461}.

\appendix

\section{Analytic results}
\label{app:analytic}

This appendix outlines how the analytic results quoted in the text are derived, for perfectly measured angles and massless decay products. We write $c=\acth$, $c_m$ for the largest accepted value of $c$ ($c_m=1$ without containment), and
\begin{equation}
P_{\mathrm{L}}(c)=\frac{1-c^2}{\NL(c_m)}\,,\qquad P_{\mathrm{T}}(c)=\frac{1+c^2}{\NT(c_m)}\,,\qquad
\NL(c)=c-\frac{c^3}{3}\,,\qquad \NT(c)=c+\frac{c^3}{3}\,,
\label{eq:app-shapes}
\end{equation}
for the longitudinal and transverse shapes of eq.~\eqref{eq:folded}, normalized on $0\le c\le c_m$.

\paragraph{Truncated shapes.} The acceptance keeps a fraction $\frac32\NL(c_m)$ of the longitudinal and $\frac34\NT(c_m)$ of the transverse bosons, and their second moments become
\begin{equation}
\langle c^2\rangle_{\mathrm{L}}=\frac{c_m^3/3-c_m^5/5}{\NL(c_m)}\,,\qquad
\langle c^2\rangle_{\mathrm{T}}=\frac{c_m^3/3+c_m^5/5}{\NT(c_m)}\,.
\label{eq:app-moments}
\end{equation}
For $c_m=0.936$, these are 0.196 and 0.345, close to the selected jets with $R=0.8$ (0.194 and 0.345), and 99.4\% of the longitudinal and 90.7\% of the transverse bosons are accepted; for $c_m=0.977$, they are 0.199 and 0.380, close to the selected jets with $R=1.5$ (0.198 and 0.380), and 99.9\% and 96.6\% are accepted (section~\ref{sec:containment}). The longitudinal fraction of the accepted bosons is then $\fL'=(\langle c^2\rangle_{\mathrm{T}}-\langle c^2\rangle)/(\langle c^2\rangle_{\mathrm{T}}-\langle c^2\rangle_{\mathrm{L}})$, which replaces eq.~\eqref{eq:fL}.

\paragraph{The ideal decay-angle tagger.} The likelihood ratio
\begin{equation}
\Lambda(c)=\frac{P_{\mathrm{L}}(c)}{P_{\mathrm{T}}(c)}=\frac{1-c^2}{1+c^2}\,\frac{\NT(c_m)}{\NL(c_m)}
\label{eq:app-lr}
\end{equation}
falls monotonically with $c$, so a cut $c<c_*$ is the most powerful selection at every working point~\cite{Neyman:1933}. The AUC is the probability that a longitudinal boson has a smaller $c$ than a transverse one,
\begin{equation}
\mathrm{AUC}(c_m)=\int_0^{c_m}P_{\mathrm{T}}(t)\,\frac{\NL(t)}{\NL(c_m)}\,\mathrm{d}t
=\frac{c_m^2/2+c_m^4/6-c_m^6/18}{\NL(c_m)\,\NT(c_m)}\,,
\label{eq:app-auc}
\end{equation}
which is $11/16=0.6875$ for $c_m=1$, 0.660 for $c_m=0.936$ and 0.677 for $c_m=0.977$. The threshold for a longitudinal efficiency $\epsL$ solves $\NL(c_*)=\epsL\NL(c_m)$, a cubic equation whose root between 0 and 1 follows with $c_*=2\sin\alpha$ and $\sin3\alpha=3\sin\alpha-4\sin^3\alpha$:
\begin{equation}
c_*=2\sin\!\left[\frac13\arcsin\!\left(\frac32\,\epsL\NL(c_m)\right)\right],\qquad
\epsT=\frac{\NT(c_*)}{\NT(c_m)}\,.
\label{eq:app-wp}
\end{equation}
Without containment, $c_*=2\sin10^\circ=0.347$ at $\epsL=0.5$, the median $\acth$ of longitudinal bosons, and $1/\epsT=3.69$; at $\epsL=0.3$, $1/\epsT=6.49$. With the containment edge $c_m=\cmax(m,\pT)$ of eq.~\eqref{eq:cmax}, eqs.~\eqref{eq:app-auc} and~\eqref{eq:app-wp} give the curves of figure~\ref{fig:ceiling}.

\paragraph{The decay azimuth.} Integrated over $c$ with $\int_0^1(1-c^2)\,\mathrm{d}c=\frac23$, eq.~\eqref{eq:azimuth} gives $\langle\cos2\phi^*\rangle=\frac12|\rho_{+-}|\cos\delta$, and positivity of $\rho$, $|\rho_{+-}|^2\le\rho_{++}\rho_{--}\le\frac14$, bounds it by $\frac14$. In the Collins--Soper frame, with polar angle $\theta$ and azimuth $\phi$, the terms in $A_0$ and $A_2$ combine with the leading term into
\begin{equation}
(1+\cos^2\theta)+\frac{A_0}{2}\,(1-3\cos^2\theta)+\frac{A_2}{2}\,\sin^2\theta\cos2\phi
\;\xrightarrow{\;A_0=A_2=1\;}\; 1+\sin^2\theta\cos^2\phi=1+(\hat n\cdot\hat x)^2\,,
\label{eq:app-cs}
\end{equation}
a transverse distribution about $\hat x$ without linear polarization.

\paragraph{Combining independent variables.} If the score of each class is normally distributed, with the same width for both classes and means $d'$ widths apart, the AUC is $\Phi(d'/\sqrt2)$. Two independent variables, combined through the product of their likelihood ratios, then reach
\begin{equation}
\mathrm{AUC}_{12}=\Phi\!\left(\sqrt{\frac{d_1'^2+d_2'^2}{2}}\right),\qquad d_i'=\sqrt2\,\Phi^{-1}(\mathrm{AUC}_i)\,.
\label{eq:app-binormal}
\end{equation}
On top of the true decay angle of the selected jets (AUC 0.660), the jet $|\eta|$ (0.519) adds 0.001. The production angle rebuilt from the jet and the leading other jet (0.594 with $R=0.8$ and 0.591 with $R=1.5$ for all selected jets, 0.595 and 0.592 on the test jets of the label scan) adds 0.044 to the weakest tagger of table~\ref{tab:scan} with $R=0.8$ and 0.024 to the true decay angle, and 0.028 and 0.020, respectively, with $R=1.5$. For two bosons with an AUC of $4/5$ each, eq.~\eqref{eq:app-binormal} gives 0.883, against the exact 0.885 derived below.

\paragraph{The factor of two.} Without containment, $\Lambda(c)\le\Lambda(0)=2$; with it, $\Lambda(0)=\NT(c_m)/\NL(c_m)$ is smaller, for example 1.82 for $c_m=0.936$. Any selection $S$ therefore keeps a fraction $P(S|\mathrm{L})=\int_S\Lambda P_{\mathrm{T}}\,\mathrm{d}c\le2P(S|\mathrm{T})$ of the longitudinal bosons and turns a longitudinal fraction $\fL$ into
\begin{equation}
\fL^S=\frac{\fL\,P(S|\mathrm{L})}{\fL\,P(S|\mathrm{L})+(1-\fL)\,P(S|\mathrm{T})}\le\frac{2\fL}{1+\fL}<2\fL\,.
\label{eq:app-double}
\end{equation}

\paragraph{The sign of the decay angle.} With $x=\cth$ measured from the quark, here the up-type quark of a $W^+$ boson, left-handed $W$ bosons decay as $(1+x)^2$ and right-handed ones as $(1-x)^2$. If a fraction $a$ of the transverse bosons is left-handed, their distribution is $\frac38(1+x^2+2bx)$ with $b=2a-1$, and within the containment edge, $|x|<c_m$,
\begin{equation}
\langle x\rangle_{\mathrm{T}}=\frac{2bc_m^2}{3+c_m^2}\,,\qquad P(x>0)=\frac12+\frac{3bc_m}{2(3+c_m^2)}\,.
\label{eq:app-signed}
\end{equation}
The transverse test jets, with $\langle x\rangle=0.164$ for $c_m=0.936$ and 0.175 for $c_m=0.977$, thus have $a=0.68$ for both radii, and the quark should lie in the forward hemisphere in 63\% of them, as observed (section~\ref{sec:limits}). The likelihood ratio, now proportional to $(1-x^2)/(1+x^2+2bx)$, is largest at $x_*=(\sqrt{1-b^2}-1)/b$, where, without containment, it reaches
\begin{equation}
\Lambda_{\max}=\frac{2}{\sqrt{1-b^2}}=\frac{1}{\sqrt{a(1-a)}}\,,
\label{eq:app-lrmax}
\end{equation}
2.14 for $a=0.68$; it is unbounded only if the transverse bosons all have one helicity. Since this likelihood ratio is not monotonic in $x$, the AUCs with the sign in table~\ref{tab:limits} are obtained by ordering the bosons in it and integrating numerically.

\paragraph{One transverse helicity.} If the transverse bosons are all left-handed, the energy fraction of the quark, $z=(1+x)/2$ in the boosted limit, follows $6z(1-z)$ for longitudinal and $3z^2$ for transverse bosons. Their likelihood ratio, $2(1-z)/z$, falls monotonically with $z$, and with the cumulative distribution $3z^2-2z^3$ of longitudinal bosons,
\begin{equation}
\mathrm{AUC}_1=\int_0^1 3z^2\,(3z^2-2z^3)\,\mathrm{d}z=\frac45\,.
\label{eq:app-auc1}
\end{equation}
Two bosons that decay independently, both longitudinal against both transverse, are best separated by the product of their likelihood ratios, $4w_1w_2$ with $w=(1-z)/z$, so that $\mathrm{AUC}_2=P(X>1)$ for $X=w_{\mathrm{L}1}w_{\mathrm{L}2}/(w_{\mathrm{T}1}w_{\mathrm{T}2})$, the ratio of the products in a longitudinal and a transverse event. The variable $w$ follows beta-prime distributions with the Mellin transforms $\langle w_{\mathrm{L}}^s\rangle=\Gamma(2+s)\,\Gamma(2-s)$ and $\langle w_{\mathrm{T}}^s\rangle=\frac12\Gamma(1+s)\,\Gamma(3-s)$, so that, with $|\Gamma(1+it)|^2=\pi t/\sinh\pi t$, the characteristic function of $\ln X$ is
\begin{equation}
\chi(t)=\big\langle X^{it}\big\rangle=\frac14\,(1+t^2)^2\,\big[(1+it)(2+it)\big]^2\left(\frac{\pi t}{\sinh\pi t}\right)^4.
\label{eq:app-charfun}
\end{equation}
The inversion formula of Gil-Pelaez~\cite{GilPelaez:1951}, $P(\ln X>0)=\frac12+\frac1\pi\int_0^\infty\mathrm{Im}\,\chi(t)\,\mathrm{d}t/t$, leads with $u=\pi t$ to the integrals
\begin{equation}
I_n=\int_0^\infty\frac{u^n}{\sinh^4u}\,\mathrm{d}u=\frac{n!}{3\cdot2^{n-2}}\,\big[\zeta(n-2)-\zeta(n)\big]\,,
\label{eq:app-In}
\end{equation}
obtained by expanding $\sinh^{-4}u$ in powers of $e^{-2u}$, and gives
\begin{equation}
\mathrm{AUC}_2=\frac12+\frac{3}{2\pi^2}\left[2I_4+\frac{3}{\pi^2}\,I_6-\frac{1}{\pi^6}\,I_{10}\right]=\frac32-\frac{24\pi^2}{385}=0.8848\,.
\label{eq:app-auc2}
\end{equation}
The same steps for one boson reproduce $4/5$. The remaining decay-angle entries of table~\ref{tab:limits} follow from the shapes of eq.~\eqref{eq:helicity}, truncated where stated, by numerical integration, with the bosons or events ordered in their likelihood ratio.

\paragraph{Statistical precision.} The statistical uncertainty of an AUC $A$ measured on $n_{\mathrm{L}}$ longitudinal and $n_{\mathrm{T}}$ transverse jets follows from the approximation of Hanley and McNeil,
\begin{equation}
\sigma^2(A)=\frac{A(1-A)+(n_{\mathrm{L}}-1)(Q_1-A^2)+(n_{\mathrm{T}}-1)(Q_2-A^2)}{n_{\mathrm{L}}\,n_{\mathrm{T}}}\,,
\label{eq:app-hm}
\end{equation}
with $Q_1=A/(2-A)$ and $Q_2=2A^2/(1+A)$. For AUCs of 0.5--0.69, it gives 0.0014--0.0016 with the 61,282 and 72,317 test jets per class of the label scan and 0.0009--0.0011 with the $1.3\times10^5$ to $1.6\times10^5$ jets per class of section~\ref{sec:production}, close to the uncertainties from bootstrap resampling of the test jets. For a rejection, the threshold is set on the longitudinal jets and fluctuates as well; since $\mathrm{d}\epsT/\mathrm{d}\epsL=1/\Lambda(c_*)$,
\begin{equation}
\sigma^2(\epsT)\simeq\frac{\epsT(1-\epsT)}{n_{\mathrm{T}}}+\frac{\epsL(1-\epsL)}{\Lambda(c_*)^2\,n_{\mathrm{L}}}\,,
\label{eq:app-rej}
\end{equation}
which, for the ideal tagger at $\epsL=0.3$ with $6\times10^4$ jets per class, amounts to a relative uncertainty of about 1\% on $1/\epsT$.

\paragraph{The three-class validation loss.} With equal numbers of longitudinal, transverse and QCD jets, a classifier that separates $W$ bosons perfectly from QCD jets but knows nothing about the polarization assigns the probabilities $(\frac12,\frac12,0)$ to $W$ jets and $(0,0,1)$ to QCD jets, with a cross entropy of $\frac23\ln2=0.462$. The ideal decay-angle tagger reduces the cross entropy of $W$ jets from $\ln2$ to the conditional entropy
\begin{equation}
H=\int_0^{c_m}\frac{P_{\mathrm{L}}(c)+P_{\mathrm{T}}(c)}{2}\;h\!\left(\frac{P_{\mathrm{L}}(c)}{P_{\mathrm{L}}(c)+P_{\mathrm{T}}(c)}\right)\mathrm{d}c\,,\qquad h(p)=-p\ln p-(1-p)\ln(1-p)\,,
\label{eq:app-entropy}
\end{equation}
which is 0.647 for $c_m=0.936$, so that the loss falls only to $\frac23\times0.647=0.432$.

\section{Simulation details}
\label{app:sim}

\paragraph{Track smearing in \Delphes.} The tracks of the stock CMS card carry no impact-parameter uncertainties, which the published pretrained models take as inputs. Adding a track-smearing module with the CMS transverse and longitudinal impact-parameter resolutions but without a momentum resolution sets the track momentum resolution to zero; the energy-flow algorithm then rescales every track in a tower without a calorimeter deposit to zero transverse momentum, and 37\% of the energy-flow tracks then have $\pT=0$. We therefore switch off the three momentum-smearing modules of the card and move the CMS charged-hadron momentum resolution into the track-smearing module; afterward, one energy-flow track in 40,015 has $\pT=0$.

\paragraph{Energy-flow candidates.} The energy-flow output contains occasional objects at pseudorapidities of several hundred, up to about $10^3$. For $350\lesssim|\eta|\lesssim710$, their momenta are finite but so large that they pass a finiteness check and make the clustering fail numerically. Candidates with non-finite momenta, zero $\pT$ or $|\eta|>8$ are therefore removed before clustering, which leaves the constituents of jets with $|\eta|<2$ untouched.

\paragraph{Generator record.} \Pythia stores the hard-process boson (status 22) followed by copies created by recoils in the shower (for example statuses 44, 51 or 52 and 62), and the decay products are attached to the last copy. A boson that radiates a photon has two daughters, of which only the one with the boson's identity continues the chain. Following that daughter is necessary: stopping at the first branching leaves a quarter of the $W$ bosons without decay quarks. For top quarks, the chain continues through the $W$ boson to give three decay quarks.

\section{Data pipeline and computing}
\label{app:computing}

\paragraph{Input throughput.} The \OL data loader reads one jet at a time from HDF5 files. With gzip compression and chunks of 1,024 jets, every random read decompresses a whole chunk and costs 14.4\,ms, too slow to feed the GPUs. With chunks of eight jets, a read takes 0.13\,ms at the same compression ratio (29\%), and an uncompressed read 0.02\,ms. With eight-jet chunks, the pretraining epochs run at the speed expected from the measured time per step (403\,s for 2,000 training and 2,000 validation steps on four GPUs), so the loader is not the bottleneck.

\paragraph{Cost.} Table~\ref{tab:cost} summarizes the computing cost. The production and conversion ran on the CPU nodes of the Perlmutter system at NERSC, and the training ran on its A100 GPU nodes.

\begin{table}[h]
\centering
\small
\begin{tabular}{@{}>{\raggedright}p{0.40\textwidth}>{\raggedright\arraybackslash}p{0.54\textwidth}@{}}
\toprule
Step & Cost \\
\midrule
Matrix elements (\MGaMC) & 26--51\,s per $10^4$ events on one node in runs of $10^4$ events; 5--7\,s in runs of $10^5$ (top-up samples: $4\times10^6$ events in 35--46\,min) \\
Shower and detector (\Pythia, \Delphes) & 3--4\,min per $10^4$ events on one core \\
Jet clustering and conversion & about 1\,min per $10^4$ events \\
\Delphes output & about 1.8\,GB per $2\times10^4$ events; about 2\,TB in total \\
HDF5 jets & at most 4.6\,kB per jet before compression \\
Training step, batch of 128 (one A100) & 157\,ms and 11\,GB; 88\,ms and 6.4\,GB in mixed precision \\
Pretraining run (\rung{pt\_cls}, four A100) & 403\,s per epoch, 3.4\,h in total \\
GPU use & four pretraining runs $\approx$ 19 node-hours (projected); 216 fine-tuning runs of the label scan, with their evaluation, about 80 GPU-hours (estimated) \\
\bottomrule
\end{tabular}
\caption{Computing cost of the study. Each fine-tuning run uses one GPU.}
\label{tab:cost}
\end{table}

\section{Tagger inputs in ATLAS Open Data}
\label{app:opendata}

The ATLAS Open Data release for education and outreach~\cite{ATLAS:2025opendata} makes it possible to look at the inputs of such taggers in real collisions. We analyzed it offline, with the reconstructed objects it provides. Of the 36.1\,fb$^{-1}$ of 13\TeV{} data in the release, we use the 3.2\,fb$^{-1}$ recorded in 2015. Events contain a $W\to\ell\nu$ candidate and one large-radius jet as the candidate for a hadronically decaying vector boson: exactly one tight, isolated electron or muon with $\pT>30\GeV$ and $|\eta|<2.5$, matched to a single-lepton trigger, no further lepton above 10\GeV, a missing transverse momentum above 40\GeV{} and a leptonic $W$ candidate with $\pT>150\GeV$, and a leading large-radius jet with $250<\pT<400\GeV$, $|\eta|<2.0$ and $40<m_J<200\GeV$, separated from the lepton by $\Delta R>1.0$ and from the missing momentum by $\Delta\phi>1.0$. The signal region requires $65<m_J<105\GeV$ and no $b$-tagged jet inside the large-radius jet. In the simulation, $W$+jets production makes up 47\% of the signal region and top-quark pair production 43\%; $W$+jets production makes up 75\% of the sideband below the signal window, and top-quark pair production 57\% of the sideband above it. A top control region requires at least one $b$-tagged jet outside the large-radius jet; top-quark pair production makes up 88\% of it and $W$+jets production 5\%. In data, 22,883 events are selected, 8,706 of them in the signal region, where the expected ratio of $WV$ signal to background is 0.028.

The release is built from reconstructed physics objects, as suits its purpose, and does not store the constituents of large-radius jets. After its overlap removal, almost no small-radius jet remains within $\Delta R=1.0$ of the large-radius jet (60 of the 22,883 events keep one), so neither $p_\theta$ nor any other estimate of the hadronic decay angle can be formed in this format. In the simulated events, reweighted to either helicity state with their generator-level decay angle, a gradient-boosted classifier on $D_2$, the jet mass and the jet $\pT$ reaches an AUC of 0.605 ($D_2$ alone 0.596), 56\% of the separation of the ideal decay-angle tagger, whose AUC of 0.6875 the same labels reproduce.

\begin{figure}[tbp]
\centering
\includegraphics[width=0.6\textwidth]{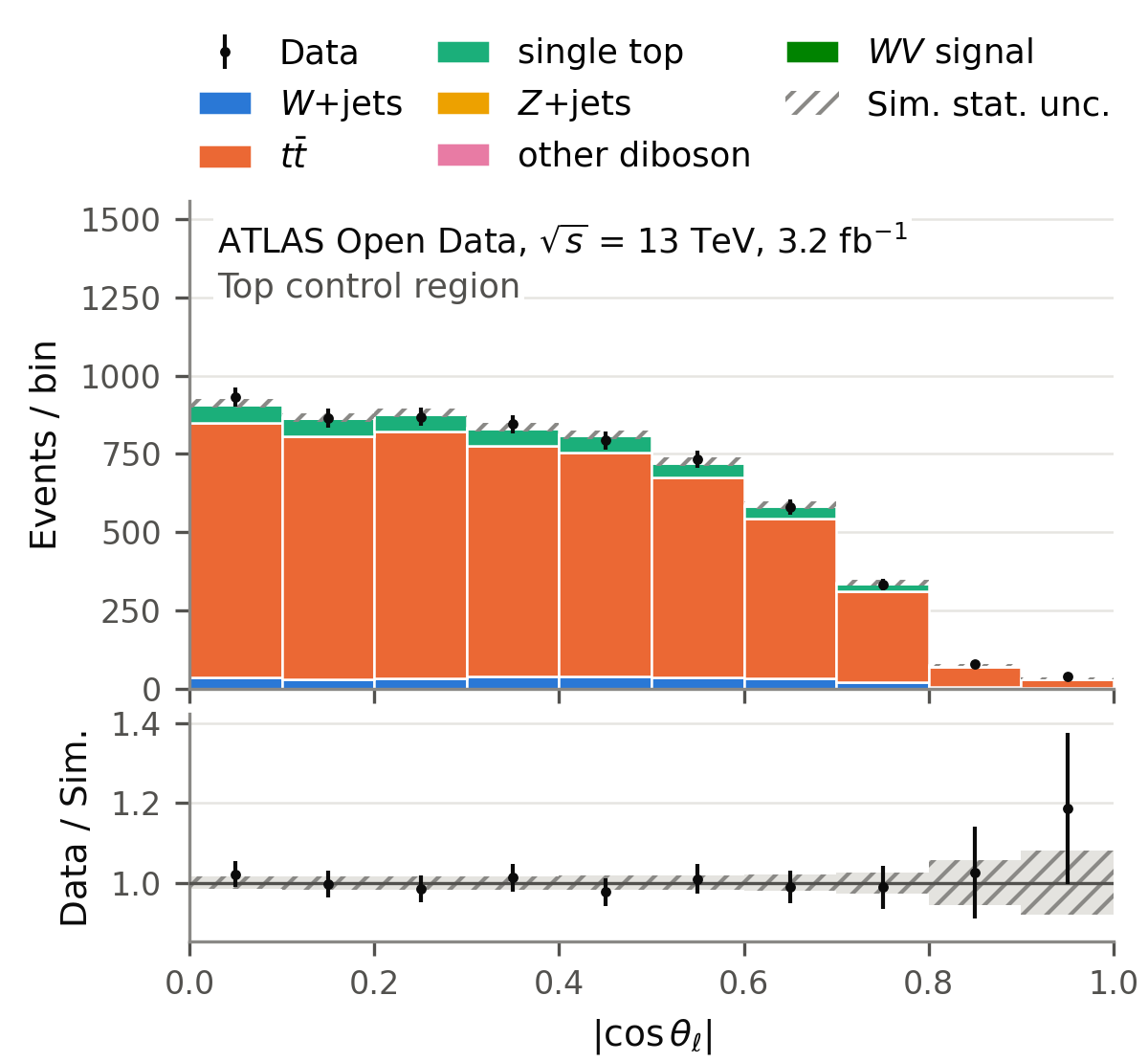}
\caption{Lepton decay angle in the top control region of ATLAS Open Data (13\TeV, 3.2\,fb$^{-1}$), the comparison with the best agreement between data and simulation (see text). The simulation, before any correction, is normalized to the number of data events (data/simulation yield ratio 0.94). The hatched bands show the statistical uncertainty of the simulation, in the ratio panel around unity, and the error bars that of the data; no systematic uncertainties are included.}
\label{fig:opendata}
\end{figure}

We compared the shapes of 14 distributions in data and simulation, in all selected events and in each of the four regions, with the simulation normalized to the data. The 67 comparisons with enough events for a test include the Poisson uncertainty of the data and the statistical uncertainty of the simulation, but no systematic uncertainties, which a dedicated analysis would derive. Figure~\ref{fig:opendata} shows the comparison with the best agreement, that is with the largest $p$-value, 0.99 ($\chi^2/\mathrm{ndf}=0.23$): the decay angle $\theta_\ell$ of the leptonic $W$ candidate in the top control region, with $\theta_\ell$ the angle between the charged lepton in the $W$ rest frame and the $W$ flight direction, computed with the longitudinal momentum of the neutrino from the $W$ mass constraint. Most of the other comparisons show no significant difference either (51 of the 67 have $p>0.01$), but some do. In all selected events, the signal region and the low-mass sideband, where $W$+jets production makes up about half of the events or more, the data exceed the simulation by 30--45\% at $|\cos\theta_\ell|>0.8$, and some other event-level distributions differ in shape as well; since they all agree in the top control region, these differences point to the modeling of $W$+jets production, or to multijet production with a misidentified lepton, which our simulation does not include, rather than to the reconstruction. The large-radius jet mass differs in all selected events, in the high-mass sideband and in the top control region, where its $W$-boson and top-quark peaks sit slightly lower in data than in the simulation, the pattern of a small difference in the jet mass scale and resolution. Since we analyze only a small part of the data, 3.2 of the 36.1\,fb$^{-1}$, this comparison is a first look; with the full dataset, the statistical uncertainties would shrink by a factor of 3.4.

Applying the taggers of this paper to data would therefore need a format that keeps the jet constituents or subjets, together with uncertainties in the scale and resolution of the large-radius jet mass, which the top control region is well suited to constrain in situ. The same region could calibrate the polarization response itself: in the top rest frame, about 70\% of the $W$ bosons from top-quark decays are longitudinal~\cite{Czarnecki:2010gb}.

\FloatBarrier
\bibliography{polfm_paper}

\end{document}